\documentclass[10pt,a4paper]{article}

\usepackage{graphicx,float,calc,ifthen}
\usepackage[fleqn]{amsmath}
\usepackage{authblk}
\usepackage{amsmath,bm}
\usepackage{hyperref}
\usepackage{lmodern} 
     \usepackage{url}
     
     \usepackage{mdframed}
\usepackage[most]{tcolorbox}
  \usepackage{subcaption}
  \usepackage{csvsimple}
  
\graphicspath{{./}{./Figures/}{./Figures/COVID19_results_17April_ALL/eps_version/}}

\input{Preamble}

\usepackage{enumitem}
\setlist{nosep} 

\title{\textbf{First-principles machine learning modelling of COVID-19}}

\author[1,2]{Luca Magri\footnote{Corresponding author: \href{mailto:lm547@cam.ac.uk}{lm547@cam.ac.uk} }}
\author[3,4]{Nguyen Anh Khoa Doan}
\affil[1]{University of Cambridge, Department of Engineering, Cambridge CB2 1PZ, United Kingdom}
\affil[2]{Institute for Advanced Study, Technical University of Munich, Garching 85748, Germany (visiting fellow)}
\affil[3]{Department of Mechanical Engineering, Technical University of Munich, Garching 85747, Germany}
\affil[4]{Institute for Advanced Study, Technical University of Munich, Garching 85748, Germany}
\date{}                     
\begin{document}
\maketitle

\section*{Abstract} 
\textbf{Background:}
The coronavirus disease 2019 (COVID-19) has changed the world since the World Health Organization declared its outbreak on 30th January 2020, recognizing the outbreak as a pandemic on 11th March 2020. 
%
As often said by politicians and scientific advisors, the objective is ``to flatten the curve'', or ``push the peak down'', or similar wording, of the virus spreading. 
Central to the official advice  are mathematical models and data, which provide estimates on the evolution of the number of infected, recovered and deaths. 
The accuracy of the models is improved day by day by inferring the contact, recovery, and death rates from data (confirmed cases). \\ 

\textbf{Methods:} 
A data-driven model trained with {\it both} data {\it and} first principles is proposed. The model can quickly be re-trained any time that new data becomes available. \\ 

\textbf{Data:} 
John Hopkins University CSSE has been collecting global data from official organizations, such as the World Health Organization, Italy Ministry of Health, and others~\cite{JHUgit}.\\

\textbf{Results:} 
The outputs of the analysis are the estimates of infected, recovered and deaths due to COVID-19, as well as the contact, recovery, death rates, basic reproduction number ($R_0$) and doubling times. The following case studies are analysed: United Kingdom, Italy, Germany, France, Spain, Belgium, USA, New York City, China, and the World.  A summary of the results is shown in Table~\ref{tab:results_peaks}. A fast exponential growth in the absence of intervention is found for all cases.\\ 

\textbf{Discussion:} %
The method can be applied to more detailed epidemic models with virtually no conceptual modification. \\


\textbf{Acknowledgements:}
L. Magri is advising the Scientific Pandemic Influenza Group on Modelling (SPI-M) through the Royal Society’s Rapid Assistance in Modelling the Pandemic (RAMP) initiative (\url{https://epcced.github.io/ramp/}). \\

\textbf{Competing interests:} The authors declare no competing interests. 

\section{Introduction}
In December 2019, a cluster of unexplained pneumonia cases in Wuhan, the capital of Hubei
province in the People's Republic of China, resulted  into a global pandemic by 11 March 2020, as declared by the World Health Organization~\cite{WHO_web}.
The disease 
is caused by a single-stranded RNA coronavirus (Severe acute respiratory syndrome coronavirus 2, SARS-CoV-2) similar to the pathogen
responsible for SARS (severe acute respiratory syndrome) and MERS (Middle East respiratory
syndrome). The disease caused by this virus has been named COVID-19 (Coronavirus Disease 2019). On 19th April 2020, 1:00 BST, the World Health Organization~\cite{WHO_web} reported 2,203,927 confirmed cases, 148,749 confirmed deaths, and 213 countries / territories with cases~\cite{WHO_web}.

To control the epidemic, aggressive measures have been implemented worldwide, for example, self-isolation of
confirmed and suspected cases, contact tracing and tracking, and social distancing. 
According to the data, the
most draconian measures have managed (or are managing) to suppress (or substantially mitigate) the epidemic. 
Examples are 
the localised
lockdown of the Hubei region in China (23rd-24th January 2020)~\cite{China_lock}; and 
the national lockdowns of Italy (9th
March 2020)~\cite{Italy_gazzetta}, Spain (14th March 2020)~\cite{Spain_lock}; the United Kingdom (24th March 2020)~\cite{UK_lock}, among others. 

Scientific advice typically relies on estimates of the contact, recovery and death rates. 
This information is summarized in the basic reproduction number, $R_0$, which is the average number of new infections generated by a single infected person within a susceptible population. Estimates of COVID-19 $R_0$ are variable due to the different methods, models
 and parameters  employed, as well as the databases used~\cite{Pellis2020}. As reported in~\cite{Pellis2020}, most official sources estimates $R_0$ to fall in the range $2-3$. 
{\it Flattening the curve} or {\it keeping the peak down}, or similar wording, which have been extensively used by governments to level with a lay audience, can be achieved by either reducing the contact rate, $\beta$, or by increasing the recovery rate, $\gamma$~\cite{Ferguson2020}. 
The latter can be achieved with a vaccine or a cure, which is not presently available.
Therefore, to {\it flatten the curve}, Governments are acting on minimising the contact rate~\cite{Ferguson2020}. 

The objective of this paper is threefold.
First, a model that optimally combines data from official databases and first principles of an epidemic model is proposed.  
Second, the model is applied to provide quantitative estimates on the contact, recovery, death rates;  the basic reproduction number, $R_0$; the doubling times; and the evolution of the number of infected, recovered, deaths, and susceptible. Ten cases are analysed: United Kingdom, Italy, Germany, France, Spain, Belgium, USA, New York City, China, and the World. 
Third, predictions of future dynamics are provided. 
Although the results are consistent with the first principles and working assumptions used, they are affected by uncertainty because of biases in the data, such as errors in reporting, changes in case definition and testing regime, \cite{Pellis2020}, and modelling assumptions. However, as argued by~\cite{Pellis2020},  the fast growth rate and large numbers likely make small biases negligible; and multiplicative corrections, such as constant under-reporting, affect the observed trend only weakly.  
The paper is structured as follows. The method is presented in Sec.~\ref{sec:ffrijmrfi4} and the results are shown in Sec.~\ref{sec:results}.

\section{Methods and data} \label{sec:ffrijmrfi4}
In first-principles machine learning modelling, we need first principles and data ({\it machine learning}) to generate a model. 
Section~\ref{sec:ff343} introduces the first principles and working assumptions, Sec.~\ref{sec:fgkjgreijfj} describes the data, and Sec.~\ref{sec:gkgkw} formulates the problem as a constrained optimization problem. The proposed solution method is presented in Sec.~\ref{sec:principled}. 

\subsection{Epidemic model: First principles}\label{sec:ff343}
\begin{figure}[!ht]
\begin{center}
\includegraphics[width=0.65\textwidth]{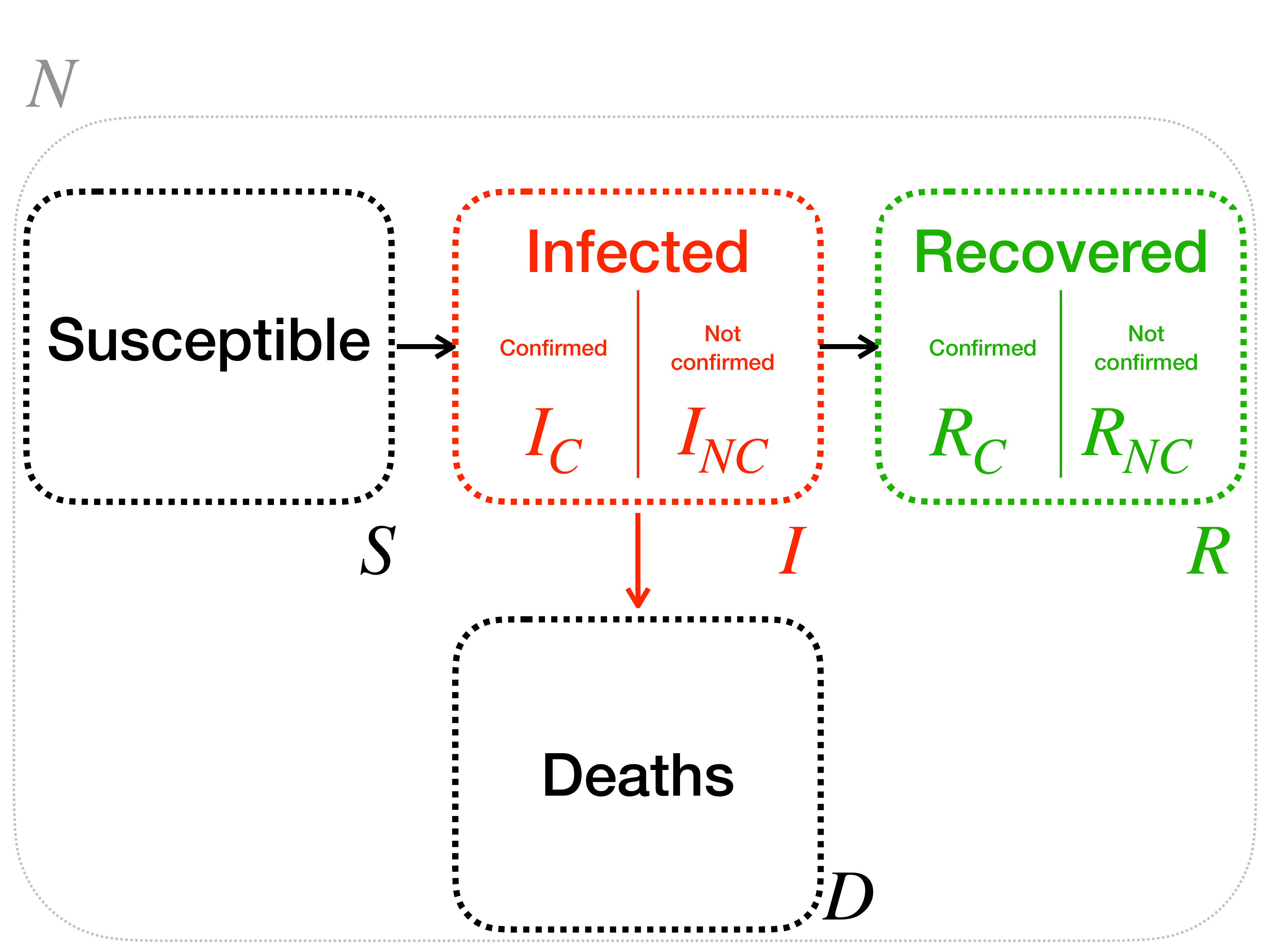}
\caption{Population and groups in the SIRD-epidemic model.\label{fig2}}
\end{center}
\end{figure}

%
%
%
The COVID-19 infectious disease is an epidemic~\cite{Grassly2008}. 
To model an epidemic, suitable groups (also known as compartments~\cite{KERMACK1991}) are defined to cover the entire population of a country.
Because 
(i) the epidemic has a (relatively) short time scale, for which the new births can be neglected; 
(ii) the number of deaths is small as compared with the entire population; and 
(iii) travel restrictions are enforced, 
the population, $N$, is assumed to be constant. 
The population of a country is divided into mutually exclusive groups: susceptible (S), infected (I), deceased (D), and recovered (R) (Fig.~\ref{fig2}). In this model, the deaths are due to COVID-19.
Every group is assumed to have the same characteristics, i.e., the groups are homogeneous. 
Every susceptible person can contract the virus (the immune group is neglected). 
These working assumptions can be relaxed in more complex models~\cite{Hethcote2000,Grassly2008}.
%
%
%

Mathematically,  
\begin{align}
N&= S + R + I + D. \label{eq:cont}
\end{align}
Equation \eqref{eq:cont} is a continuity equation. The population $N$, which does not vary in time, is the sum of the groups $S$, $I$, $R$, $D$, which  vary in time. 
This compartmental approach is known as the SIR-epidemic model with vital dynamics and constant population~\cite{KERMACK1991,wiki_SIR}. The model  will be called the SIRD-model for brevity. 
The working assumptions and first principles are mathematically expressed by four ordinary differential equations (ODEs) with time-varying parameters (non-autonomous dynamical system) 
\begin{align}
\dot{S} & = -\beta \frac{I}{N} S, \label{eq:s}\\
 \dot{I} &= \beta \frac{I}{N} S  - (\mu + \gamma) I \label{eq:fjfjkf}\\
 \dot{R} & = \gamma I \\ 
 \dot{D} & = \mu I \label{eq:d}
 \end{align}
subject to initial conditions $S_0$, $I_0$, $R_0$ and $D_0$. The symbol $\dot{}$ denotes the time derivative, $d/dt$. 
In compact form
\begin{align}
\dot{{\bm q}} & = \mathbf{F}({\bm q}; \boldsymbol{\alpha) }\label{eq:1} \\ 
{\bm q} &= {\bm q}_0\;\;\;\textrm{at}\;\;\;t=0 \label{eq:2}
\end{align}
where $\mathbf{F}$ is the model (i.e., the SIRD equations), and 
\begin{align}
{\bm q} & \equiv \left[S, I, R, D\right]^T, \\
{\bm \alpha} & \equiv \left[\beta,  \gamma, \mu\right]^T, \label{eq:eere}
\end{align}
are the column vectors of the state and parameters, respectively. 
$I/N$ is the probability to come into contact with an infected individual; 
$\beta$ is the average number of contacts per person per unit of time weighed by the transmissibility (contact rate); 
$\gamma$ is the average number of recovered people per unit of time (recovery rate); 
$\mu$ is the average number of deaths due to COVID-19 per unit of time (death rate). 
These parameters are time dependent and depend on several variables, such as governmental policies (lockdown, school/university closures, social distancing, etc.), heterogeneity in the population (age, life style, herd immunity, hygiene standards, etc.), and  properties of the epidemic (virus genome, spreading mechanisms, etc.).  
The SIRD parameters estimate the epidemic time scales: $1/\gamma$ is the average time to recover; 
$1/\beta$ is the average time between one contact (with an infected) and another; 
and $1/\mu$ is the average time to decease (for those who do not recover). 
The basic reproduction ratio\footnote{$R_0\equiv\beta/(\gamma+\mu)$. Because $\mu$ is sufficiently small, it will be neglected.}, $R_0\equiv\beta/\gamma$, is the expected number of secondary infections from a single infection entering a population where all members are susceptible~\cite{Grassly2008}.  If $R_0>1$, the number of infected increases (Eq.~\eqref{eq:fjfjkf}). If $R_0<1$, the disease does not grow on average.  The total number of new cases per unit of time due to the contact of $S$ susceptible people with infected people is $\beta I/N\cdot S$. This is the only nonlinear term of the equations. (Other nonlinearities are hidden in the time dependence of the parameters $\beta$, $\gamma$ and $\mu$.)\\

Equations~\eqref{eq:s}-\eqref{eq:d} are interpreted as follows.
The first equation is the rate of change of the susceptible group. 
The number of susceptible, $S$, changes faster in time if there are more infected people, $I$ and more susceptible that can be infected, $S$. 
Clearly, the susceptible group is constant in time if the contact rate of the virus is zero, and/or if the number of infected is zero, and/or if the number of susceptible is zero. 
The second equation is the time derivative of the continuity equation.
It expresses the fact that, in this epidemic model, the population $N$ is assumed to be constant. 
The third equation is the rate of change of recovered people. 
The number of recovered is proportional to the number of infected, $I$, because a recovered person must have been infected.
The fourth equation is the rate of change of the deceased group.
The number of deaths is proportional to the number of infected, $I$,  because a deceased individual must have been infected (in this model). 

\subsection{Data sources}\label{sec:fgkjgreijfj}
The reliable data is about the number of confirmed infected, $I_c$, and confirmed deaths, $D_c$, which are arranged in a vector
\begin{align}
{\bm q}_c \equiv \left[
 I_c, 
 D_c   
\right]^T.
\end{align}
The data on the confirmed recovered, $R_c$, was discontinued because it was deemed inaccurate\footnote{\url{https://github.com/CSSEGISandData/COVID-19/tree/master/csse_covid_19_data/csse_covid_19_time_series}}. The data  used here is publicly available in the CSSEGISandData/COVID-19 GitHub repository\footnote{\url{https://github.com/CSSEGISandData/COVID-19/tree/master/csse_covid_19_data/csse_covid_19_daily_reports}}, which collects data from official sources and organizations.

\subsection{Problem formulation}\label{sec:gkgkw}
%
%
The calculation of the groups' dynamics and time-varying epidemic parameters is a constrained optimization problem: \\

\begin{tcolorbox}
\texttt{Calculate } 
\begin{align}
\mathbf{q},{\bm \alpha} 
\end{align}
\texttt{to minimize}
\begin{align}
E_d \equiv \lambda_1\lvert\lvert I-I_c \lvert\lvert^2+ \lambda_2\lvert\lvert D-D_c \lvert\lvert^2
\end{align}
\texttt{subject to} 
\begin{align}
\textrm{an epidemic model}. 
\end{align}
\end{tcolorbox}
The epidemic model used in this paper is provided by Eqs.~\eqref{eq:1} and \eqref{eq:2}, however, more detailed models can be used. 
 $\lvert\lvert\cdot\lvert\lvert$ is a norm, $\lambda_1$ and $\lambda_2$ are user-defined normalization factors.  
The loss function, $E_d$, measures the error between the candidate solution ($I$, $D$) and the data ($I_c$, $D_c$). Among all the possible candidate solutions, only the solutions that fulfil the epidemic model (Eqs.~\eqref{eq:1} and \eqref{eq:2}) will be accepted. 
%
The cumulative confirmed number of cases is the dataset used. This is a quantity to be preferred over the daily increase of confirmed cases because it is smoother, i.e., it is not significantly affected by random fluctuations, in contrast with the daily increase. The algorithm that solves this constrained optimization problem is presented in Sec.~\ref{sec:principled}.

\subsection{First-principles machine learning epidemic modelling} \label{sec:principled}
A data-driven model combined with first principles is proposed. This is referred to as {\it first-principles machine learning} for brevity.  The data-driven algorithm is an optimal interpolator,  while the epidemic model helps to obtain parameters that are consistent with the model. This synergistic combination helps to reduce the uncertainty in the predictions, which are as good as the employed epidemic model and the accuracy of the data. \\

\begin{figure}[!ht]
    \centering
    \includegraphics[width=0.75\textwidth]{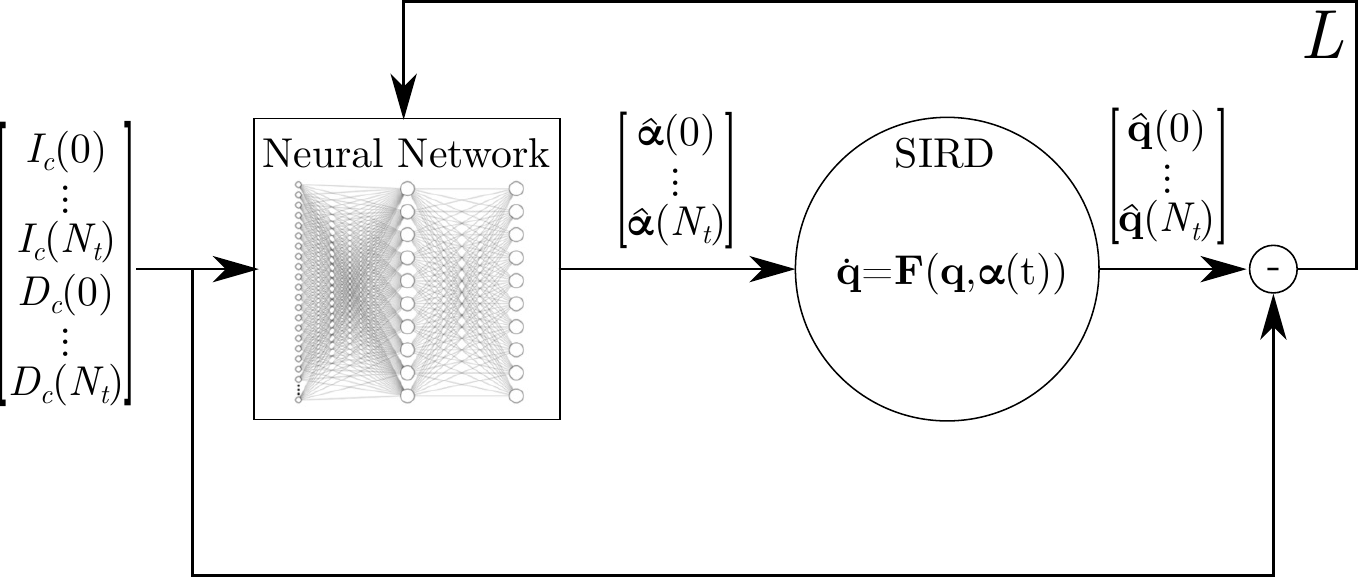}
    \caption{First-principles machine learning architecture for epidemic modelling. The graph of the neural network is pictorial.}
    \label{fig:NN_architecture}
\end{figure}

The first-principles machine learning epidemic modelling is based on the combination of an ODE-solver, which time-advances the SIRD model in Eqs.~\eqref{eq:s}-\eqref{eq:d} (first principles), and a feedforward neural network (machine learning), which performs the assimilation of data with the epidemic model to learn the parameters' vector ${\bm \alpha}(t)$ (Fig. \ref{fig:NN_architecture}) and predict the state, ${\bm q}(t)$. 
The Neural Network (NN) receives as an input the entire time series of total confirmed infected cases $\{I_c(t)\}_{t=0}^{N_t}$ and total confirmed deceased $\{D_c(t)\}_{t=0}^{N_t}$ up until the 17th of April 2020. The time $t=0$ corresponds to the day when the first  infection was recorded, and $N_t$ is the number of days from $t=0$ to the 17th of April 2020.
From the time series, $\{I_c(t)\}_{t=0}^{N_t}$ and $\{D_c(t)\}_{t=0}^{N_t}$, the NN infers the time evolution of the parameters of the SIRD model, i.e. $\{\widehat{\beta}(t)\}_{t=0}^{N_t}$, $\{\widehat{\gamma}(t)\}_{t=0}^{N_t}$ and $\{\widehat{\mu}(t)\}_{t=0}^{N_t}$, where  $\,\widehat{}\,$ denotes the quantity estimated by the neural network. 
Consistently with~\eqref{eq:eere}, the parameters $\widehat{\beta}$, $\widehat{\gamma}$, $\widehat{\mu}$ are cast in the vector $\widehat{\bm{\alpha}}\equiv [\widehat{\beta},\widehat{\gamma},\widehat{\mu}]^T$. 
Subsequently, $\{\widehat{\bm{\alpha}}\}$ is fed into the time-integration of the SIRD model with initial condition $\bm{q}_0 = [N_0-I_0-D_0,I_0,0,D_0]^T$ where $N_0$ is the population of the country analysed (Table \ref{tab:country_data}), whereas $I_0$ and $D_0$ are the confirmed infections and deaths on the day of the first confirmed cases, respectively. Finally, the time-integration of the SIRD model provides the state $\widehat{\bm{q}}$.

\begin{figure}[!ht]
    \centering
    \includegraphics[width=0.6\textwidth]{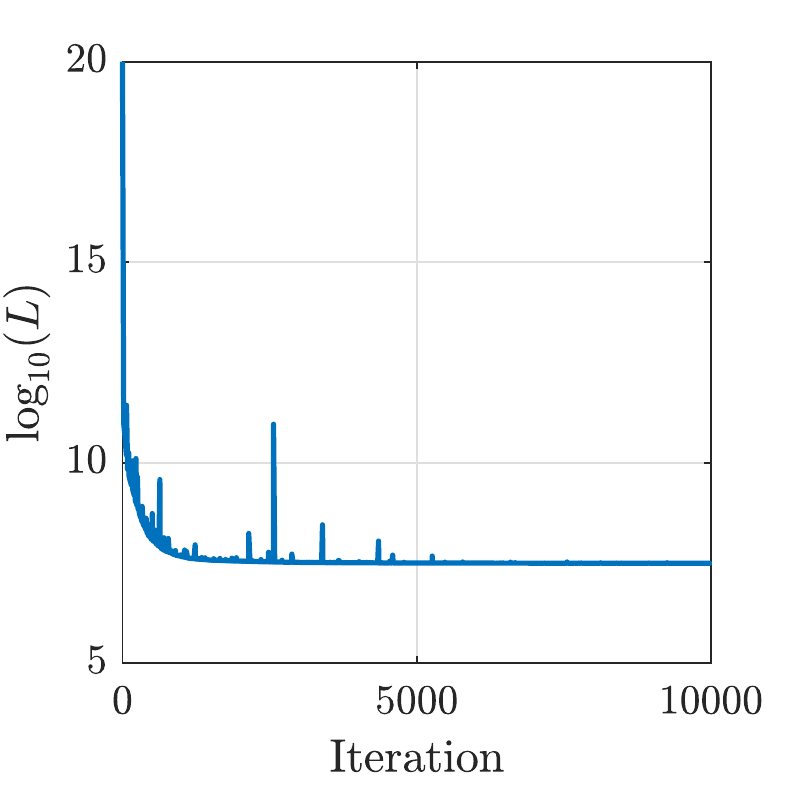}
    \caption{Typical evolution of the loss function during the o-eps-converted-to.pdfptimization process (step~\ref{item3}). World data.}
    \label{fig:OptConv}
\end{figure}

Algorithmically, the architecture is trained as follows:
\begin{tcolorbox}[breakable]
\begin{enumerate}
    \item \texttt{First guess on the parameters.} 
    
    From the dataset $\{I_c(t)\}_{t=0}^{N_t}$ and $\{D_c(t)\}_{t=0}^{N_t}$, a set of constant parameters $\bm{\alpha}_0\equiv[\beta_0,\gamma_0,\mu_0]^T$ is obtained by nonlinear regression of the data, $I_c$ and $D_c$, during the initial exponential growth only. This time window is $[0, t=\textrm{Regr}]$ (Table \ref{tab:country_data}). \label{secitem1}\\

    \item \texttt{Initialization of the neural network.} 
    
    The neural network is pre-trained to output the set of constant parameters, $\bm{\alpha}_0$. This set of parameters ensures that the initial state of the neural network is consistent with the initial exponential growth, which makes the time integration of the SIRD model robust. 
Unless otherwise specified, the neural network consists of $1$ layer with $8$ neurons \footnote{Architectures with $4$ to $64$ neurons provide the same accuracy (result not shown).}. The time evolution of the SIRD parameters is obtained by nonlinear combination of the neurons with a sigmoid activation. \label{item2}\\
    
    \item \texttt{Training of the neural network.}
    
    The entire architecture, which consists of the neural network and the SIRD time-integrator, is optimized by a gradient-based optimizer (L-BFGS-B optimizer~\cite{Press2007}) to minimize the loss function 
    \begin{align}
        &L = \nonumber\\
        & \underbrace{\sum_{t=0}^{N_t} \left( (\log(I_c(t)) - \log(\widehat{I}(t)))^2 + (\log(D_c(t)) - \log(\widehat{D}(t)))^2 \right)}_{E_{d1}} + \nonumber \\
        & \underbrace{0.01 \frac{\log(\max(I_c))}{\max(I_c)} \sum_{t=0}^{N_t} \left( (I_c(t) - \widehat{I}(t))^2 + (D_c(t) - \widehat{D}(t))^2 \right)}_{E_{d2}} + \nonumber\\
         & \underbrace{100 \frac{\log(\max(I_c))}{\max(\bm{\alpha}_0)}\sum_{t=0}^{N_t-1} \left( (\widehat{\beta}(t)-\widehat{\beta}(t+1))^2 + (\widehat{\gamma}(t)-\widehat{\gamma}(t+1))^2 + 100(\widehat{\mu}(t)-\widehat{\mu}(t+1))^2 \right)}_{E_r} + \nonumber \\
         & \underbrace{100 \frac{\log(\max(I_c))}{\max(\bm{\alpha}_0)} \left( (\widehat{\beta}(0)-\beta_0)^2 + (\widehat{\gamma}(0)-\gamma_0)^2 + 100(\widehat{\mu}(0)-\mu_0)^2 \right) }_{E_0} \label{eq:f30rjfl2}
    \end{align}
The loss function is composed of four terms, which can be interpreted as follows: 
\begin{itemize}
    \item $E_{d1}$ is the error in a log-scale between the prediction and the available data (infected and deaths). This removes noisy fluctuations from the solution. 
    \item $E_{d2}$ is the error in a linear scale between the prediction and the available data (infected and deaths). 
    \item $E_r$ is a regularization term, which prevents large discontinuities in the time-variation of the SIRD parameters from occurring, making the evolution smoother. The regularization factor before the sum in $E_r$ is an empirical scaling factor to ensure that the orders of magnitude of $E_r$ and $E_d$ are comparable. The  factor $100$ before the terms with $\widehat{\mu}$ ensures that the parameters of the SIRD model have a comparable order of magnitude.
    \item $E_0$ constrains the initial values of $\widehat{\bm{\alpha}}$ to be close to the first guess obtained at step \ref{secitem1}. This ensures that, in the early stage of the epidemic, the growth is largely exponential with parameters that are nearly constant. A typical convergence of the optimizer is shown in Fig. \ref{fig:OptConv}.\label{item3}
\end{itemize}
\end{enumerate}
\end{tcolorbox}
\begin{table}[!ht]
    \centering
    \begin{tabular}{l|cccccccccc}
          & NY & Italy & Germany & UK & Spain & USA & France & China & Belgium & World \\ \hline
         $N_0$ [Mil] & 5.8 & 60.36 & 83.02 & 66.56 & 46.94 & 327.2 & 66.99 & 1386 & 11.46 & 7777.06 \\
         Regr & 20 & 40 & 60 & 60 & 50 & 70 & 60 & 15 & 55& 20 
    \end{tabular}
    \caption{Country populations, $N_0$ (from Census databases on Google) and time (days) of initial exponential growth, $\textrm{Regr}$.}
    \label{tab:country_data}    
\end{table}

\section{Results}\label{sec:results}
\newcommand{\UKlabel}{United Kingdom (day 0 = 31st January 2020):}
\newcommand{\Italylabel}{Italy (day 0 = 31st January 2020):}
\newcommand{\Germanylabel}{Germany (day 0 = 27th January 2020):}
\newcommand{\Francelabel}{France (day 0 = 24th January 2020):}
\newcommand{\Spainlabel}{Spain (day 0 = 1st February 2020):}
\newcommand{\Belgiumlabel}{Belgium (day 0 = 4th February 2020):}
\newcommand{\USlabel}{USA (day 0 = 22nd January 2020):}
\newcommand{\NYlabel}{New York City (day 0 = 2nd March 2020 ):}
\newcommand{\Chinalabel}{China (day 0 = 22nd January 2020):}
\newcommand{\Worldlabel}{World (day 0 = 22nd January 2020:}
%

\newcommand{\captionFIGVAL}{First and second rows: Validation of first-principles machine learning epidemic modelling. Third and fourth rows: Inference of recovered and susceptible. The vertical dotted lines indicate the day of lockdown.}
\newcommand{\subcaptionFIGaVAL}{Cumulative quantities.}
\newcommand{\subcaptionFIGbVAL}{Daily rates.}

\newcommand{\captionFIGalpha}{SIRD parameters. Neural network trained with the $\log$ (solid line) and without the $\log$ (dashed line) in the loss function~\eqref{eq:f30rjfl2}. The vertical dotted lines indicate the day of lockdown.}
\newcommand{\subcaptionFIGaalpha}{Time-varying contact rate ($\beta$), recovery rate ($\gamma$), and death rate ($\mu$). }
\newcommand{\subcaptionFIGbalpha}{ Basic reproduction number. The blue (red) curve corresponds to the left (right) vertical axis. }

\newcommand{\captionFIGextrap}{ Extrapolated trends of the SIRD parameters. The vertical dotted lines indicate the day of lockdown.}
\newcommand{\subcaptionFIGaextrap}{Extrapolated trends of the time-varying contact rate ($\beta$), recovery rate ($\gamma$), and death rate ($\mu$) with average slope over the last seven days (dotted lines) and fourteen days (dashed lines).   }
\newcommand{\subcaptionFIGbextrap}{ Extrapolated trend of the basic reproduction number with average slope over the last seven days (dotted lines) and fourteen days (dashed lines).  }
%
\newcommand{\captionFIGdoubling}{Doubling time with the $\log$ (solid line) and without the $\log$ (dashed line) in the loss function~\eqref{eq:f30rjfl2}. The doubling time is calculated as $t=\log(2)/\beta(t)$. (To take into account the time derivative of $\beta(t)$, semi-parametric methods, e.g.~\cite{Pellis2020}, can be used.)}
\newcommand{\captionFIGextraptwo}{From the top: Blue lines  indicate the extrapolated trends of the percentage of infected, recovered, deaths, and susceptible. Estimates with average slope over the last seven days (dotted lines),  fourteen days (dashed lines), and with values of the parameters assumed to be constant and equal to the last day (solid lines). Black lines: The left vertical dotted line represents the day of  lockdown, the right vertical line is the last day of the training data set, hence, the starting day for extrapolation. }

\clearpage
\subsection{United Kingdom}

\begin{figure}[ht]
     \centering
     \begin{subfigure}[t]{0.48\textwidth}
         \centering
        \includegraphics[width=\textwidth]{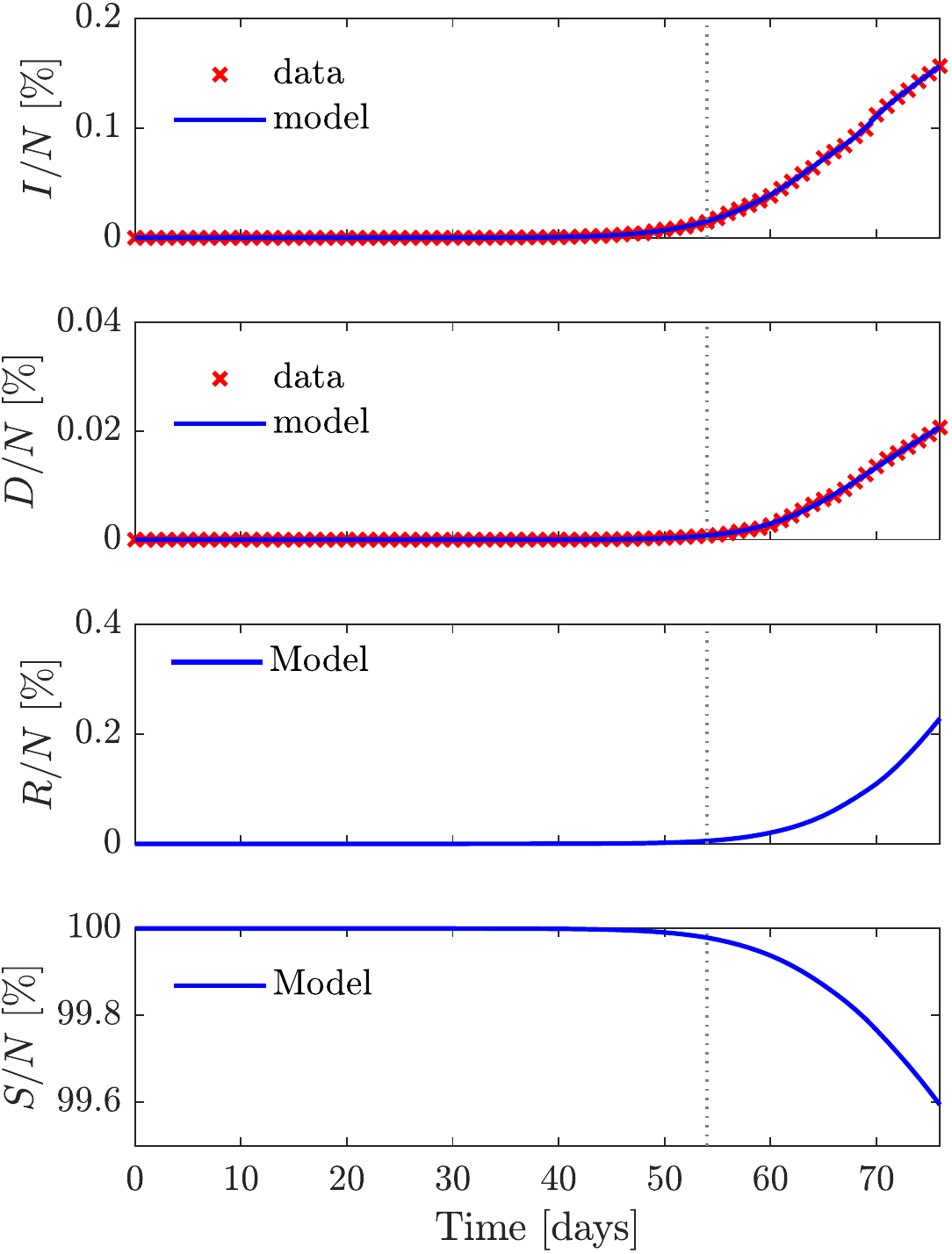}
         \caption{\subcaptionFIGaVAL}
         \label{fig:validation1_UK}
     \end{subfigure}
     \hfill
     \begin{subfigure}[t]{0.48\textwidth}
         \centering
        \includegraphics[width=\textwidth]{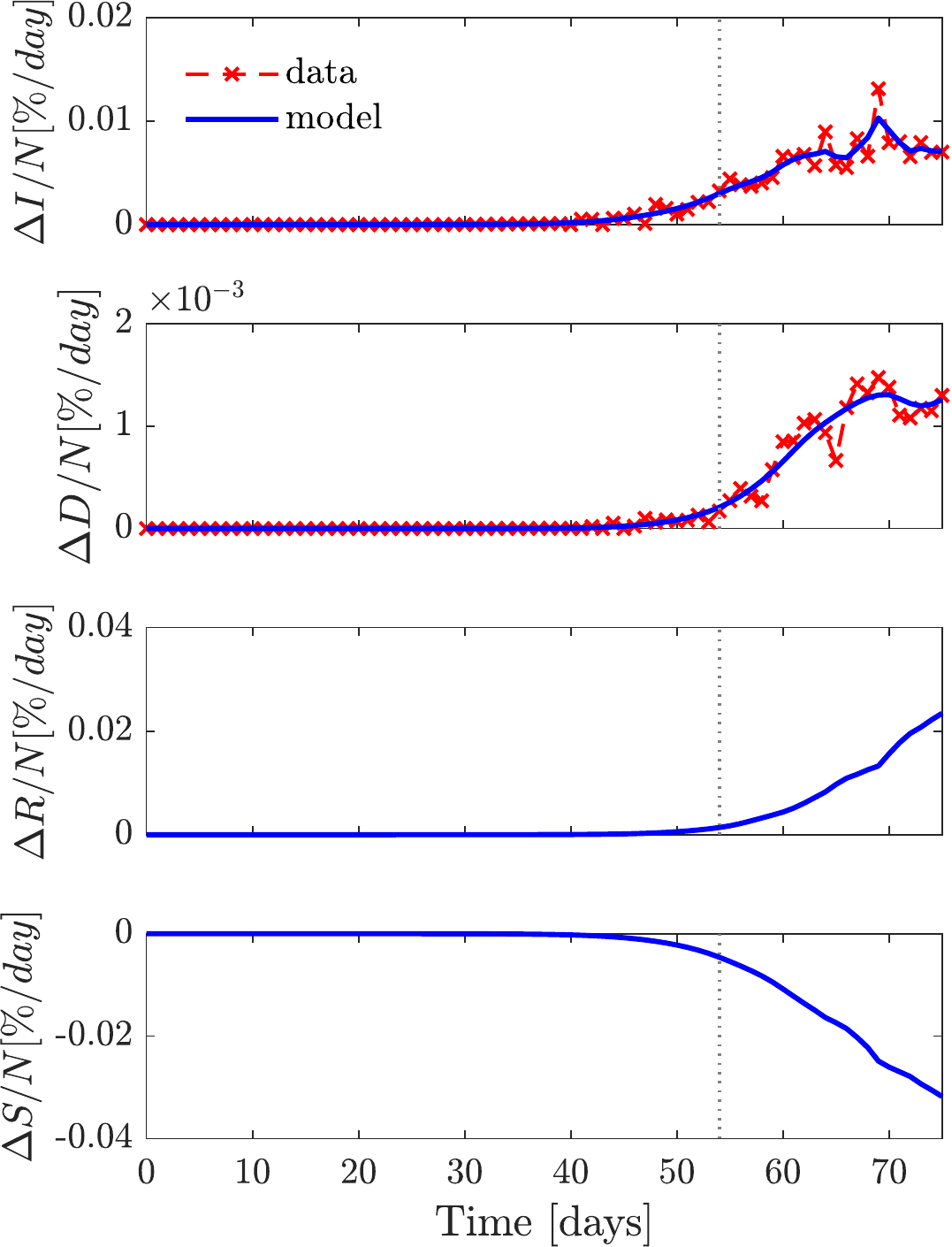}
         \caption{\subcaptionFIGbVAL}
         \label{fig:validation2_UK}
     \end{subfigure}
     \caption{\UKlabel \, \captionFIGVAL}
\end{figure}
%
%
\begin{figure}[ht]
     \centering
     \begin{subfigure}[t]{0.48\textwidth}
         \centering
        \includegraphics[width=\textwidth]{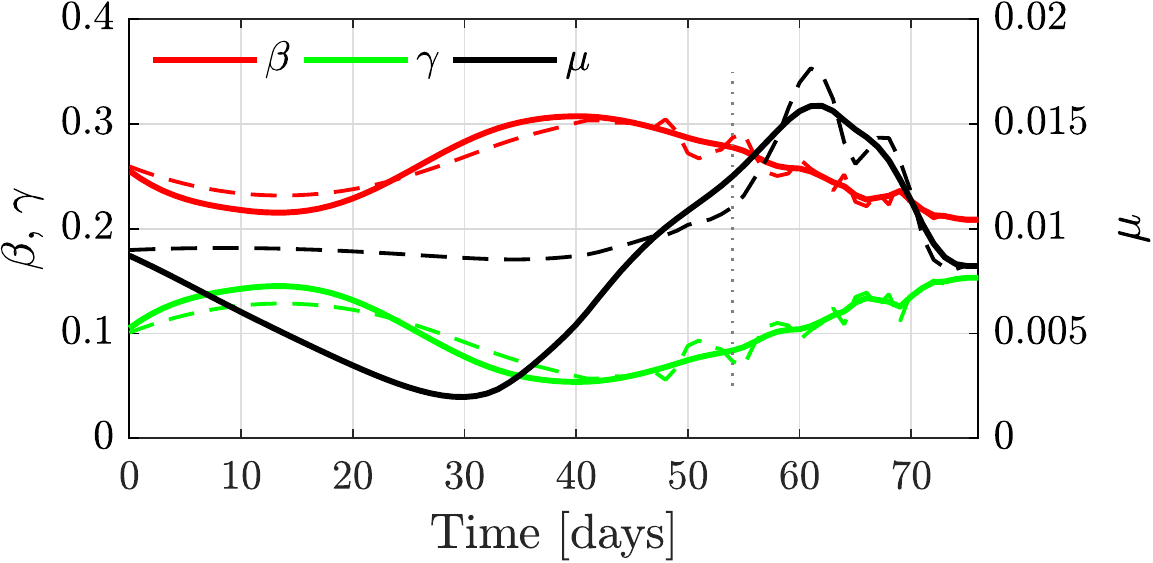}
         \caption{\subcaptionFIGaalpha}
         \label{fig:alpha_UK}
     \end{subfigure}
     \hfill
     \begin{subfigure}[t]{0.48\textwidth}
         \centering
        \includegraphics[width=\textwidth]{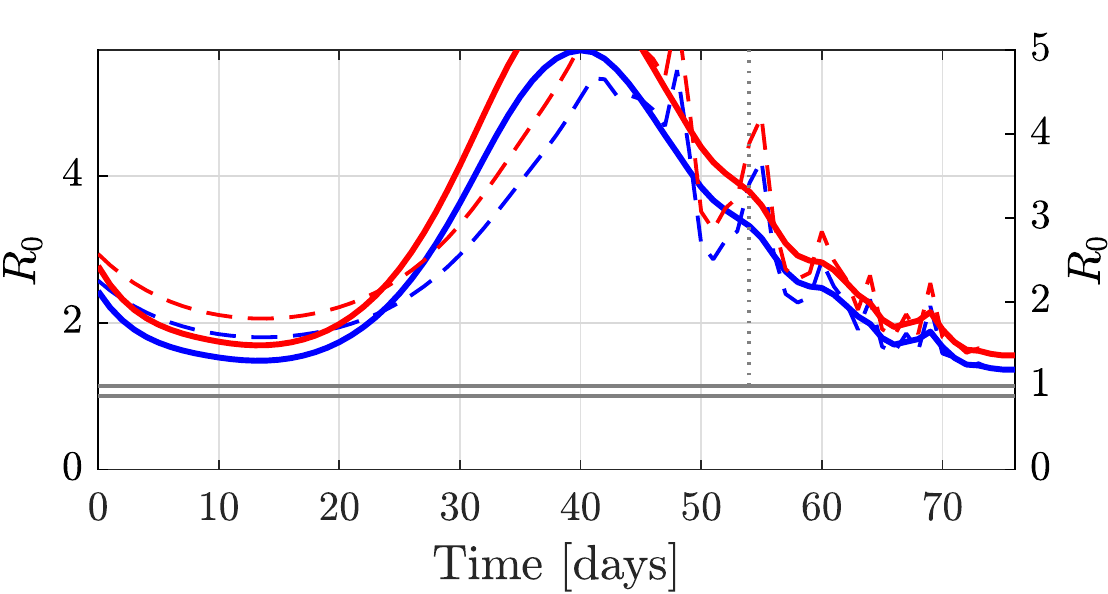}
         \caption{\subcaptionFIGbalpha}
         \label{fig:R0_UK}
     \end{subfigure}
     \caption{\UKlabel \,\captionFIGalpha}
\end{figure}
\begin{figure}[ht]
     \centering
     \begin{subfigure}[t]{0.48\textwidth}
         \centering
        \includegraphics[width=\textwidth]{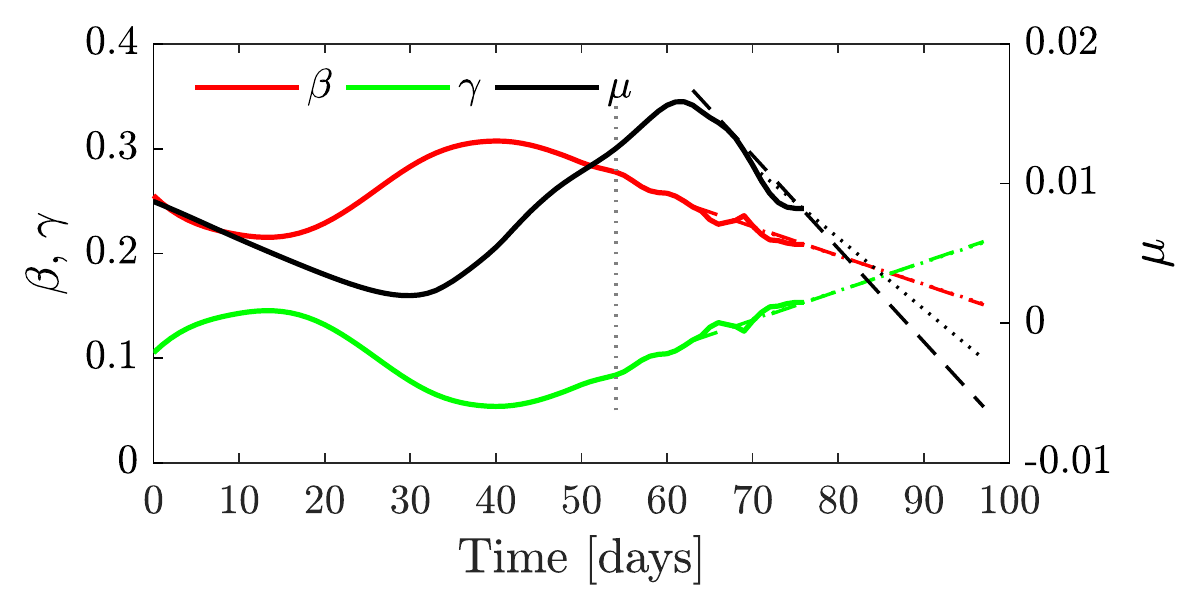}
         \caption{\subcaptionFIGaextrap}
         \label{fig:extrap_alpha_UK}
     \end{subfigure}
     \hfill
     \begin{subfigure}[t]{0.48\textwidth}
         \centering
        \includegraphics[width=\textwidth]{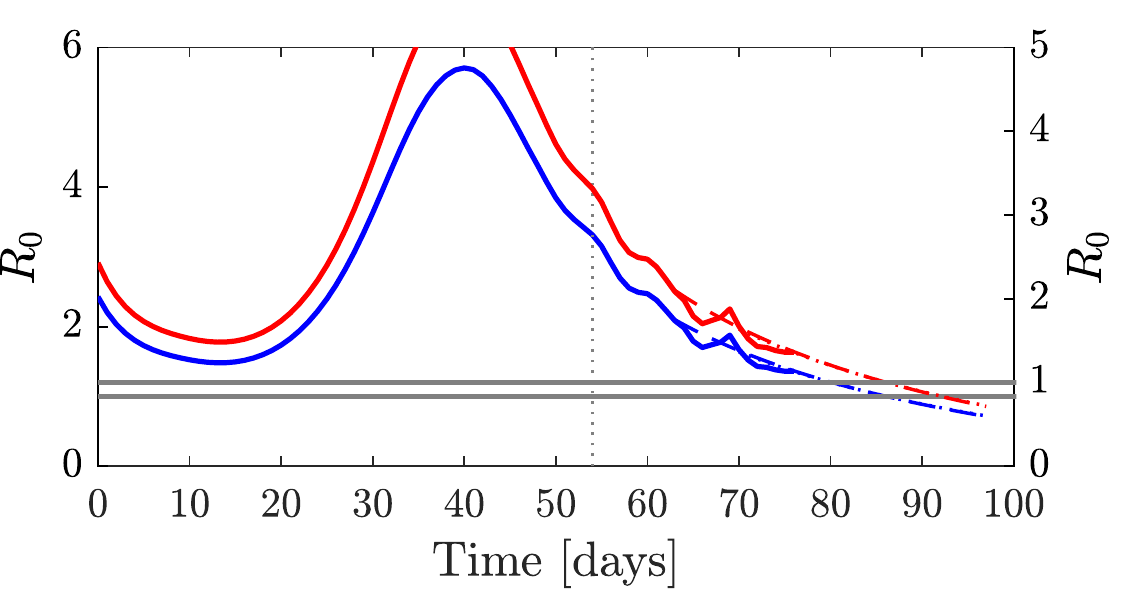}
         \caption{\subcaptionFIGbextrap}
         \label{fig:extrap_R0_UK}
     \end{subfigure}
     \caption{\UKlabel \,\captionFIGextrap}
\end{figure}

     \begin{figure}[h] 
         \centering
        \includegraphics[width=0.48\textwidth]{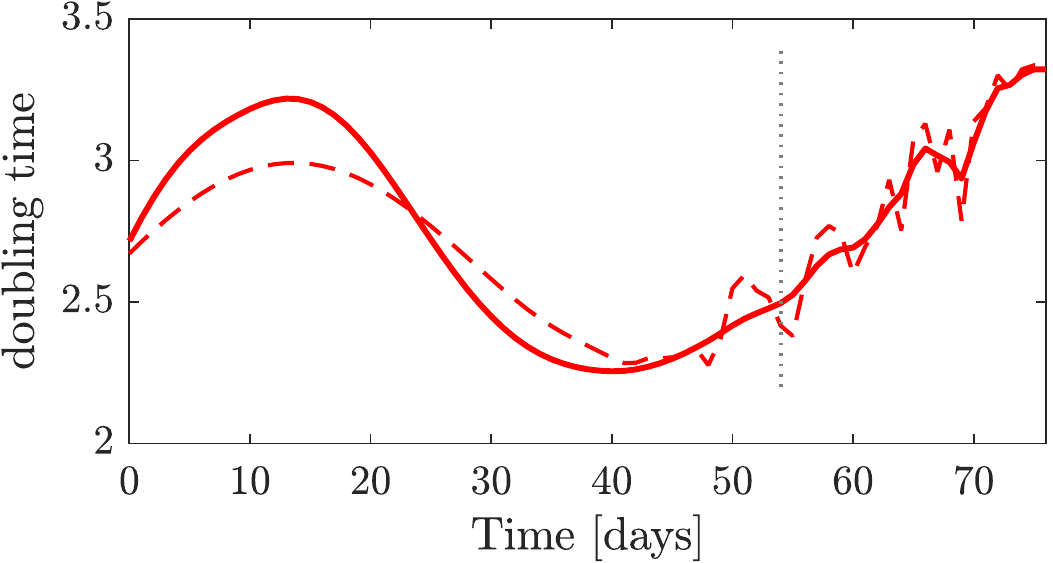}
         \caption{\UKlabel \, \captionFIGdoubling }
         \label{fig:doubling_UK}
     \end{figure}
     \begin{figure}[h] 
         \centering
        \includegraphics[width=0.48\textwidth]{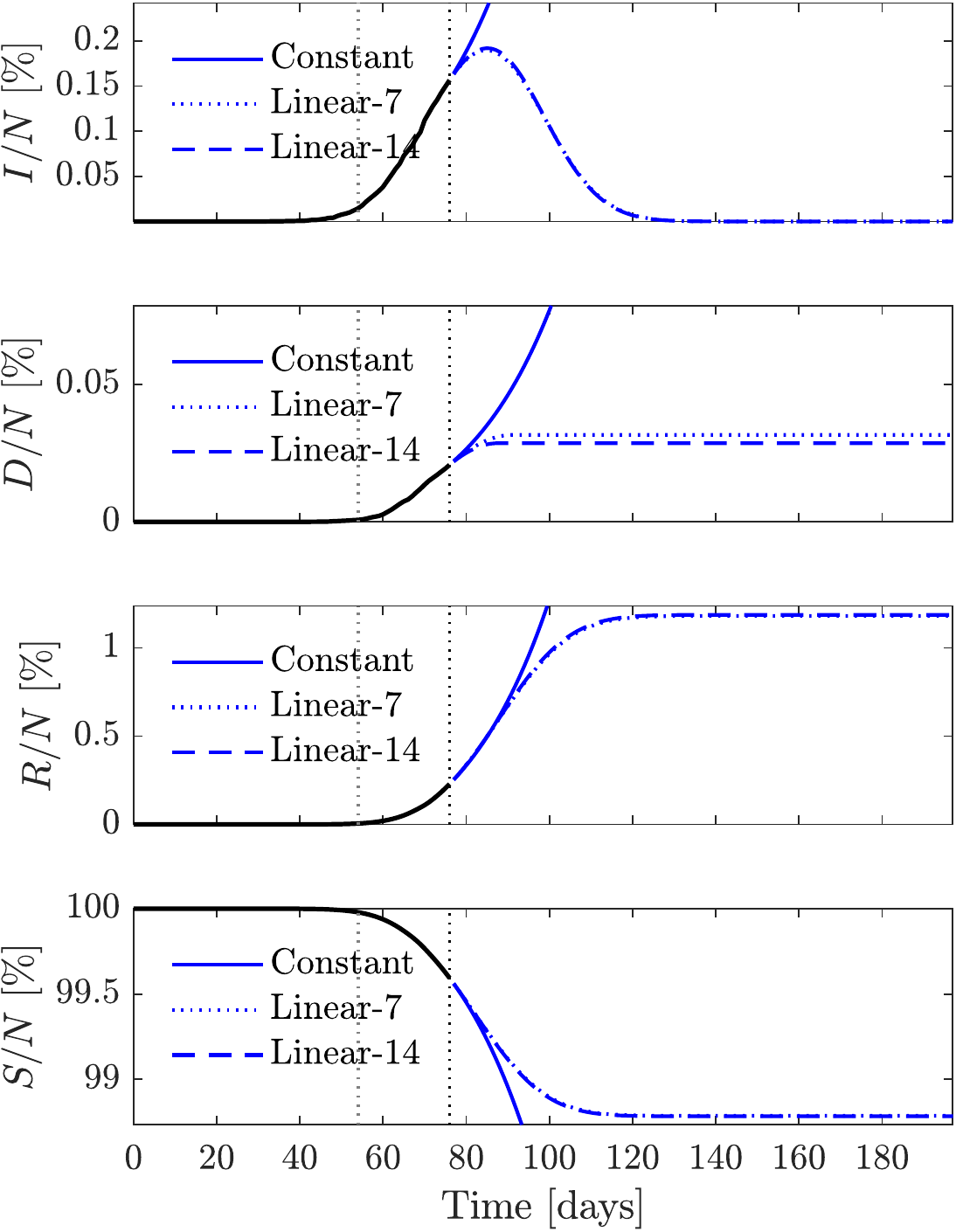}
         \caption{\UKlabel \, \captionFIGextraptwo The  black solid lines are taken from Fig.~\ref{fig:validation1_UK}. }
         \label{fig:extrap_state_UK}
     \end{figure}

\clearpage
\subsection{Italy}

\begin{figure}[ht]
     \centering
     \begin{subfigure}[t]{0.48\textwidth}
         \centering
        \includegraphics[width=\textwidth]{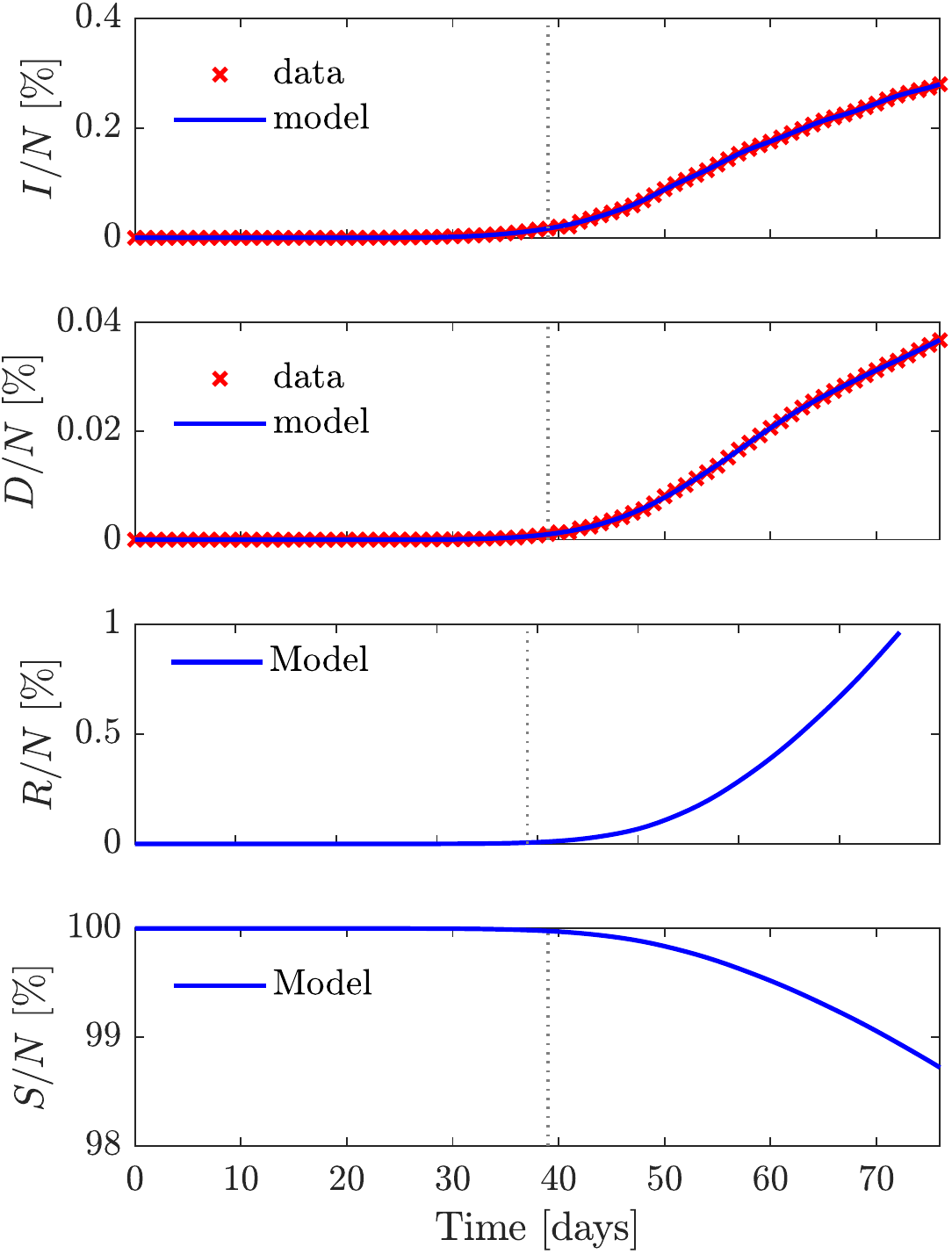}
         \caption{\subcaptionFIGaVAL}
         \label{fig:validation1_Italy}
     \end{subfigure}
     \hfill
     \begin{subfigure}[t]{0.48\textwidth}
         \centering
        \includegraphics[width=\textwidth]{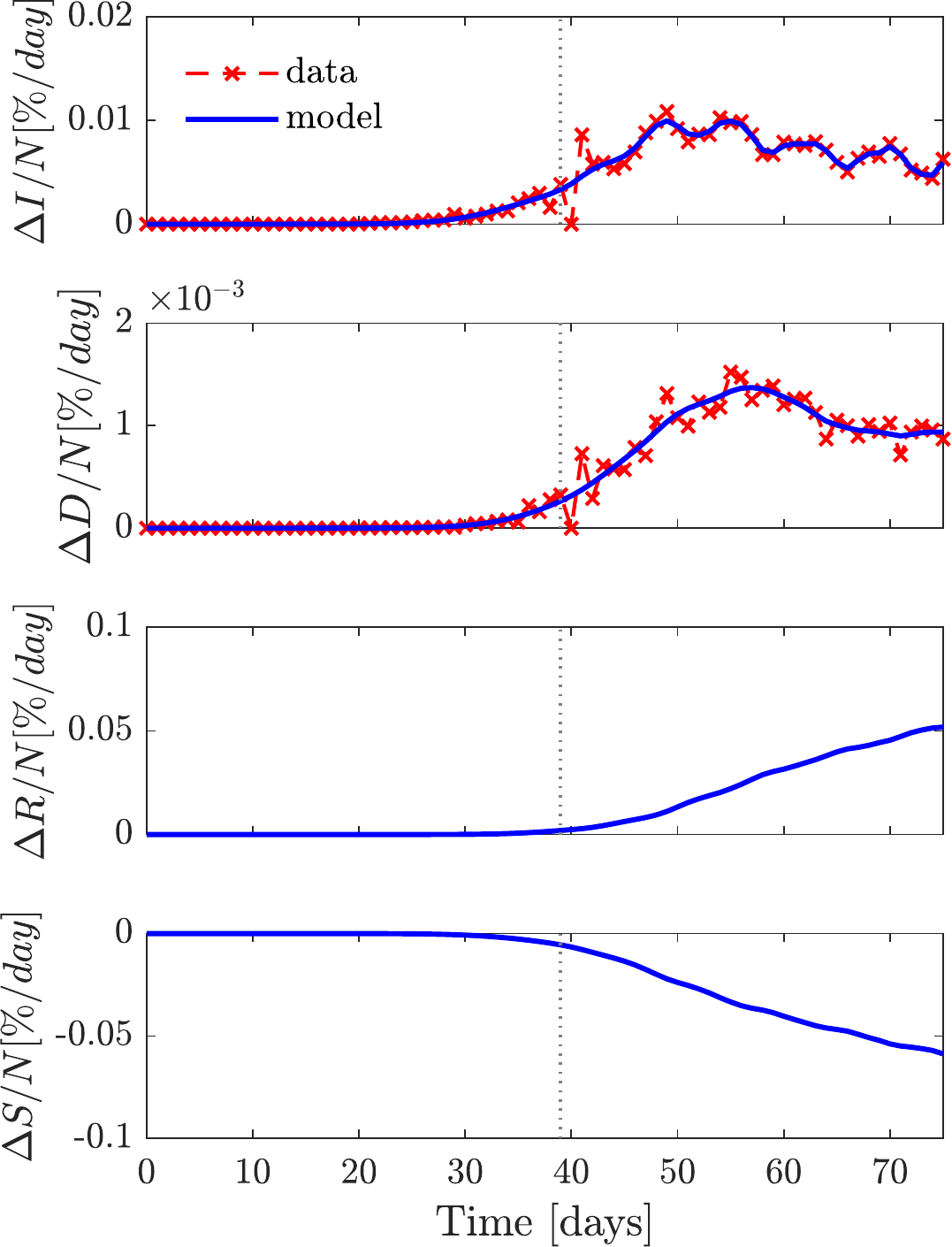}
         \caption{\subcaptionFIGbVAL}
         \label{fig:validation2_Italy}
     \end{subfigure}
     \caption{\Italylabel \, \captionFIGVAL}
\end{figure}
%
%
\begin{figure}[ht]
     \centering
     \begin{subfigure}[t]{0.48\textwidth}
         \centering
        \includegraphics[width=\textwidth]{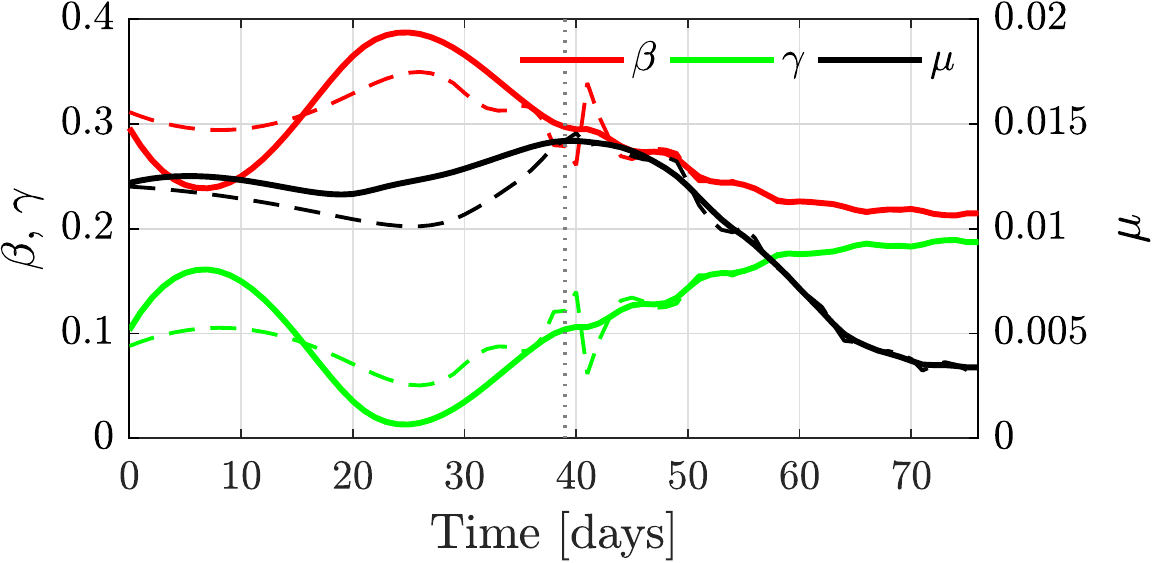}
         \caption{\subcaptionFIGaalpha}
         \label{fig:alpha_Italy}
     \end{subfigure}
     \hfill
     \begin{subfigure}[t]{0.48\textwidth}
         \centering
        \includegraphics[width=\textwidth]{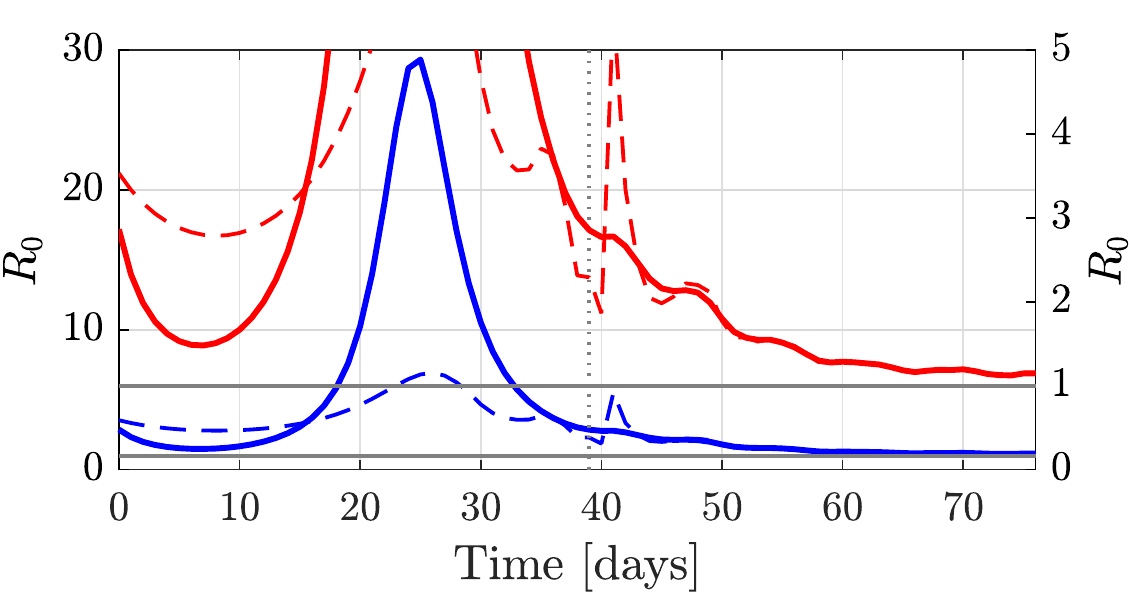}
         \caption{\subcaptionFIGbalpha}
         \label{fig:R0_Italy}
     \end{subfigure}
     \caption{\Italylabel \,\captionFIGalpha}
\end{figure}
\begin{figure}[ht]
     \centering
     \begin{subfigure}[t]{0.48\textwidth}
         \centering
        \includegraphics[width=\textwidth]{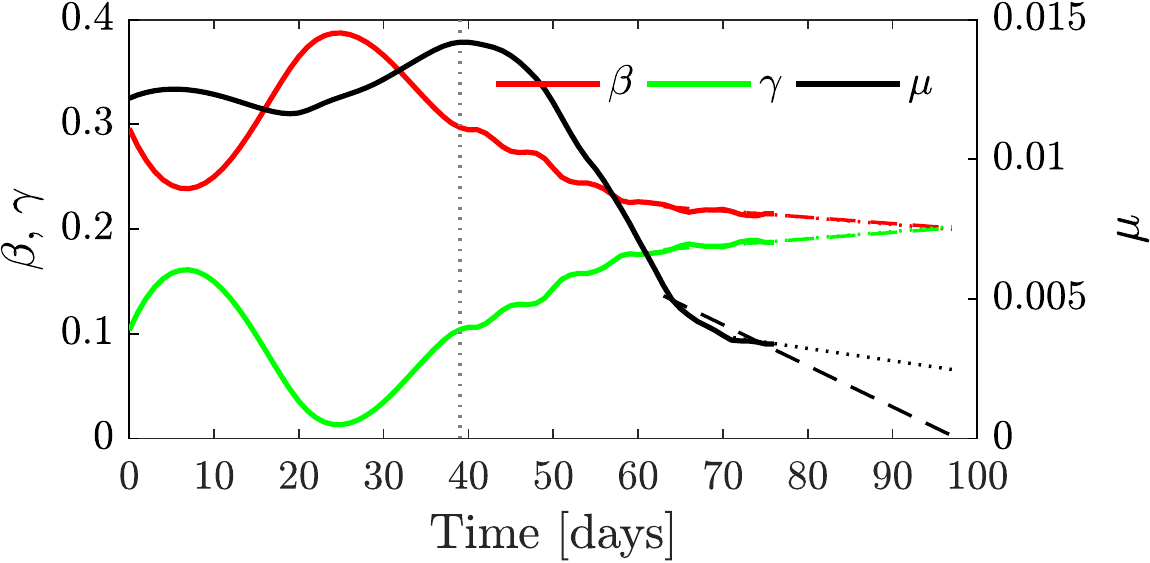}
         \caption{\subcaptionFIGaextrap }
         \label{fig:extrap_alpha_Italy}
     \end{subfigure}
     \hfill
     \begin{subfigure}[t]{0.48\textwidth}
         \centering
        \includegraphics[width=\textwidth]{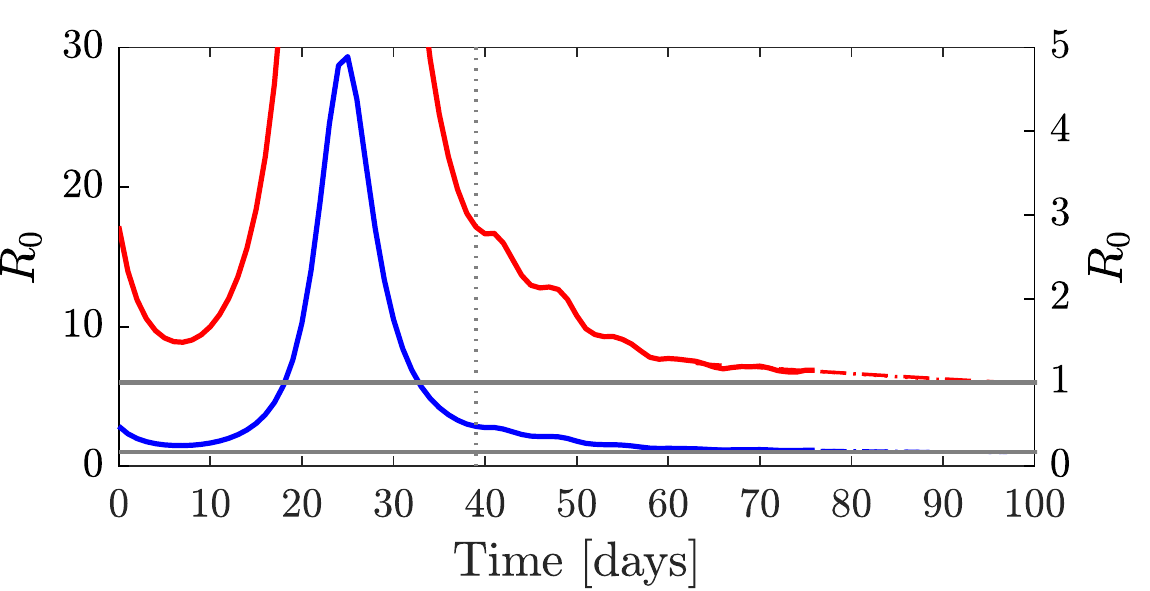}
         \caption{\subcaptionFIGbextrap }
         \label{fig:extrap_R0_Italy}
     \end{subfigure}
     \caption{\Italylabel \,\captionFIGextrap}
\end{figure}

     \begin{figure}[h] 
         \centering
        \includegraphics[width=0.48\textwidth]{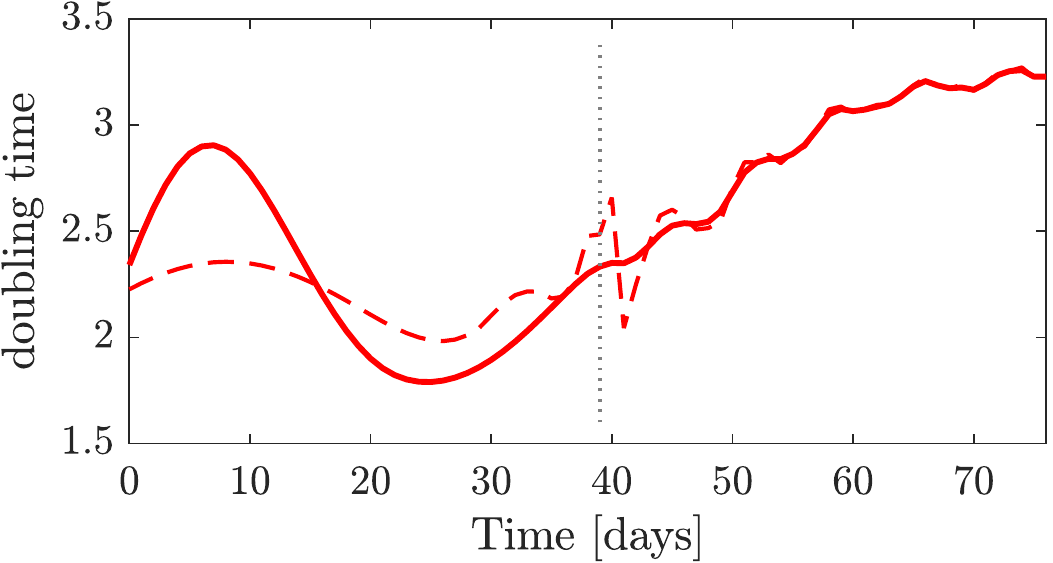}
         \caption{\Italylabel \, \captionFIGdoubling }
         \label{fig:doubling_Italy}
     \end{figure}
     \begin{figure}[h] 
         \centering
        \includegraphics[width=0.48\textwidth]{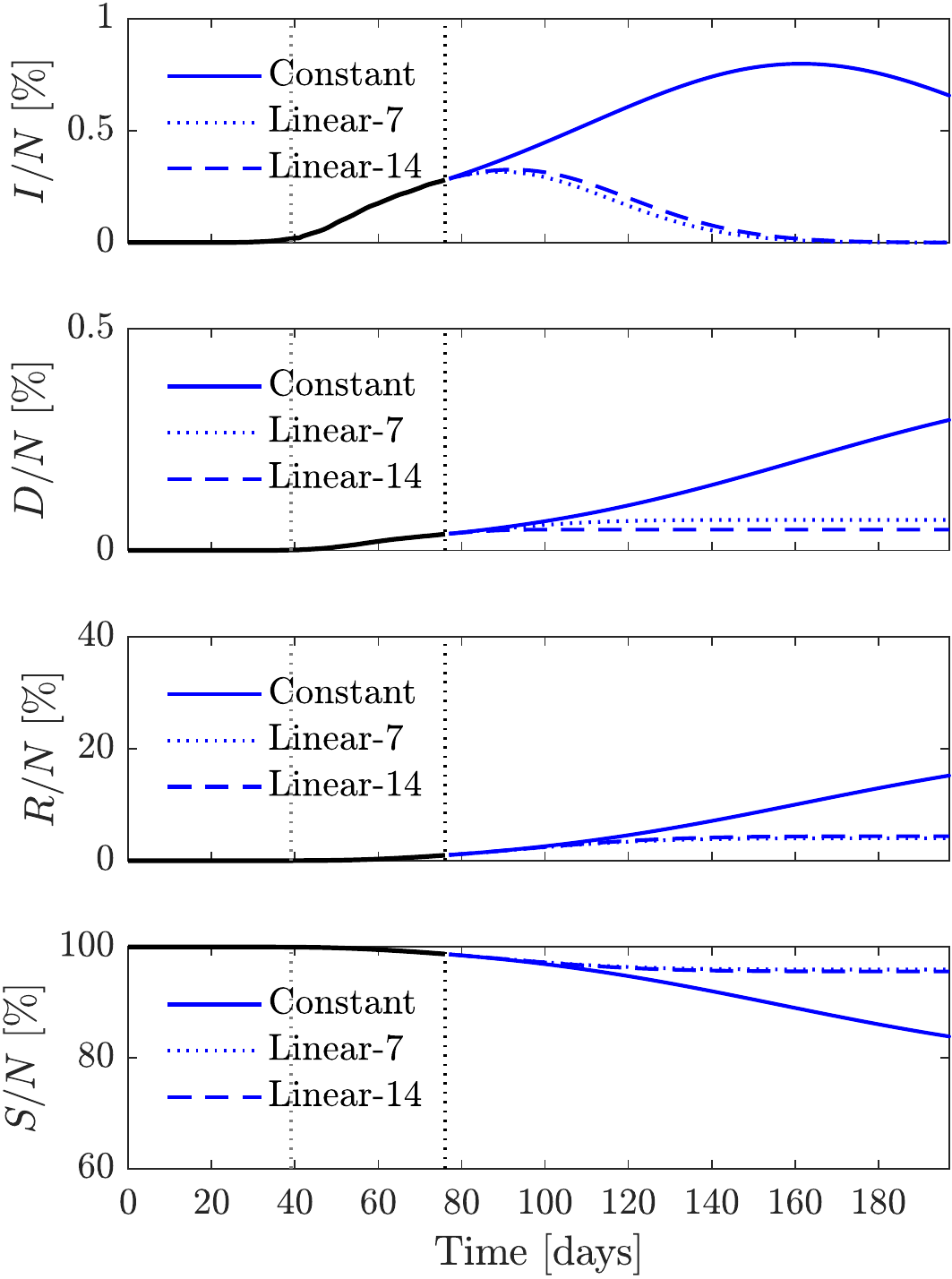}
         \caption{\Italylabel \, \captionFIGextraptwo The black solid lines are taken from Fig.~\ref{fig:validation1_Italy}. }
         \label{fig:extrap_state_Italy}
     \end{figure}
     
     \clearpage
\subsection{Germany}

\begin{figure}[ht]
     \centering
     \begin{subfigure}[t]{0.48\textwidth}
         \centering
        \includegraphics[width=\textwidth]{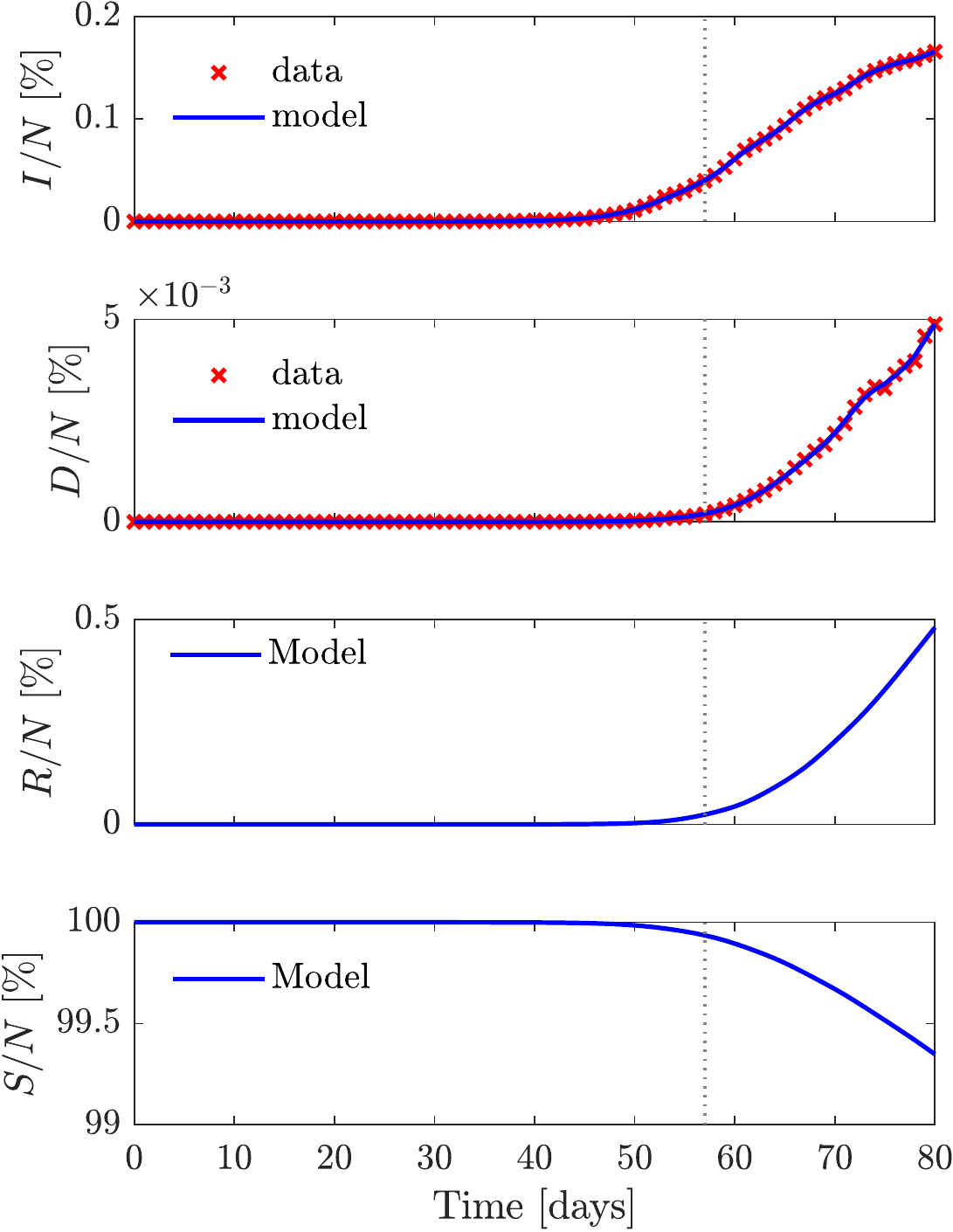}
         \caption{\subcaptionFIGaVAL}
         \label{fig:validation1_Germany}
     \end{subfigure}
     \hfill
     \begin{subfigure}[t]{0.48\textwidth}
         \centering
        \includegraphics[width=\textwidth]{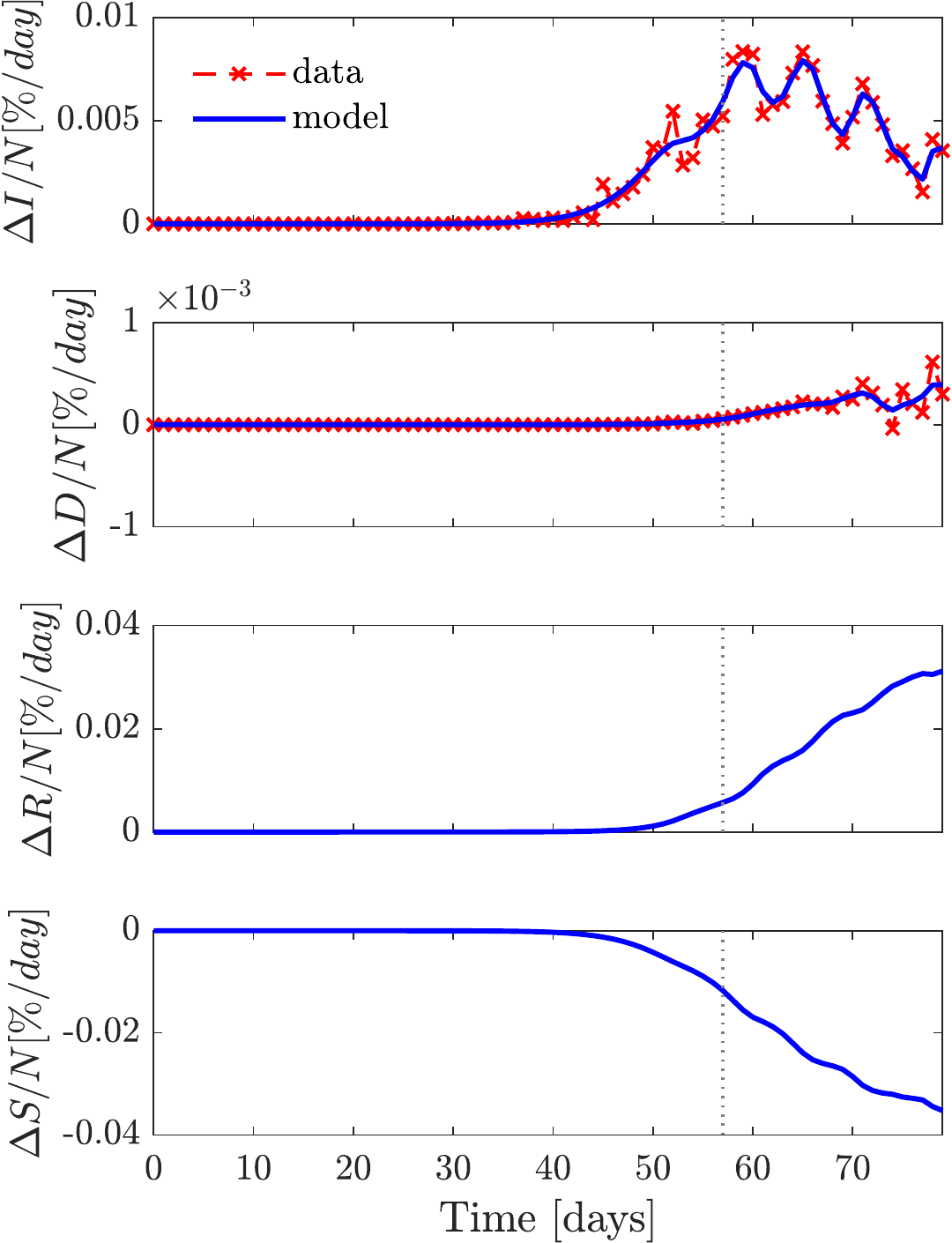}
         \caption{\subcaptionFIGbVAL}
         \label{fig:validation2_Germany}
     \end{subfigure}
     \caption{\Germanylabel \, \captionFIGVAL}
\end{figure}
%
%
\begin{figure}[ht]
     \centering
     \begin{subfigure}[t]{0.48\textwidth}
         \centering
        \includegraphics[width=\textwidth]{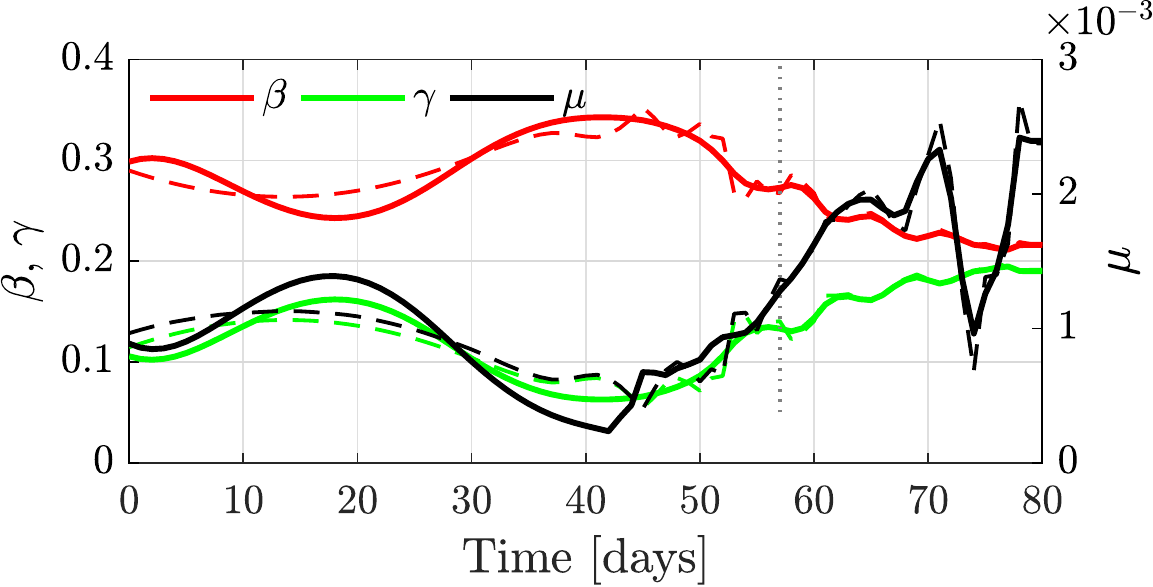}
         \caption{\subcaptionFIGaalpha}
         \label{fig:alpha_Germany}
     \end{subfigure}
     \hfill
     \begin{subfigure}[t]{0.48\textwidth}
         \centering
        \includegraphics[width=\textwidth]{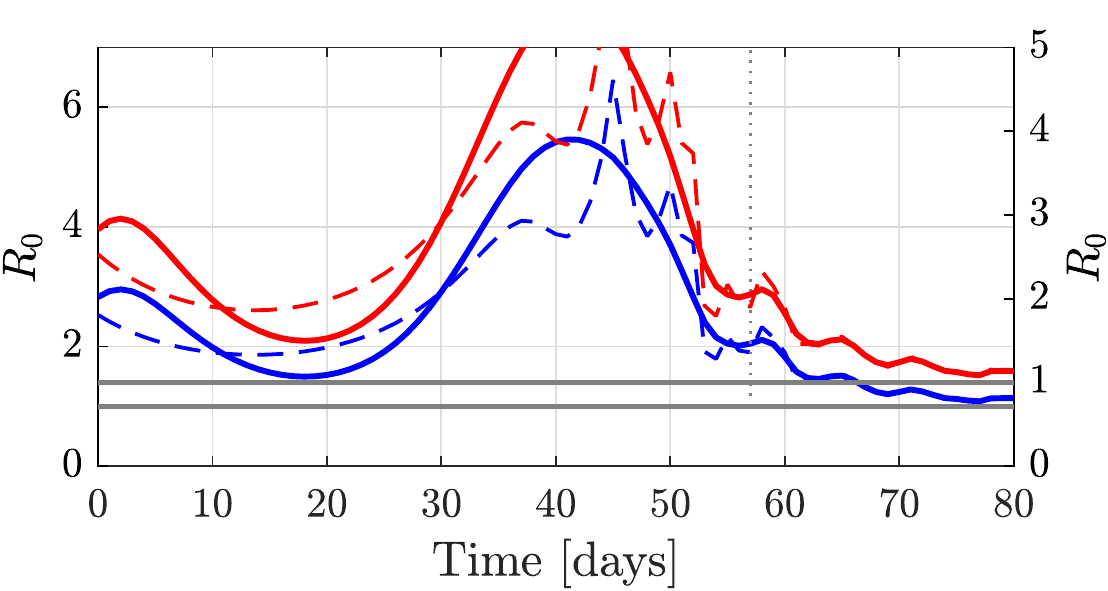}
         \caption{\subcaptionFIGbalpha}
         \label{fig:R0_Germany}
     \end{subfigure}
     \caption{\Germanylabel \,\captionFIGalpha}
\end{figure}
\begin{figure}[ht]
     \centering
     \begin{subfigure}[t]{0.48\textwidth}
         \centering
        \includegraphics[width=\textwidth]{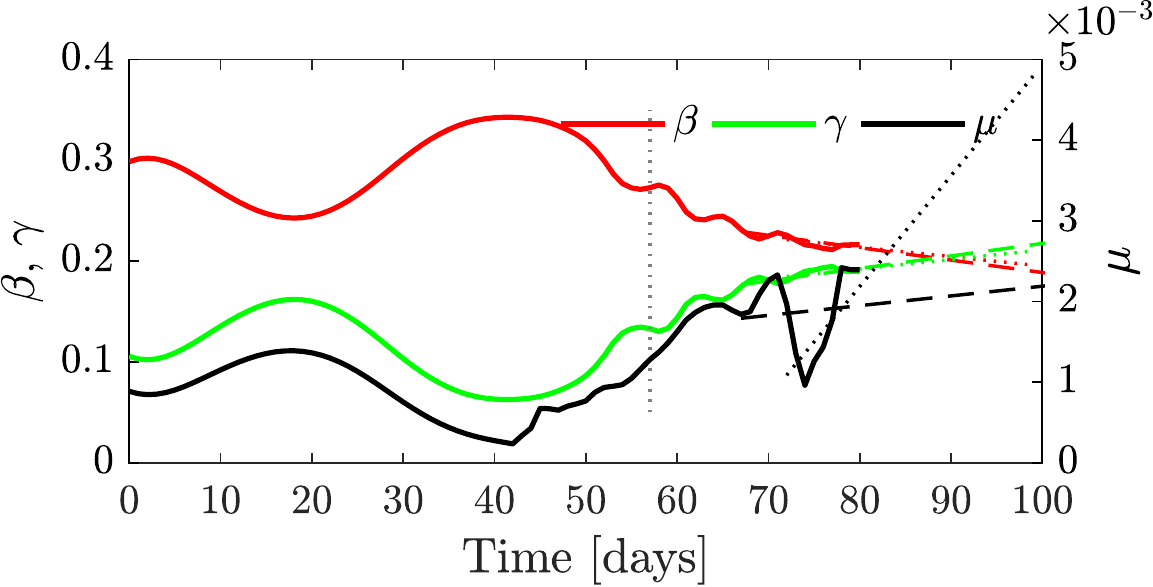}
         \caption{Extrapolated trends of the time-varying contact rate ($\beta$), recovery rate ($\gamma$), and death rate ($\mu$) with average slope over the last nine days (dotted lines) and fourteen days (dashed lines).   The positive slope of the death rate is a consequence of an anomaly in the data on confirmed deaths. The cause of the anomaly is not known to the authors.}
         \label{fig:extrap_alpha_Germany}
     \end{subfigure}
     \hfill
     \begin{subfigure}[t]{0.48\textwidth}
         \centering
    \includegraphics[width=\textwidth]{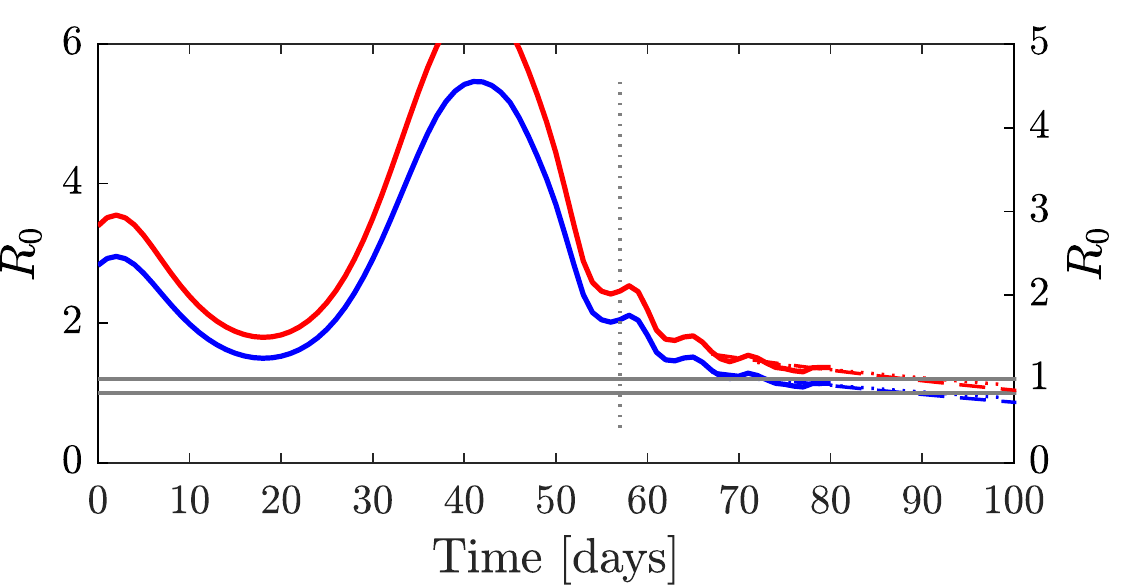}
         \caption{Extrapolated trend of the basic reproduction number with average slope over the last nine days (dotted lines) and fourteen days (dashed lines). }
         \label{fig:extrap_R0_Germany}
     \end{subfigure}
     \caption{\Germanylabel \,\captionFIGextrap}
\end{figure}

     \begin{figure}[h] 
         \centering
        \includegraphics[width=0.48\textwidth]{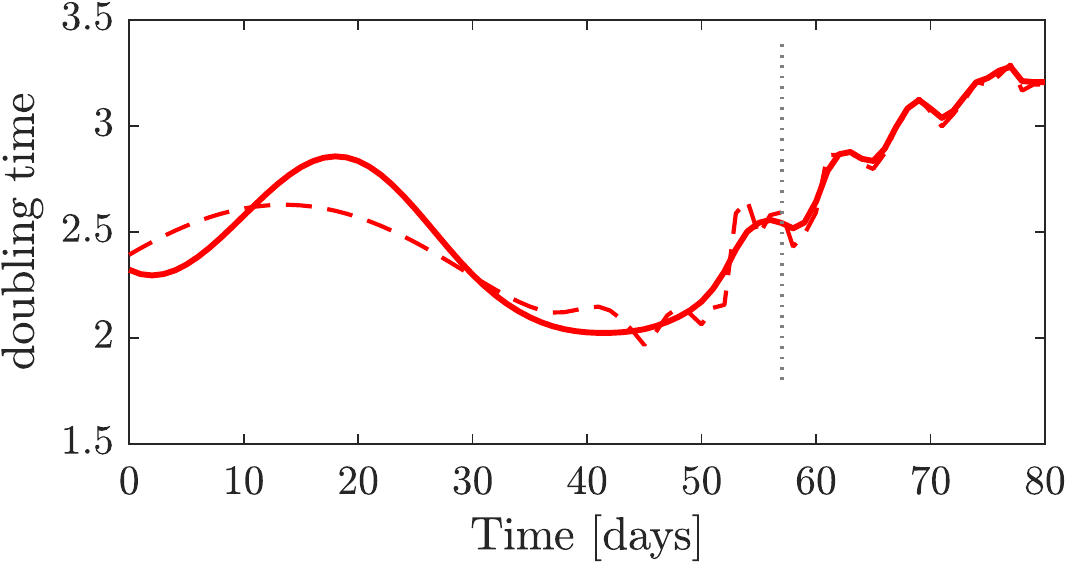}
         \caption{\Germanylabel \, \captionFIGdoubling }
         \label{fig:doubling_Germany}
     \end{figure}
     \begin{figure}[h] 
         \centering
        \includegraphics[width=0.48\textwidth]{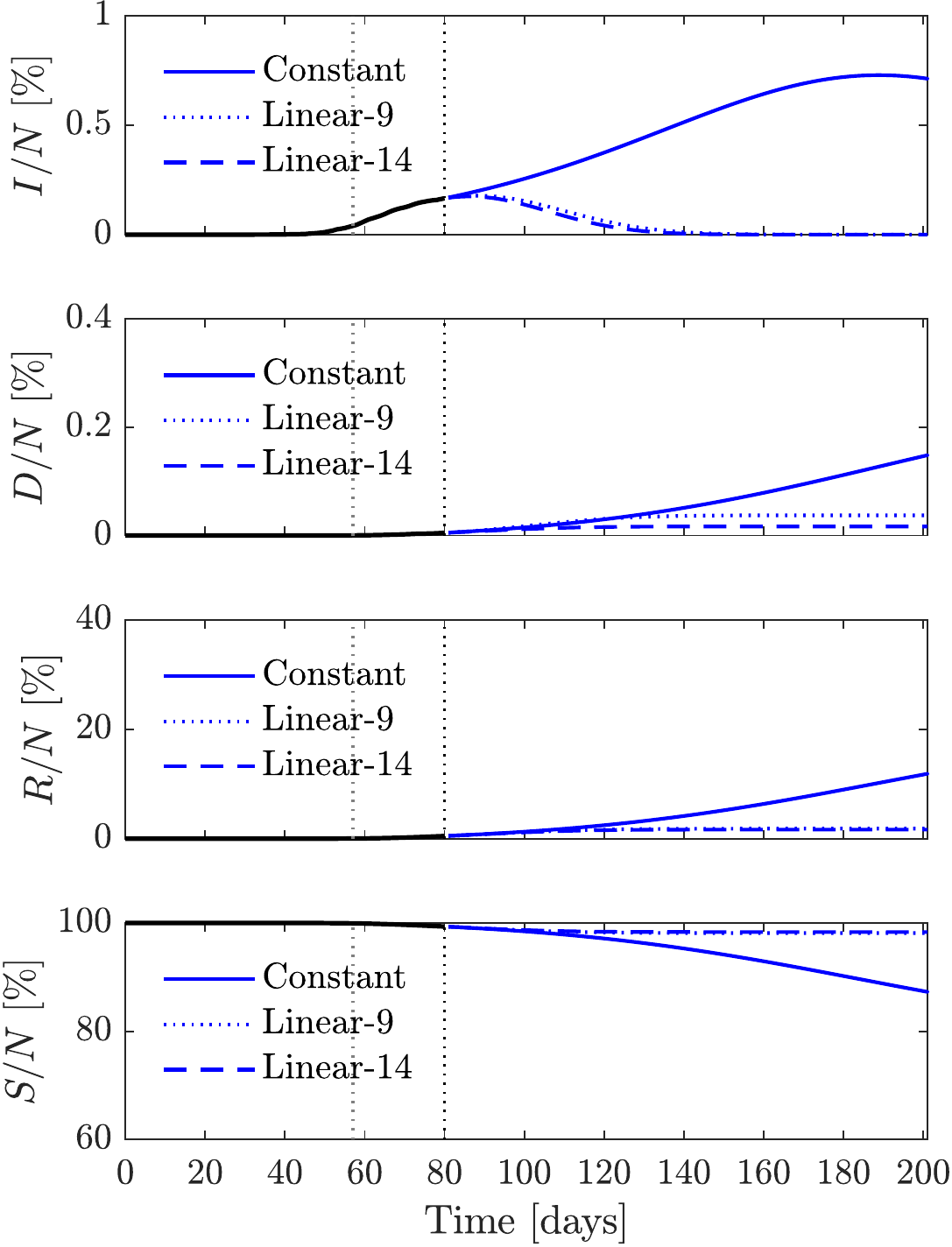}
         \caption{\Germanylabel \, From the top: Blue lines  indicate the extrapolated trends of the percentage of infected, recovered, deaths, and susceptible. Estimates with average slope over the last nine days (dotted lines),  fourteen days (dashed lines), and with values of the parameters assumed to be constant and equal to the last day (solid lines). Black lines: The left vertical dotted line represents the day of  lockdown, the right vertical line is the last day of the training data set, hence, the starting day for extrapolation. The black solid lines are taken from Fig.~\ref{fig:validation1_Germany}. }
         \label{fig:extrap_state_Germany}
     \end{figure}

     \clearpage
     
\subsection{France}

\begin{figure}[ht]
     \centering
     \begin{subfigure}[t]{0.48\textwidth}
         \centering
        \includegraphics[width=\textwidth]{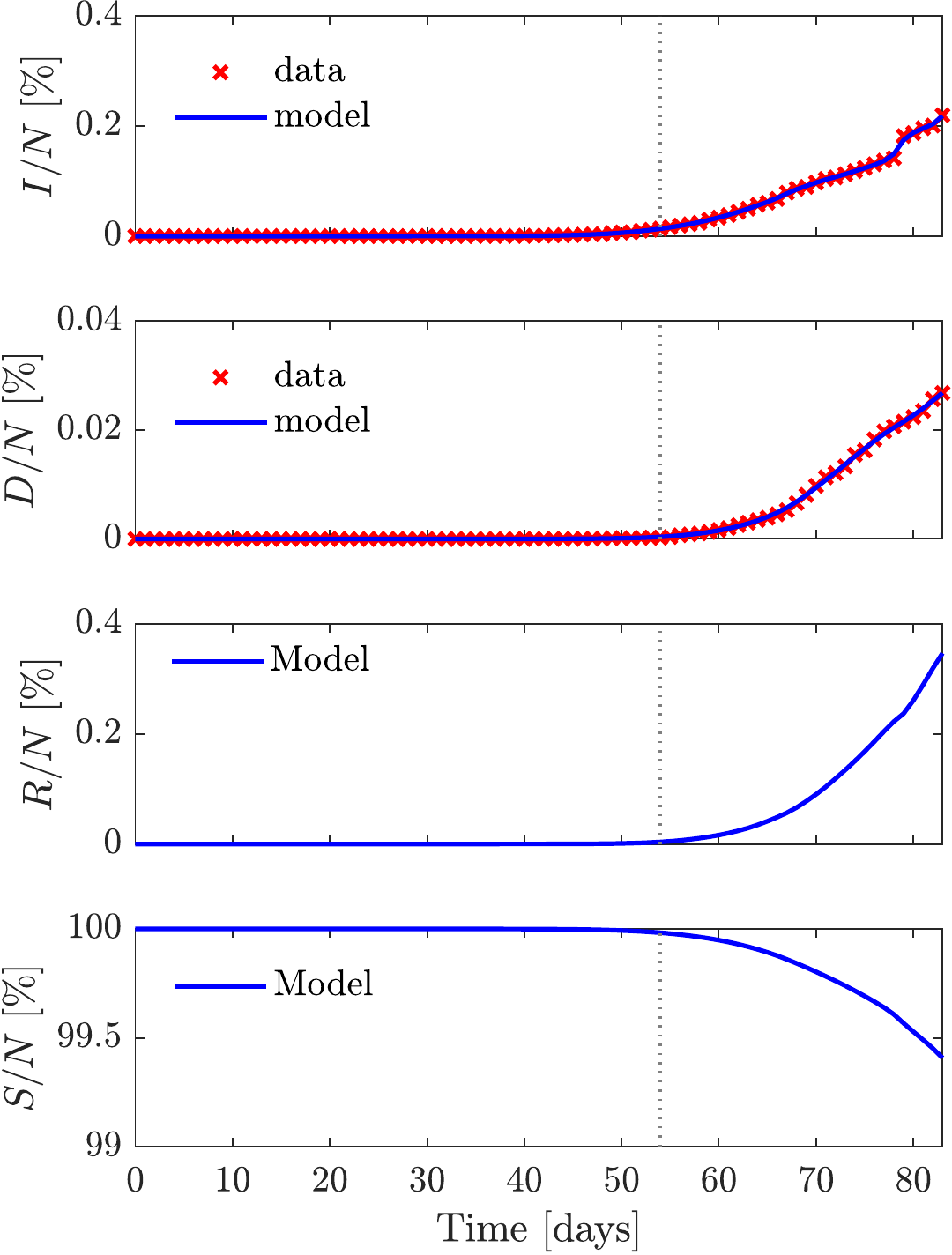}
         \caption{\subcaptionFIGaVAL}
         \label{fig:validation1_France}
     \end{subfigure}
     \hfill
     \begin{subfigure}[t]{0.48\textwidth}
         \centering
        \includegraphics[width=\textwidth]{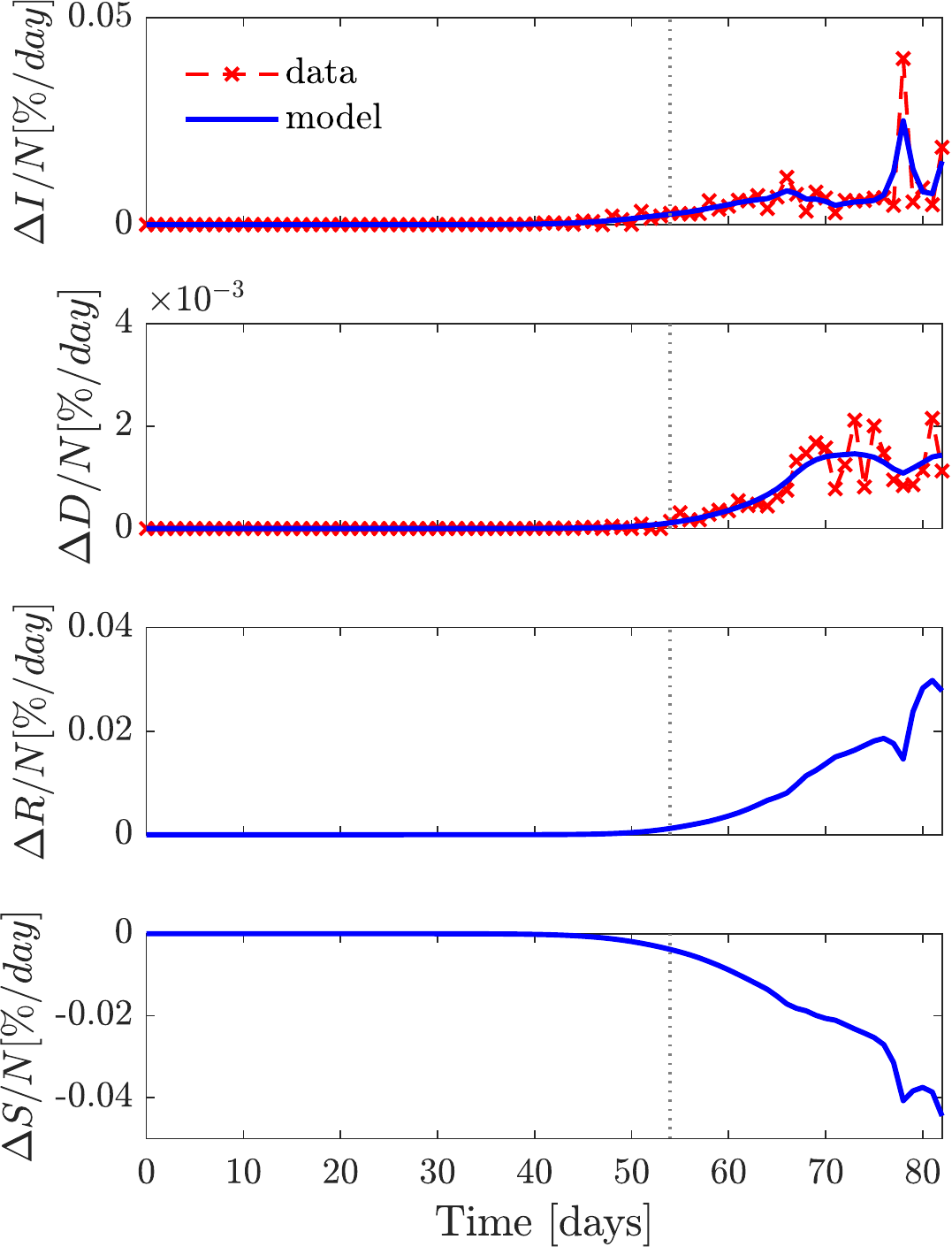}
         \caption{\subcaptionFIGbVAL}
         \label{fig:validation2_France}
     \end{subfigure}
     \caption{\Francelabel \, \captionFIGVAL}
\end{figure}
%
%
\begin{figure}[ht]
     \centering
     \begin{subfigure}[t]{0.48\textwidth}
         \centering
        \includegraphics[width=\textwidth]{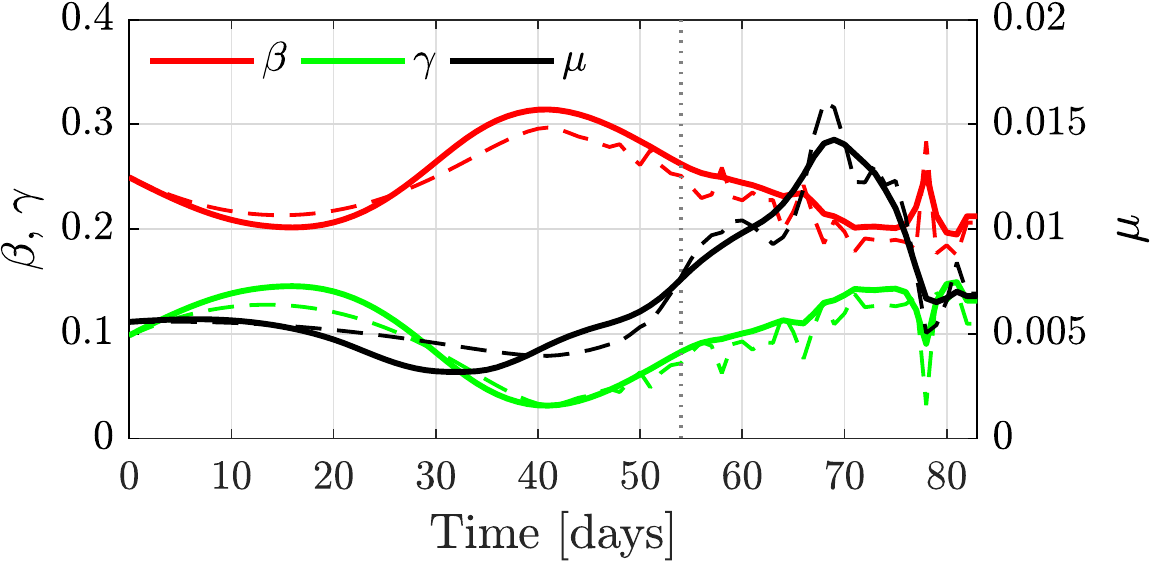}
         \caption{\subcaptionFIGaalpha}
         \label{fig:alpha_France}
     \end{subfigure}
     \hfill
     \begin{subfigure}[t]{0.48\textwidth}
         \centering
        \includegraphics[width=\textwidth]{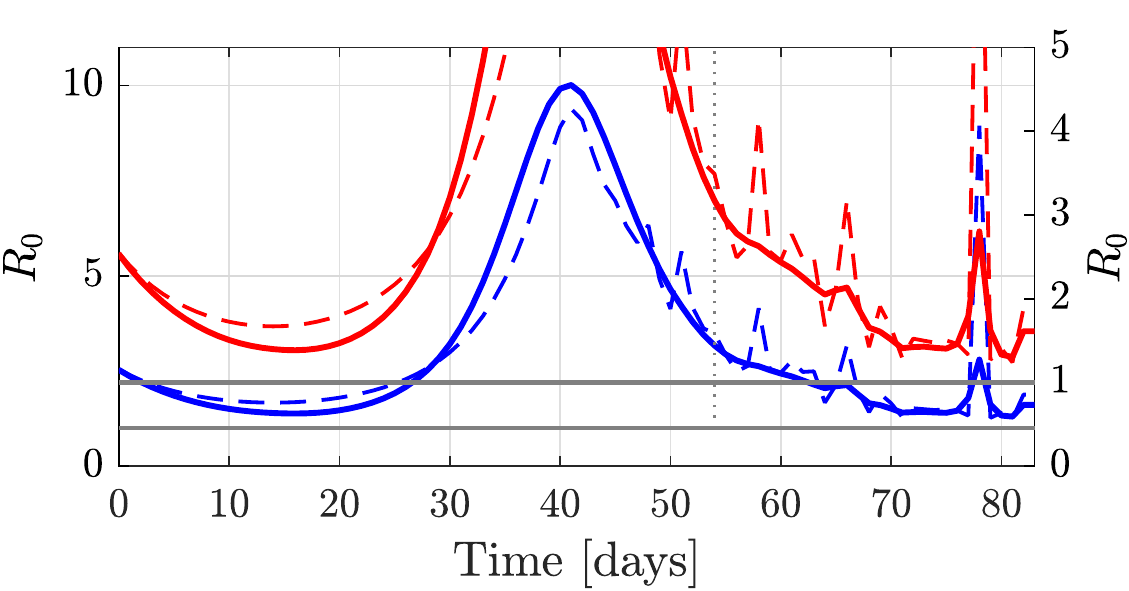}
         \caption{\subcaptionFIGbalpha}
         \label{fig:R0_France}
     \end{subfigure}
     \caption{\Francelabel \,\captionFIGalpha}
\end{figure}
\begin{figure}[ht]
     \centering
     \begin{subfigure}[t]{0.48\textwidth}
         \centering
        \includegraphics[width=\textwidth]{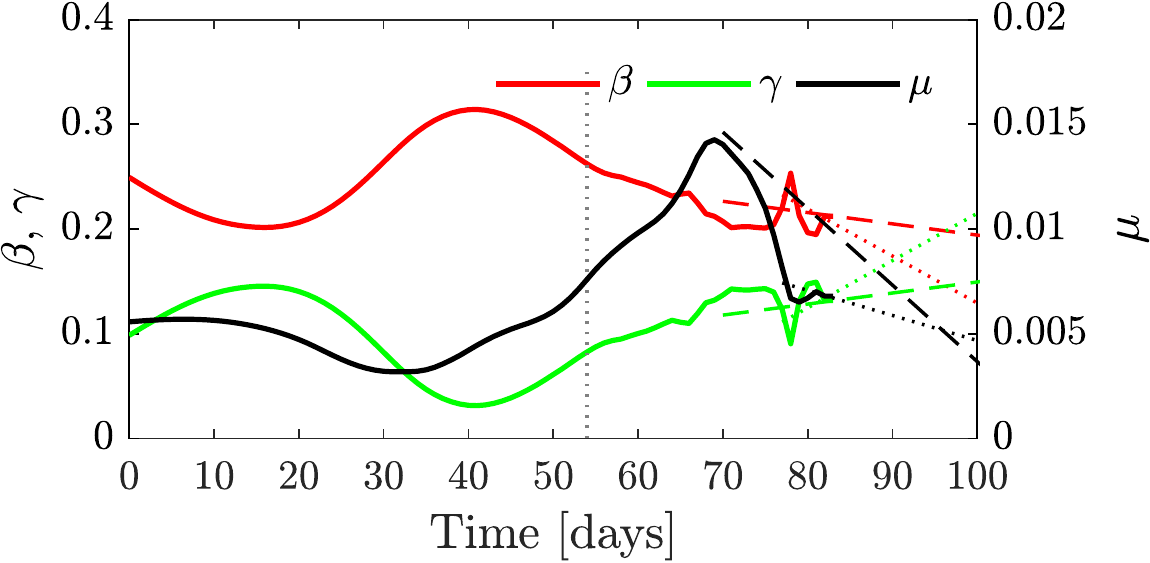}
         \caption{Extrapolated trends of the time-varying contact rate ($\beta$), recovery rate ($\gamma$), and death rate ($\mu$) with average slope over the last seven days (dotted lines) and twenty-one days (dashed lines).      }
         \label{fig:extrap_alpha_France}
     \end{subfigure}
     \hfill
     \begin{subfigure}[t]{0.48\textwidth}
         \centering
        \includegraphics[width=\textwidth]{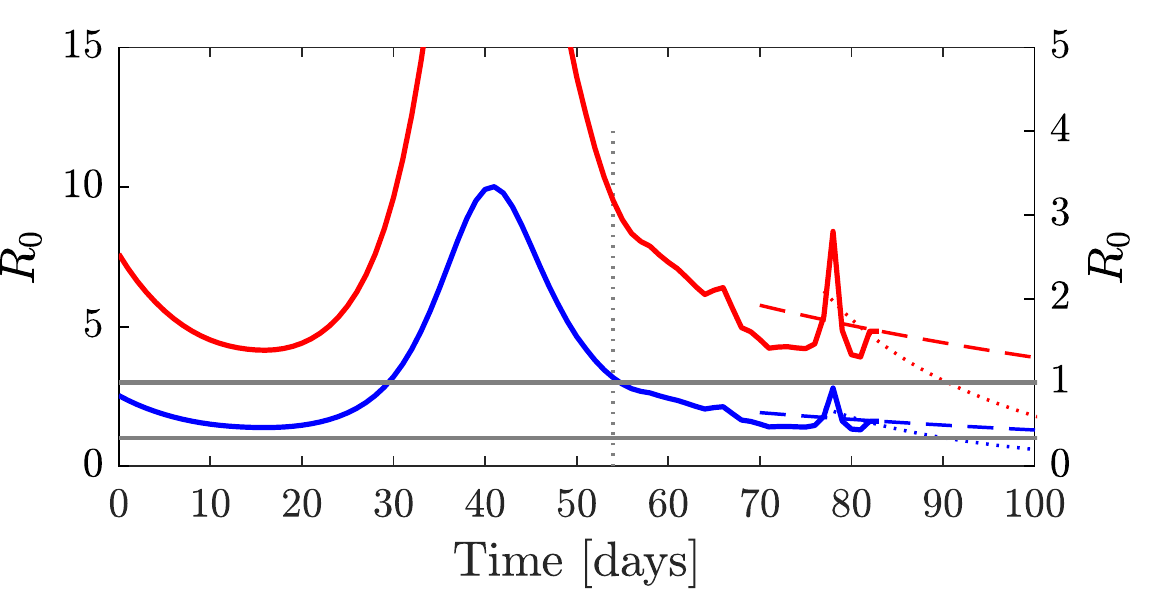}
         \caption{Extrapolated trend of the basic reproduction number with average slope over the last seven days (dotted lines) and twenty-one days (dashed lines). }
         \label{fig:extrap_R0_France}
     \end{subfigure}
     \caption{\Francelabel \,\captionFIGextrap}
\end{figure}

     \begin{figure}[h] 
         \centering
        \includegraphics[width=0.48\textwidth]{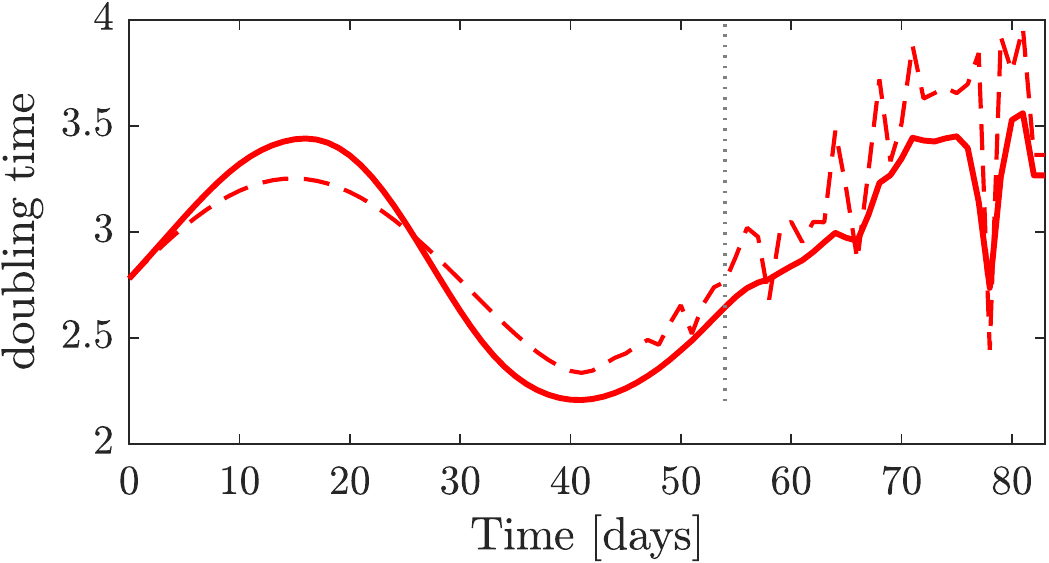}
         \caption{\Francelabel \, \captionFIGdoubling }
         \label{fig:doubling_France}
     \end{figure}
     \begin{figure}[h] 
         \centering
        \includegraphics[width=0.48\textwidth]{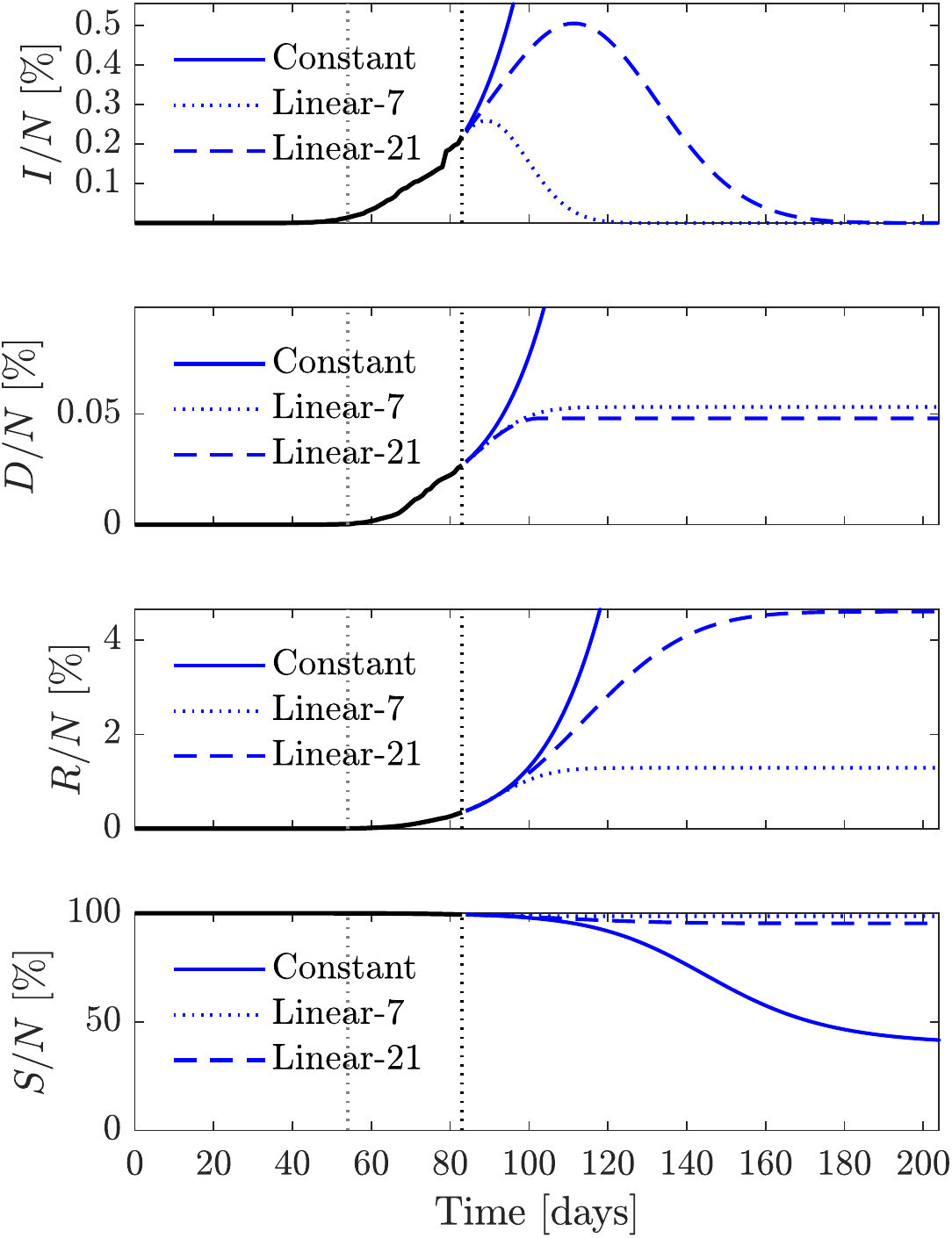}
         \caption{\Francelabel \, From the top: Blue lines  indicate the extrapolated trends of the percentage of infected, recovered, deaths, and susceptible. Estimates with average slope over the last seven days (dotted lines),  twenty-one days (dashed lines), and with values of the parameters assumed to be constant and equal to the last day (solid lines). Black lines: The left vertical dotted line represents the day of  lockdown, the right vertical line is the last day of the training data set, hence, the starting day for extrapolation. The black solid lines are taken from Fig.~\ref{fig:validation1_France}. }
         \label{fig:extrap_state_France}
     \end{figure}

     \clearpage

\subsection{Spain}

\begin{figure}[ht]
     \centering
     \begin{subfigure}[t]{0.48\textwidth}
         \centering
        \includegraphics[width=\textwidth]{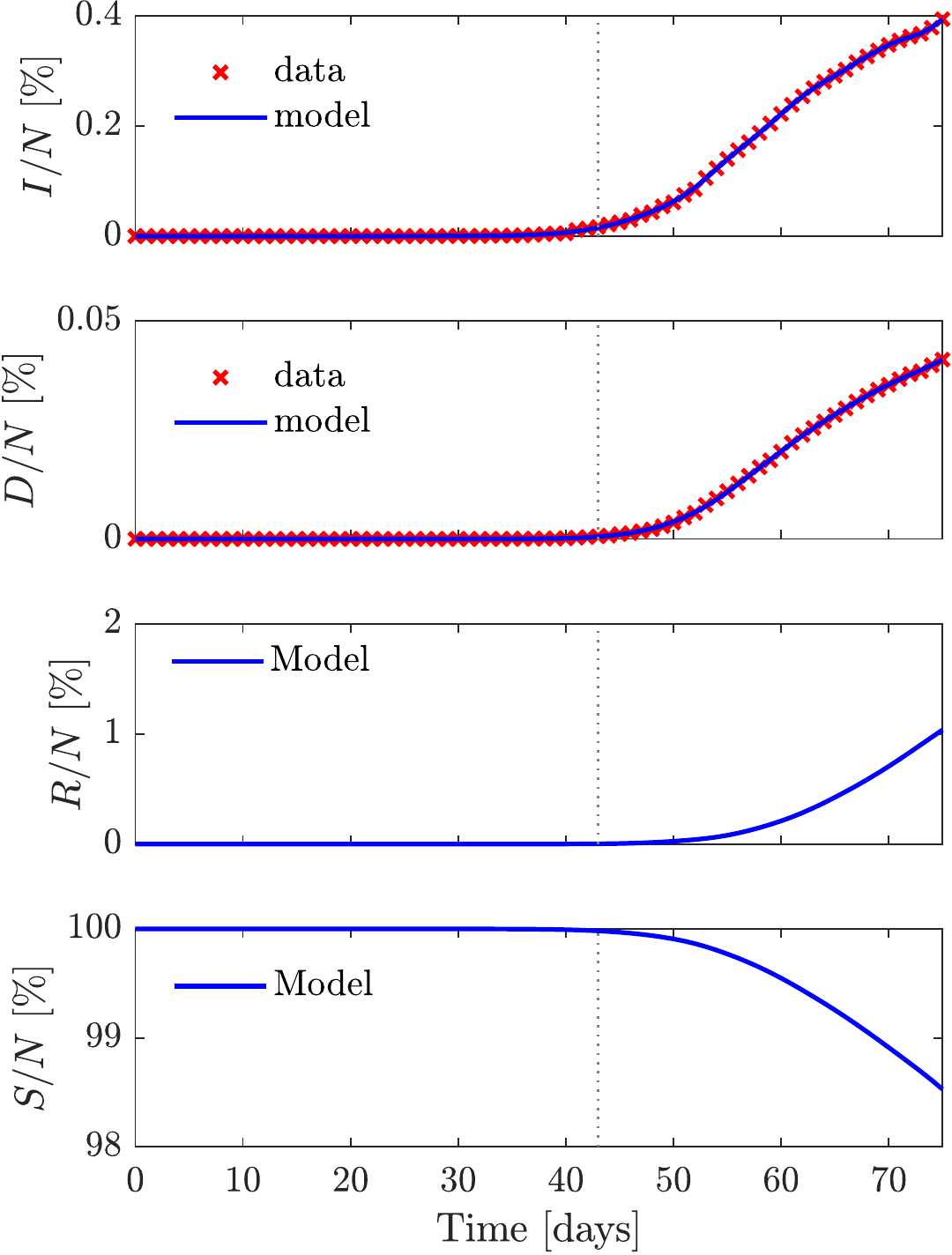}
         \caption{\subcaptionFIGaVAL}
         \label{fig:validation1_Spain}
     \end{subfigure}
     \hfill
     \begin{subfigure}[t]{0.48\textwidth}
         \centering
        \includegraphics[width=\textwidth]{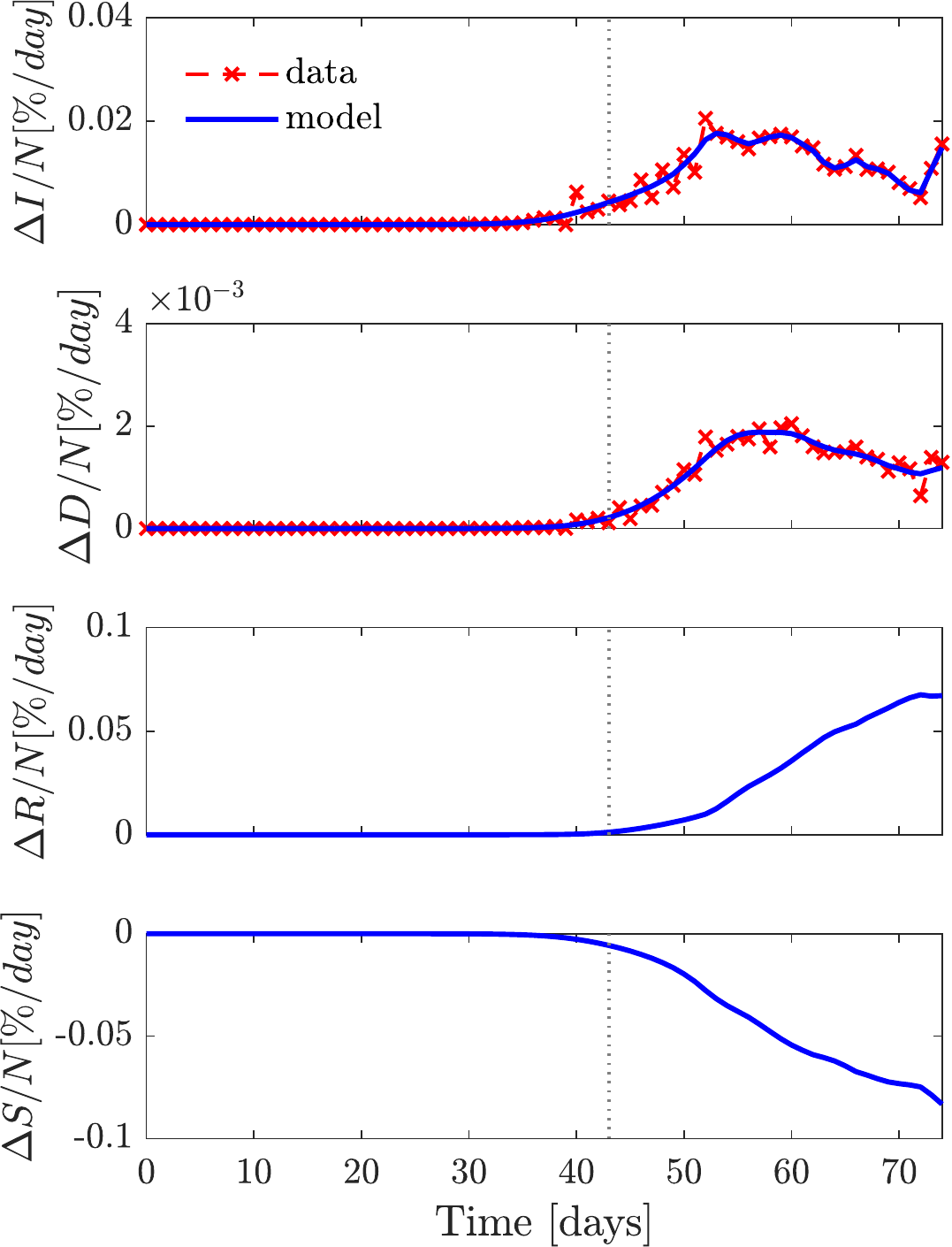}
         \caption{\subcaptionFIGbVAL}
         \label{fig:validation2_Spain}
     \end{subfigure}
     \caption{\Spainlabel \, \captionFIGVAL}
\end{figure}
%
%
\begin{figure}[ht]
     \centering
     \begin{subfigure}[t]{0.48\textwidth}
         \centering
        \includegraphics[width=\textwidth]{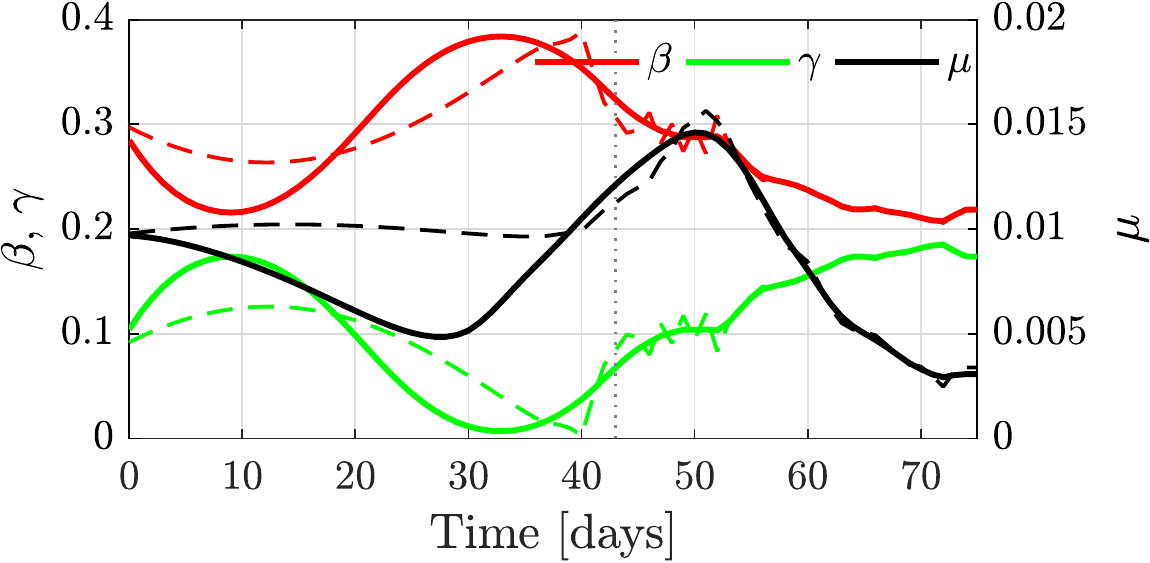}
         \caption{\subcaptionFIGaalpha}
         \label{fig:alpha_Spain}
     \end{subfigure}
     \hfill
     \begin{subfigure}[t]{0.48\textwidth}
         \centering
        \includegraphics[width=\textwidth]{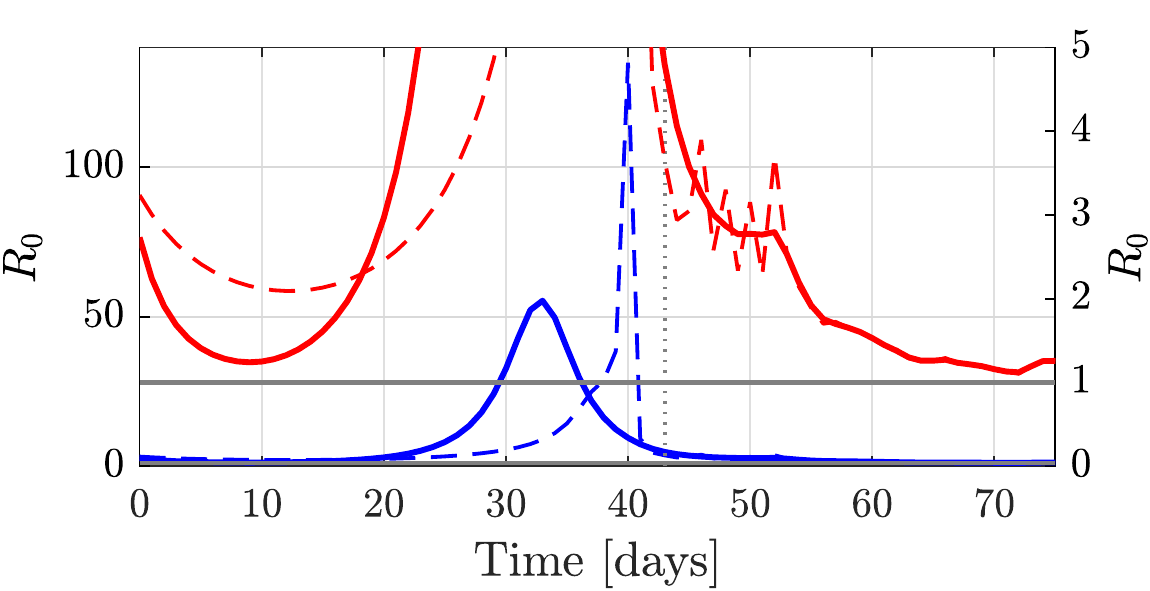}
         \caption{\subcaptionFIGbalpha}
         \label{fig:R0_Spain}
     \end{subfigure}
     \caption{\Spainlabel \,\captionFIGalpha}
\end{figure}
\begin{figure}[ht]
     \centering
     \begin{subfigure}[t]{0.48\textwidth}
         \centering
        \includegraphics[width=\textwidth]{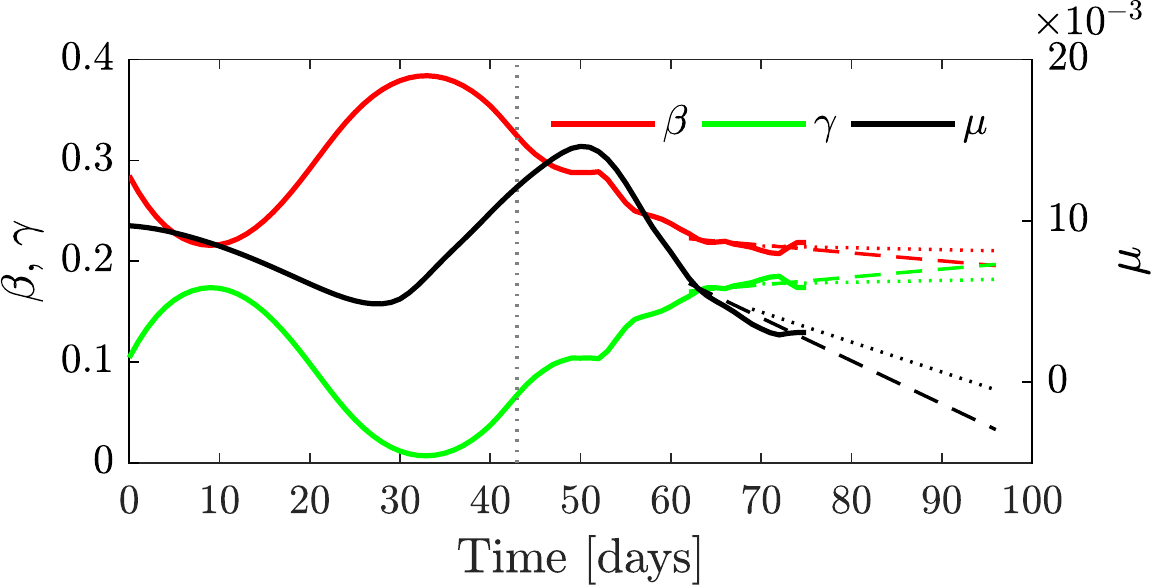}
         \caption{Extrapolated trends of the time-varying contact rate ($\beta$), recovery rate ($\gamma$), and death rate ($\mu$) with average slope over the last ten days (dotted lines) and fourteen days (dashed lines).     }
         \label{fig:extrap_alpha_Spain}
     \end{subfigure}
     \hfill
     \begin{subfigure}[t]{0.48\textwidth}
         \centering
        \includegraphics[width=\textwidth]{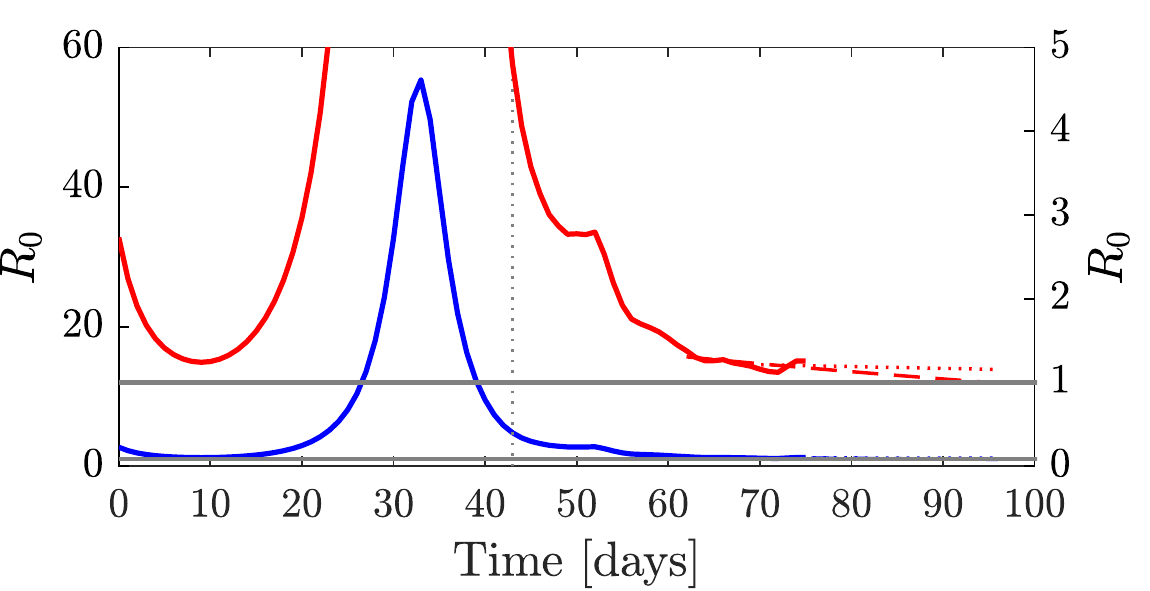}
         \caption{Extrapolated trend of the basic reproduction number with average slope over the last ten days (dotted lines) and fourteen days (dashed lines).  }
         \label{fig:extrap_R0_Spain}
     \end{subfigure}
     \caption{\Spainlabel \,\captionFIGextrap}
\end{figure}

     \begin{figure}[h] 
         \centering
        \includegraphics[width=0.48\textwidth]{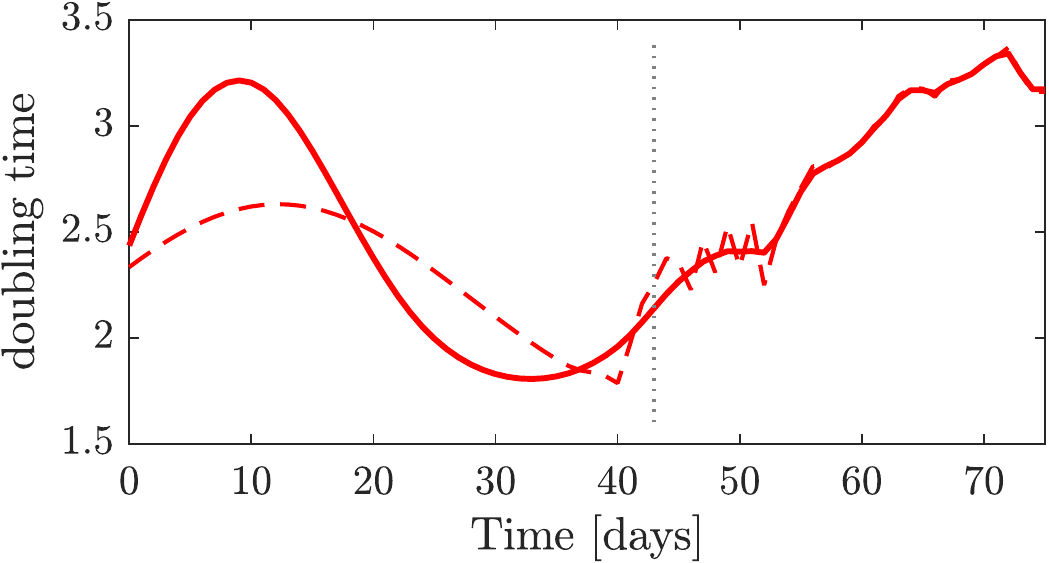}
         \caption{\Spainlabel \, \captionFIGdoubling }
         \label{fig:doubling_Spain}
     \end{figure}
     
     \begin{figure}[h] 
         \centering
        \includegraphics[width=0.48\textwidth]{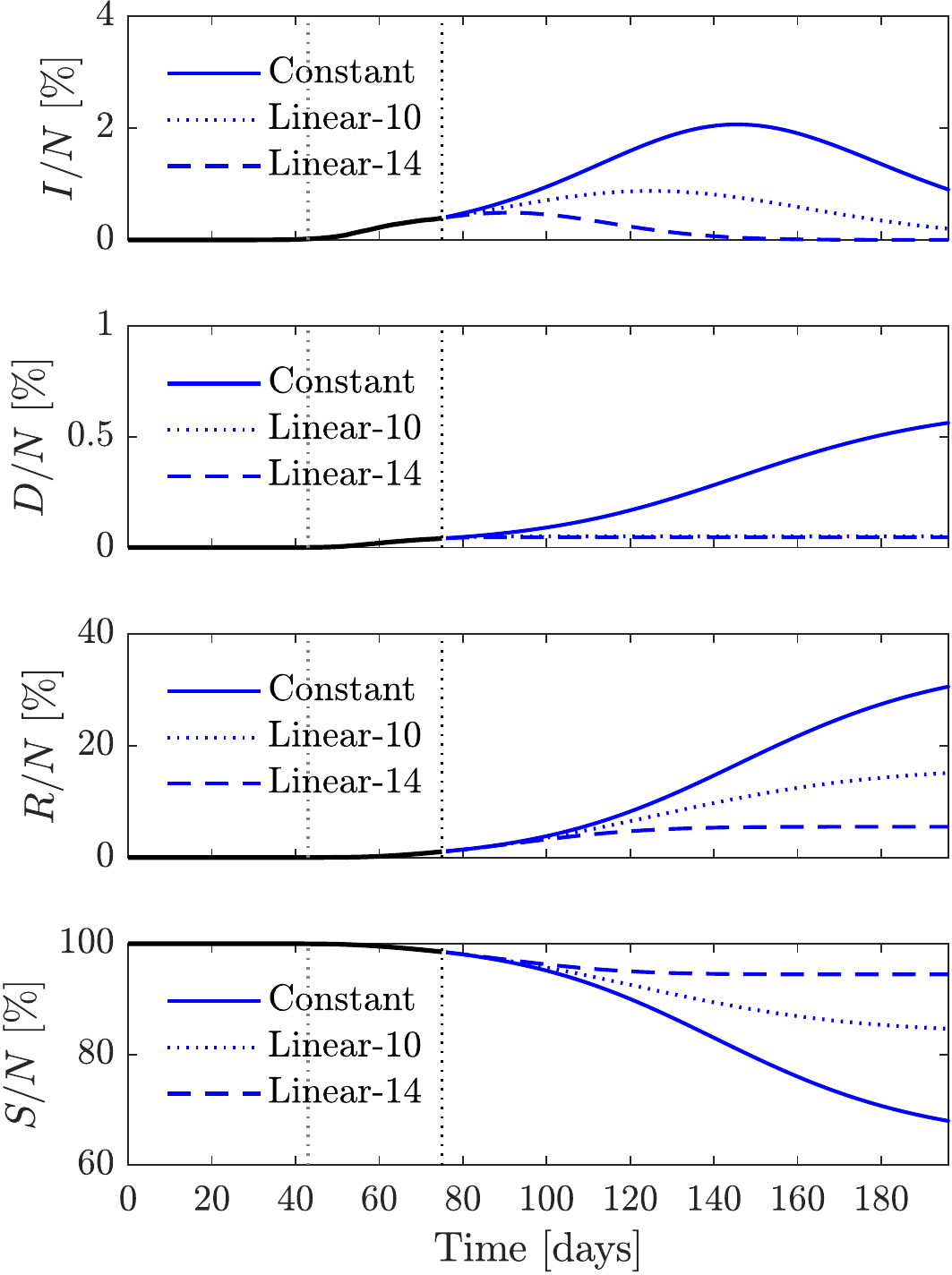}
         \caption{\Spainlabel \, From the top: Blue lines  indicate the extrapolated trends of the percentage of infected, recovered, deaths, and susceptible. Estimates with average slope over the last ten days (dotted lines),  twn days (dashed lines), and with values of the parameters assumed to be constant and equal to the last day (solid lines). Black lines: The left vertical dotted line represents the day of  lockdown, the right vertical line is the last day of the training data set, hence, the starting day for extrapolation.  The black solid lines are taken from Fig.~\ref{fig:validation1_Spain}. }
         \label{fig:extrap_state_Spain}
     \end{figure}

     \clearpage

\subsection{Belgium}

\begin{figure}[ht]
     \centering
     \begin{subfigure}[t]{0.48\textwidth}
         \centering
        \includegraphics[width=\textwidth]{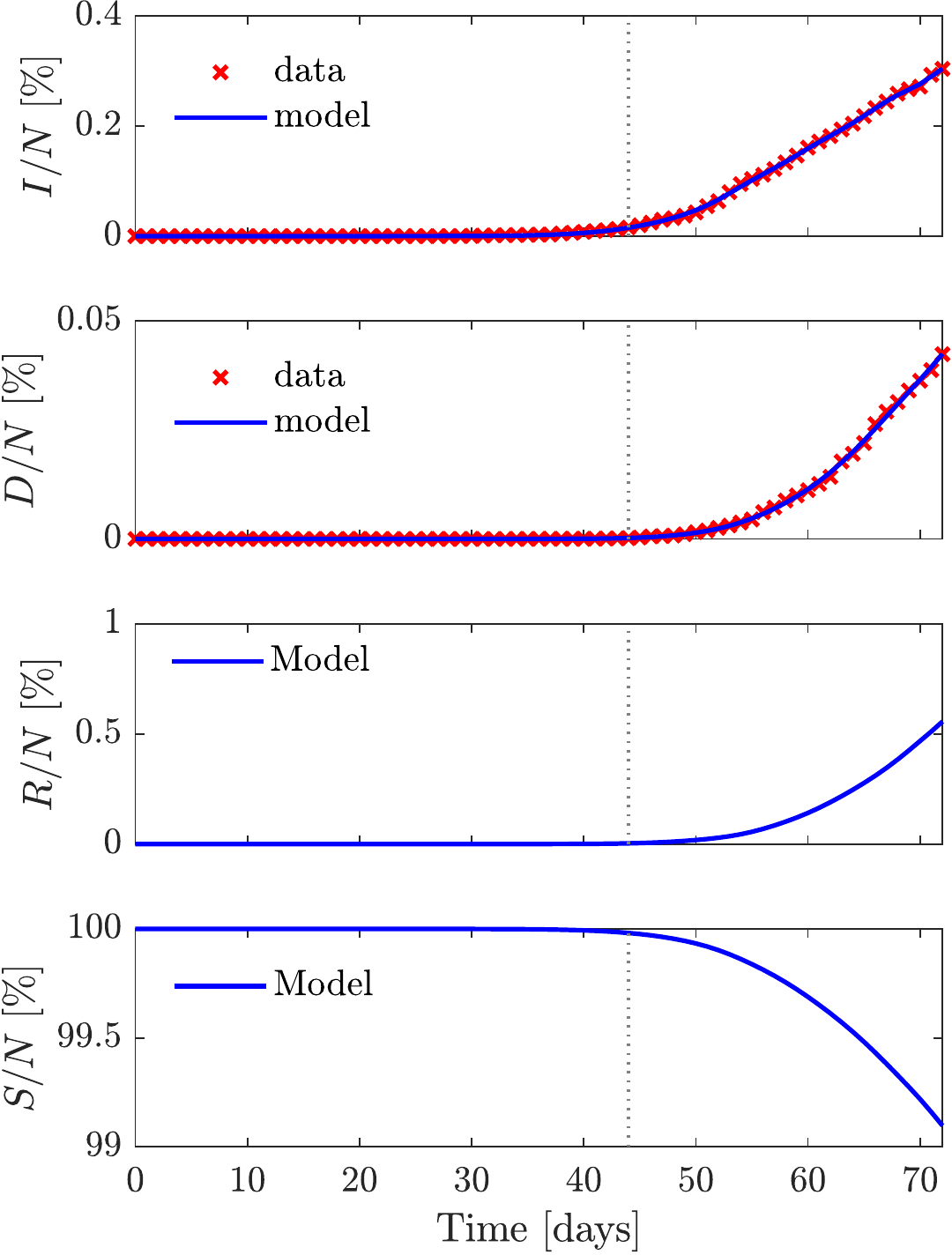}
         \caption{\subcaptionFIGaVAL}
         \label{fig:validation1_Belgium}
     \end{subfigure}
     \hfill
     \begin{subfigure}[t]{0.48\textwidth}
         \centering
        \includegraphics[width=\textwidth]{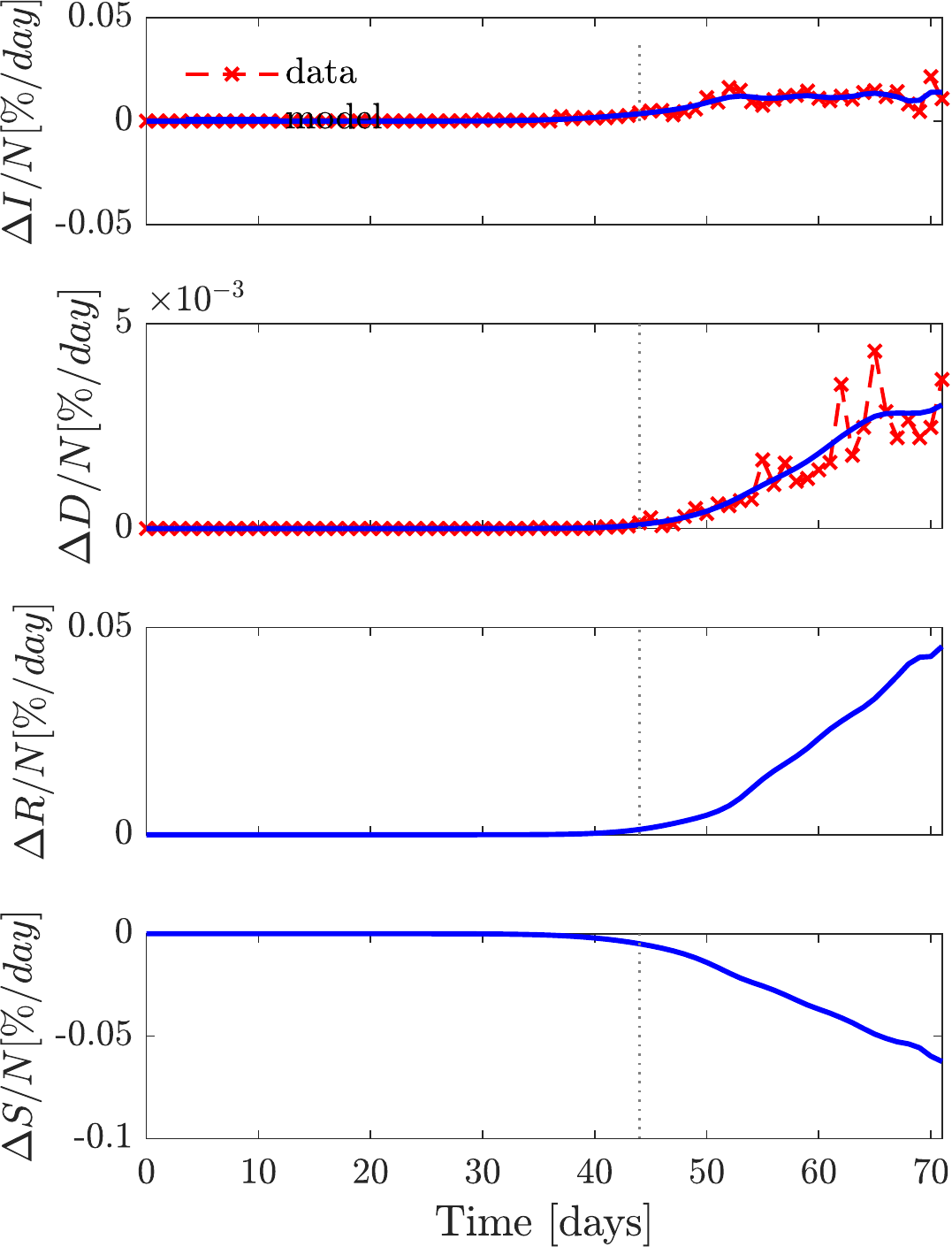}
         \caption{\subcaptionFIGbVAL}
         \label{fig:validation2_Belgium}
     \end{subfigure}
     \caption{\Belgiumlabel \, \captionFIGVAL}
\end{figure}
%
%
\begin{figure}[ht]
     \centering
     \begin{subfigure}[t]{0.48\textwidth}
         \centering
        \includegraphics[width=\textwidth]{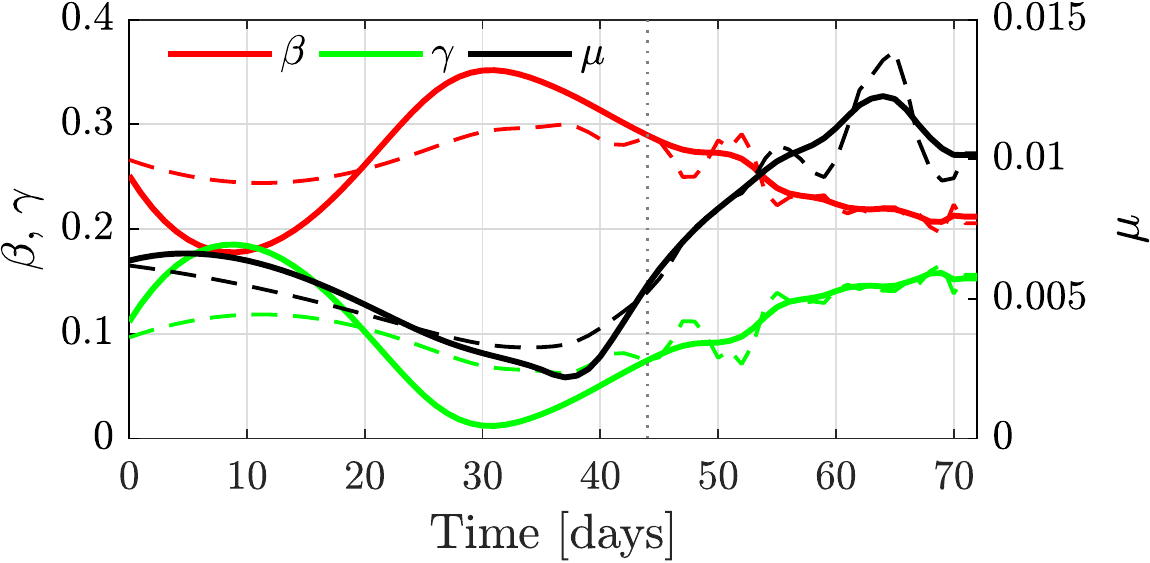}
         \caption{\subcaptionFIGaalpha}
         \label{fig:alpha_Belgium}
     \end{subfigure}
     \hfill
     \begin{subfigure}[t]{0.48\textwidth}
         \centering
        \includegraphics[width=\textwidth]{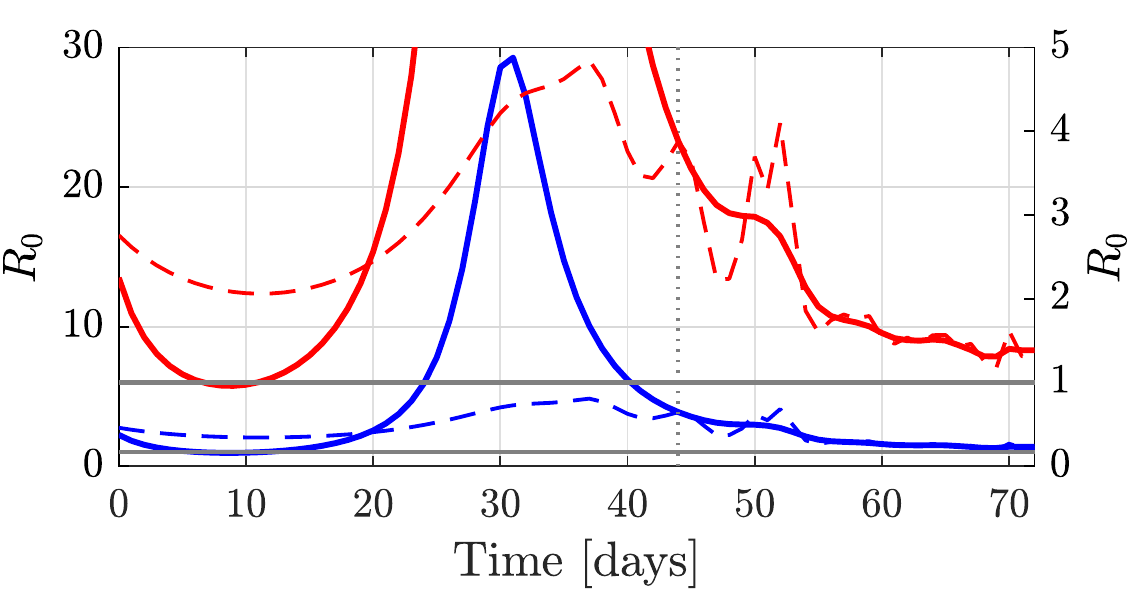}
         \caption{\subcaptionFIGbalpha}
         \label{fig:R0_Belgium}
     \end{subfigure}
     \caption{\Belgiumlabel \,\captionFIGalpha}
\end{figure}
\begin{figure}[ht]
     \centering
     \begin{subfigure}[t]{0.48\textwidth}
         \centering
        \includegraphics[width=\textwidth]{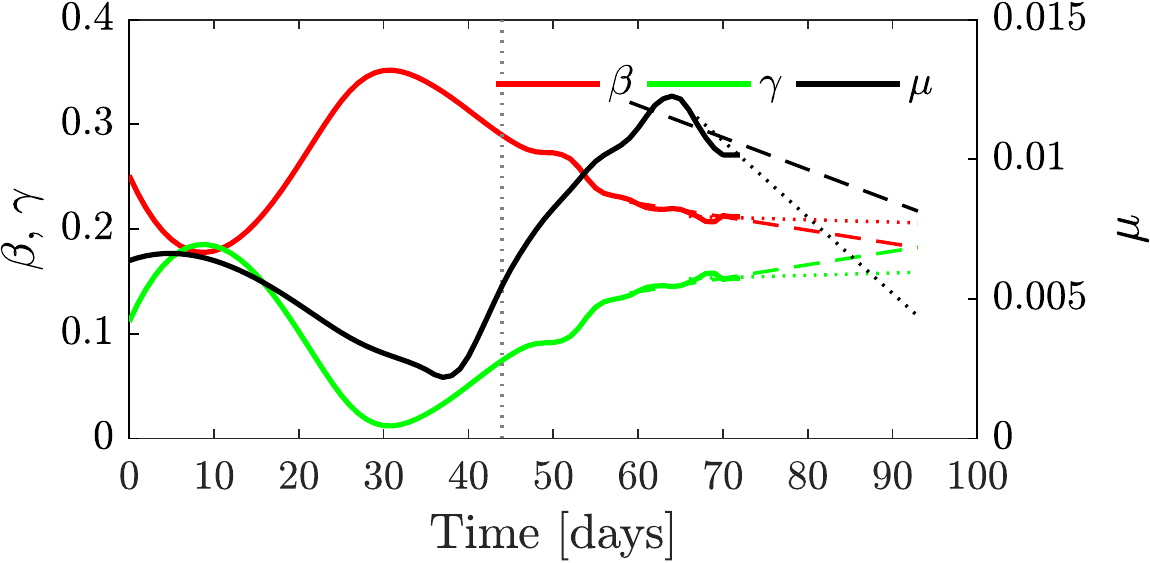}
         \caption{\subcaptionFIGaextrap   }
         \label{fig:extrap_alpha_Belgium}
     \end{subfigure}
     \hfill
     \begin{subfigure}[t]{0.48\textwidth}
         \centering
        \includegraphics[width=\textwidth]{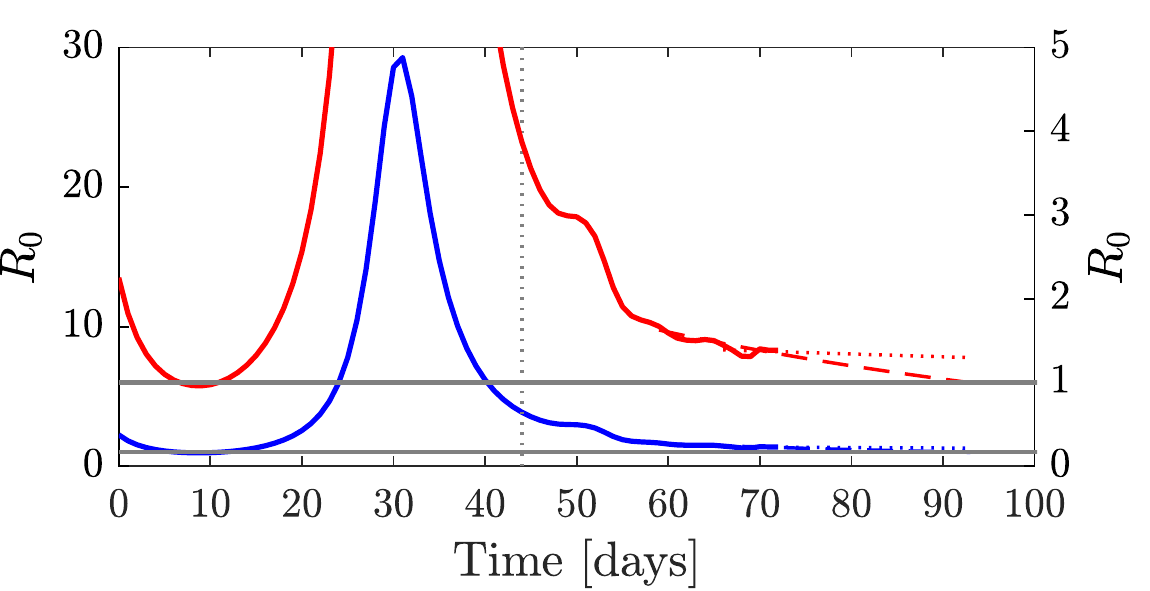}
         \caption{\subcaptionFIGbextrap }
         \label{fig:extrap_R0_Belgium}
     \end{subfigure}
     \caption{\Belgiumlabel \,\captionFIGextrap}
\end{figure}

     \begin{figure}[h] 
         \centering
        \includegraphics[width=0.48\textwidth]{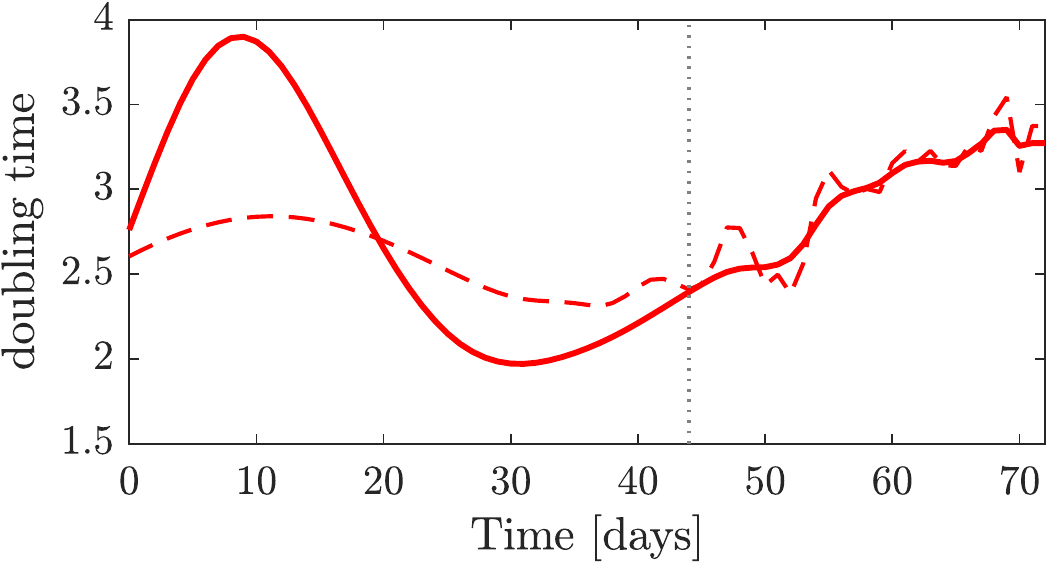}
         \caption{\Belgiumlabel \, \captionFIGdoubling }
         \label{fig:doubling_Belgium}
     \end{figure}
     \begin{figure}[h] 
         \centering
        \includegraphics[width=0.48\textwidth]{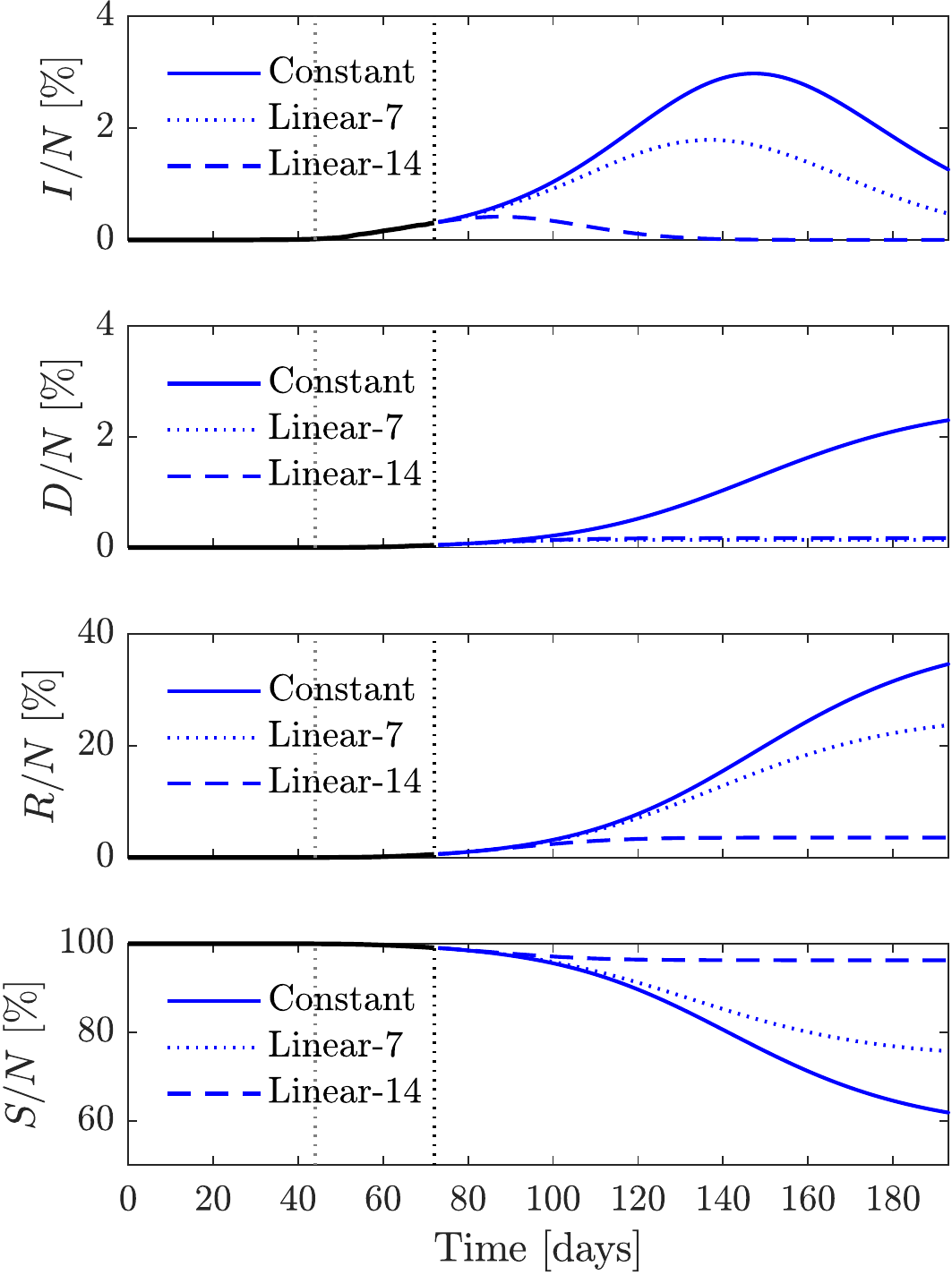}
         \caption{\Belgiumlabel \, \captionFIGextraptwo The black solid lines are taken from Fig.~\ref{fig:validation1_Belgium}. }
         \label{fig:extrap_state_Belgium}
     \end{figure}

     \clearpage

\subsection{USA}

\begin{figure}[ht]
     \centering
     \begin{subfigure}[t]{0.48\textwidth}
         \centering
        \includegraphics[width=\textwidth]{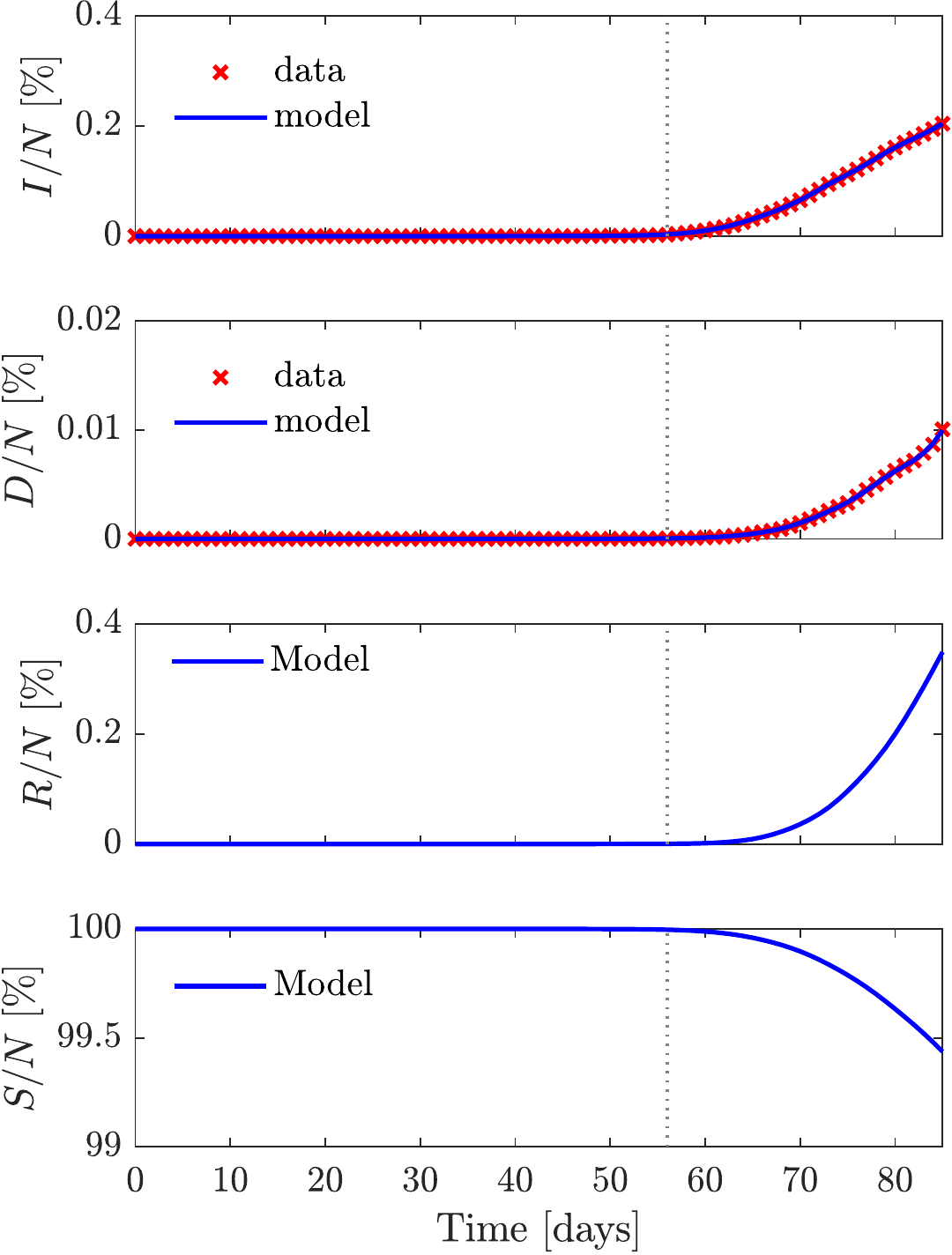}
         \caption{\subcaptionFIGaVAL}
         \label{fig:validation1_US}
     \end{subfigure}
     \hfill
     \begin{subfigure}[t]{0.48\textwidth}
         \centering
        \includegraphics[width=\textwidth]{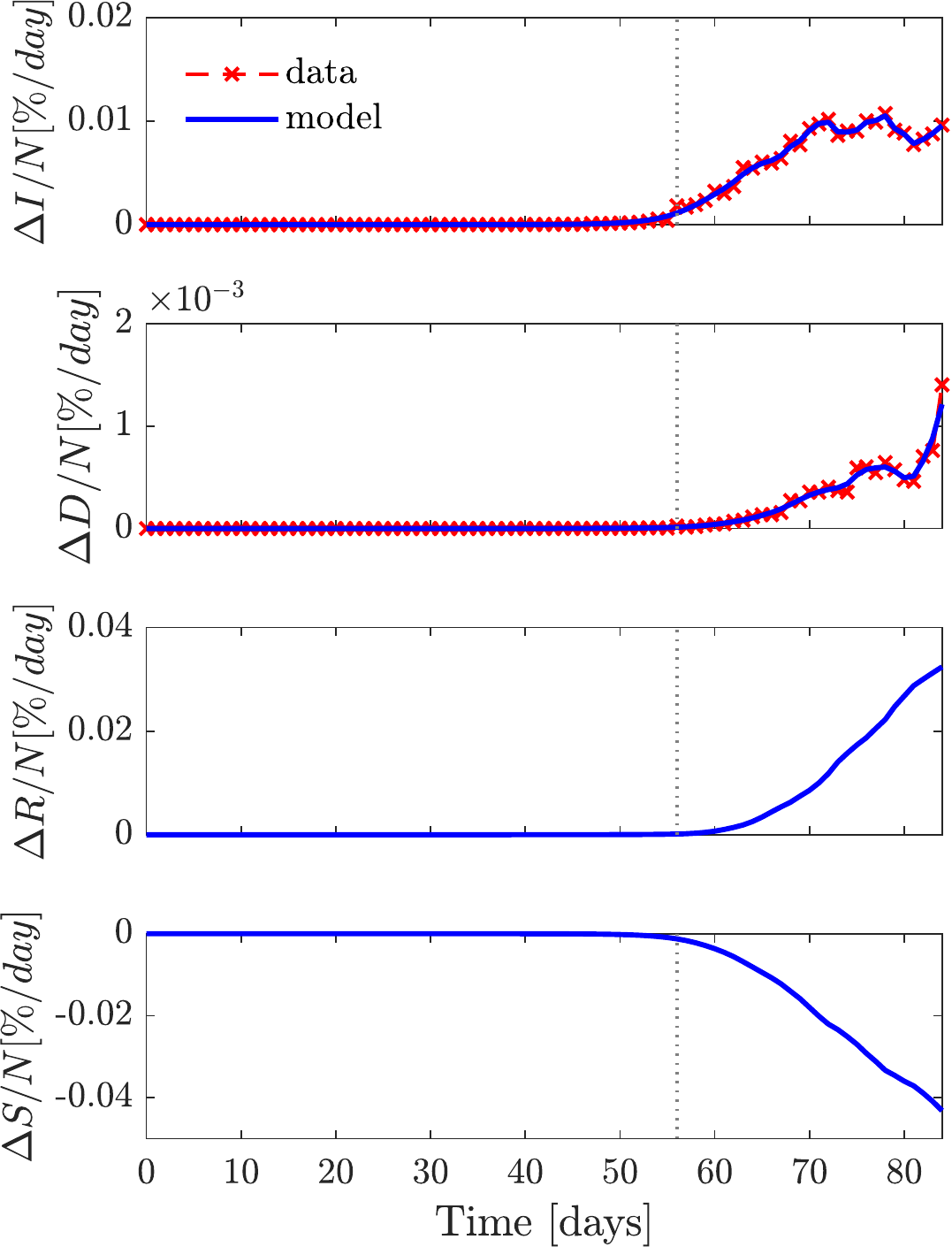}
         \caption{\subcaptionFIGbVAL}
         \label{fig:validation2_US}
     \end{subfigure}
     \caption{\USlabel \, \captionFIGVAL}
\end{figure}
%
%
\begin{figure}[ht]
     \centering
     \begin{subfigure}[t]{0.48\textwidth}
         \centering
        \includegraphics[width=\textwidth]{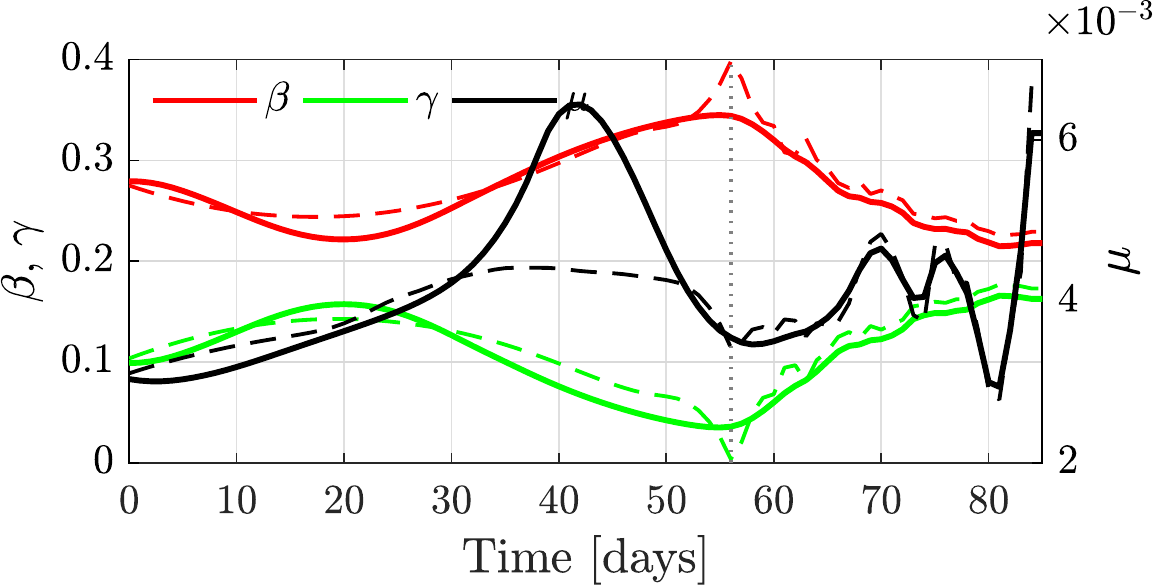}
         \caption{\subcaptionFIGaalpha}
         \label{fig:alpha_US}
     \end{subfigure}
     \hfill
     \begin{subfigure}[t]{0.48\textwidth}
         \centering
        \includegraphics[width=\textwidth]{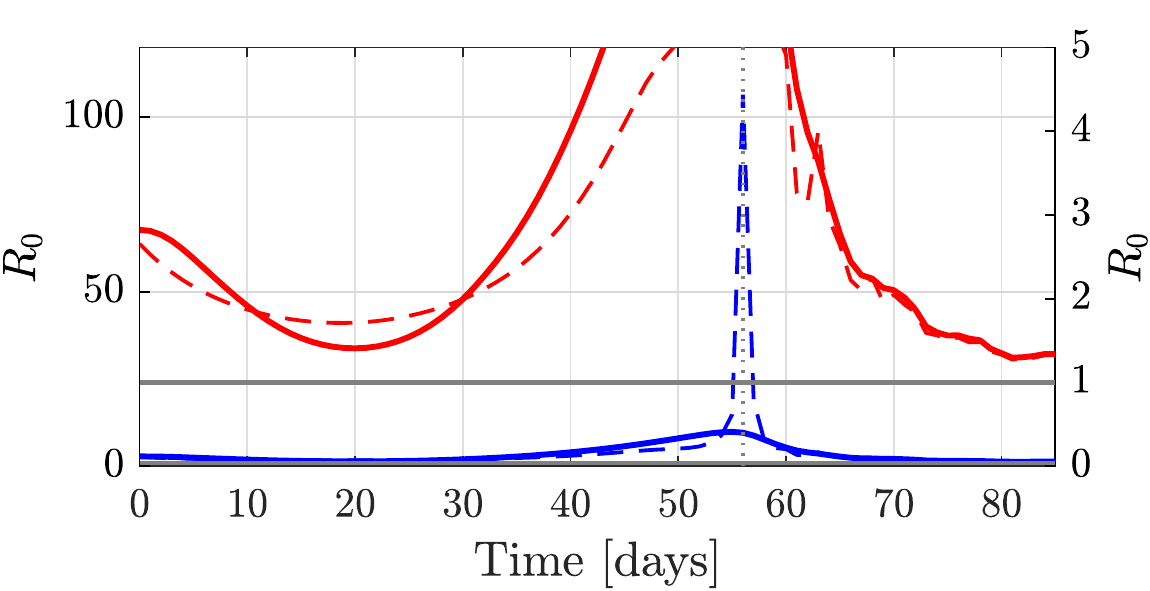}
         \caption{\subcaptionFIGbalpha}
         \label{fig:R0_US}
     \end{subfigure}
     \caption{\USlabel \,\captionFIGalpha}
\end{figure}
\begin{figure}[ht]
     \centering
     \begin{subfigure}[t]{0.48\textwidth}
         \centering
        \includegraphics[width=\textwidth]{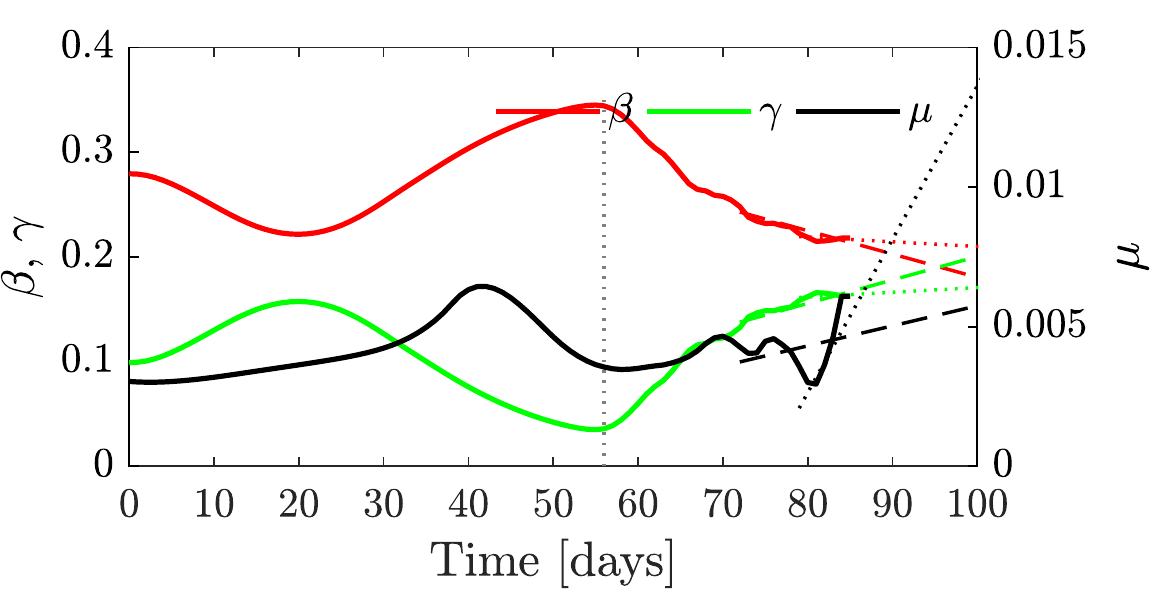}
         \caption{\subcaptionFIGaextrap  The positive slope of the death rate is a consequence of an anomaly in the data on confirmed deaths. The cause of the anomaly is not known to the authors. }
         \label{fig:extrap_alpha_US}
     \end{subfigure}
     \hfill
     \begin{subfigure}[t]{0.48\textwidth}
         \centering
       \includegraphics[width=\textwidth]{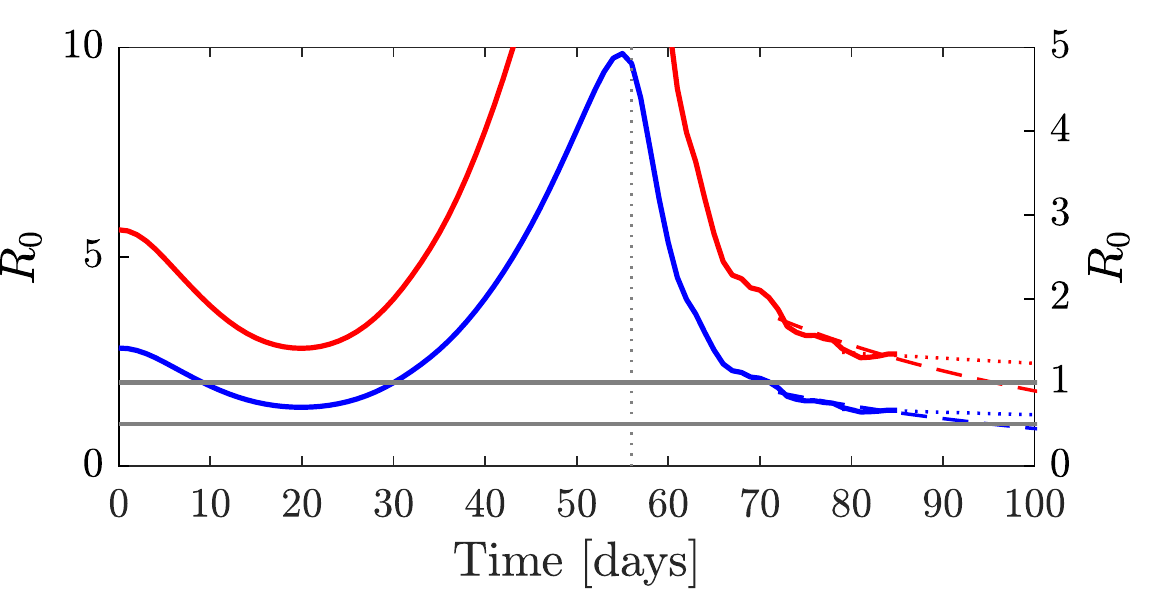}
         \caption{\subcaptionFIGbextrap }
         \label{fig:extrap_R0_US}
     \end{subfigure}
     \caption{\USlabel \,\captionFIGextrap}
\end{figure}

     \begin{figure}[h] 
         \centering
        \includegraphics[width=0.48\textwidth]{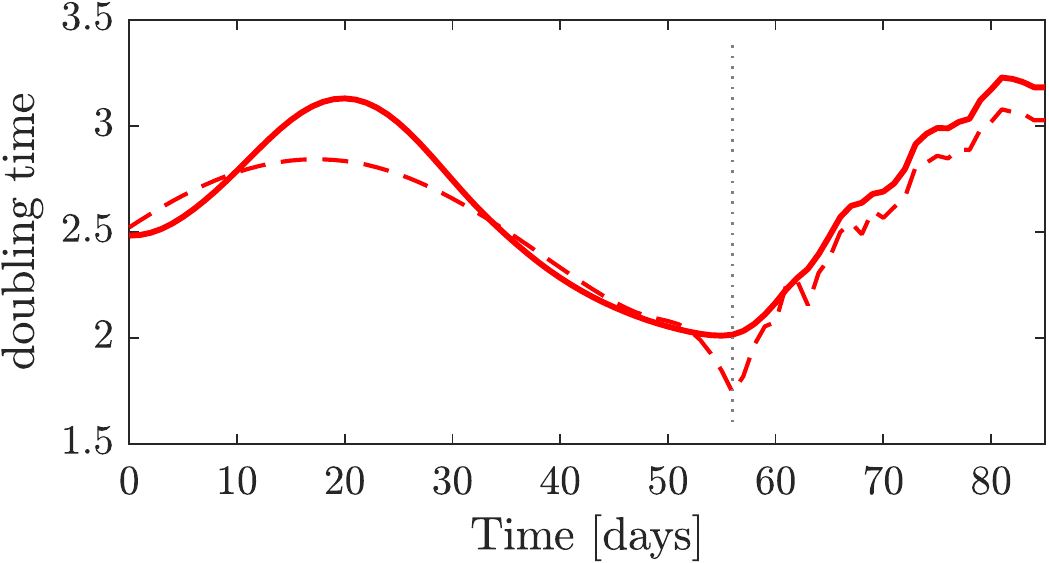}
         \caption{\USlabel \, \captionFIGdoubling }
         \label{fig:doubling_US}
     \end{figure}
     \begin{figure}[h] 
         \centering
        \includegraphics[width=0.48\textwidth]{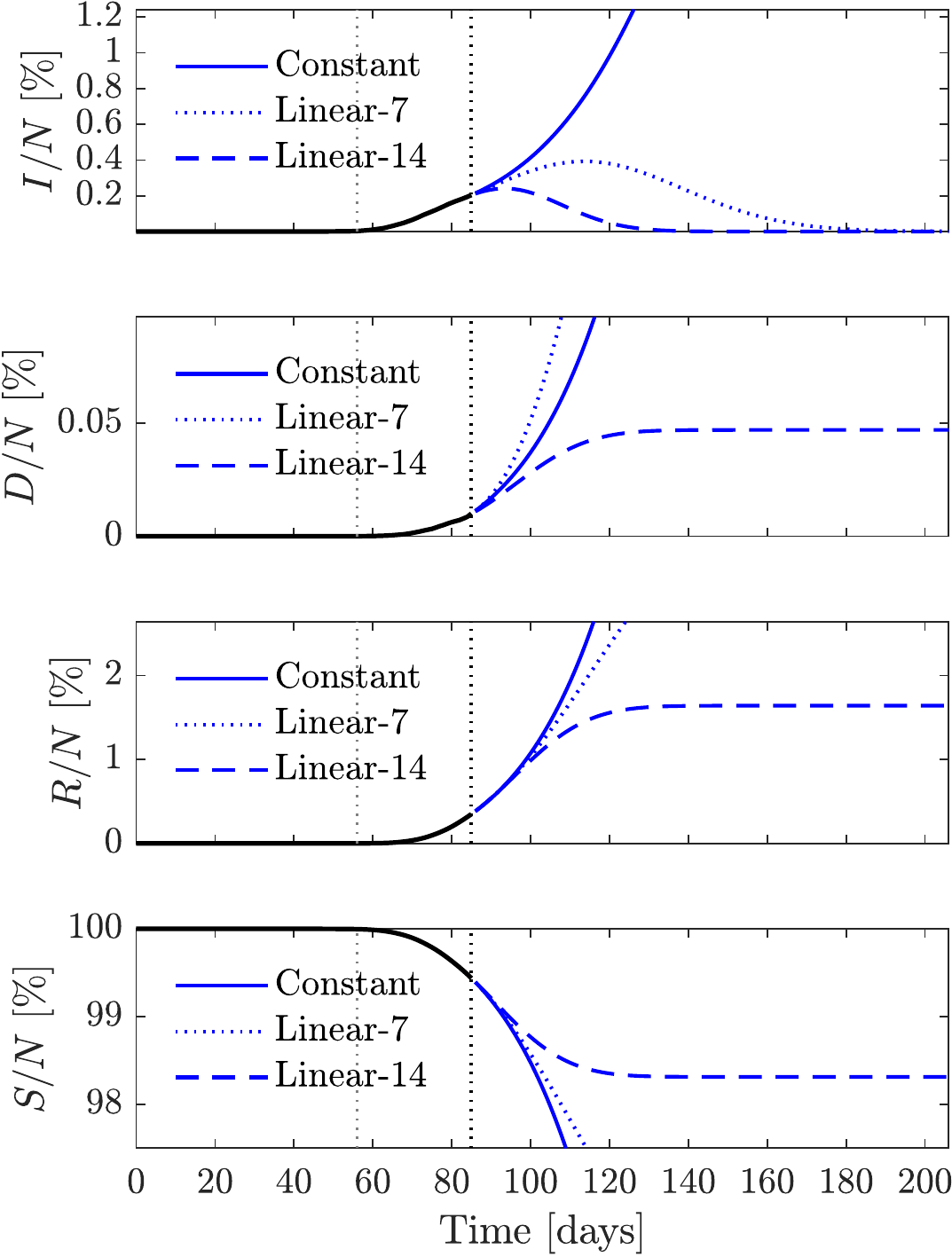}
         \caption{\USlabel \, \captionFIGextraptwo The black solid lines are taken from Fig.~\ref{fig:validation1_US}. }
         \label{fig:extrap_state_US}
     \end{figure}

     \clearpage

\subsection{New York City}

\begin{figure}[ht]
     \centering
     \begin{subfigure}[t]{0.48\textwidth}
         \centering
        \includegraphics[width=\textwidth]{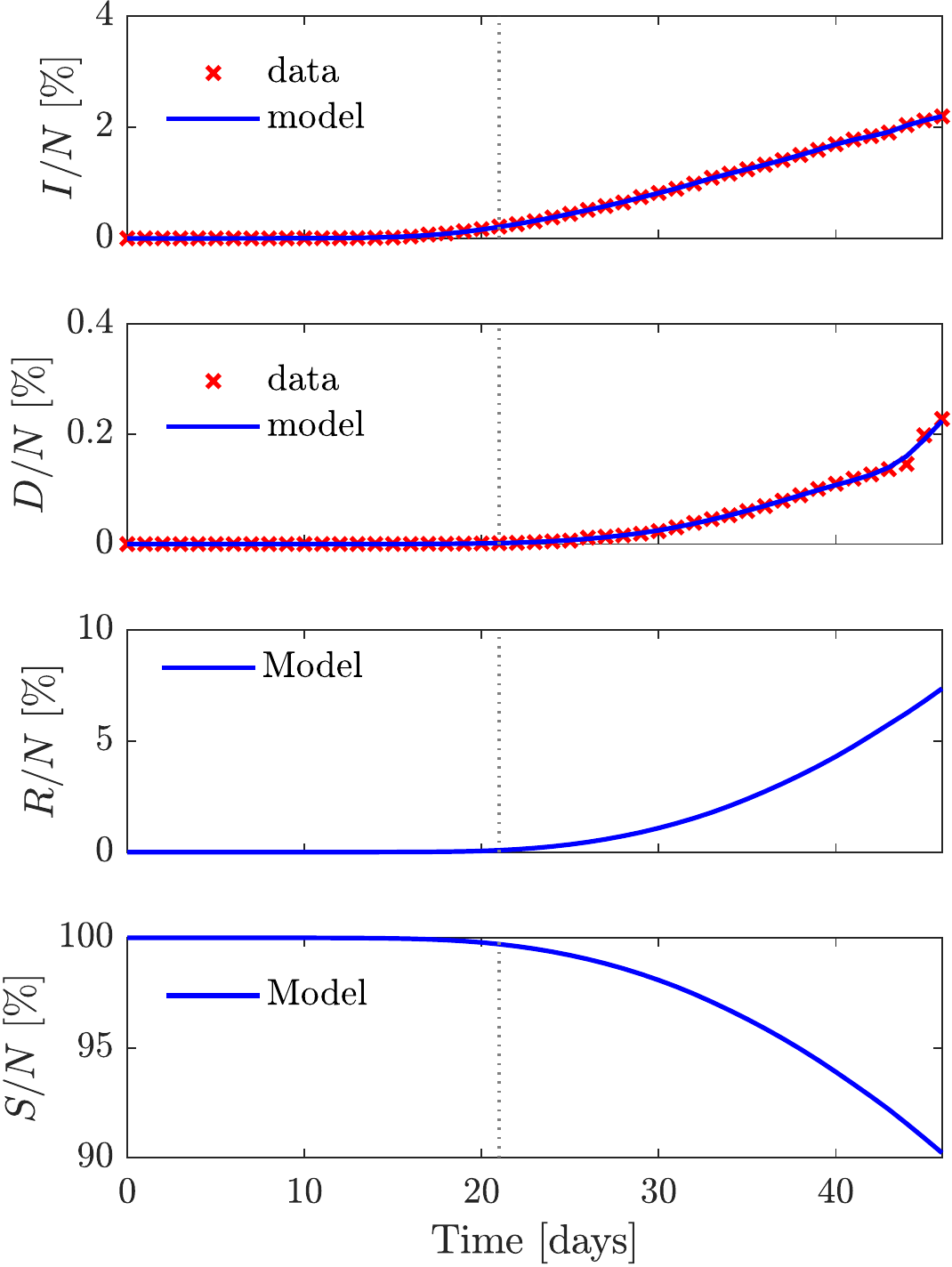}
         \caption{\subcaptionFIGaVAL}
         \label{fig:validation1_NY}
     \end{subfigure}
     \hfill
     \begin{subfigure}[t]{0.48\textwidth}
         \centering
        \includegraphics[width=\textwidth]{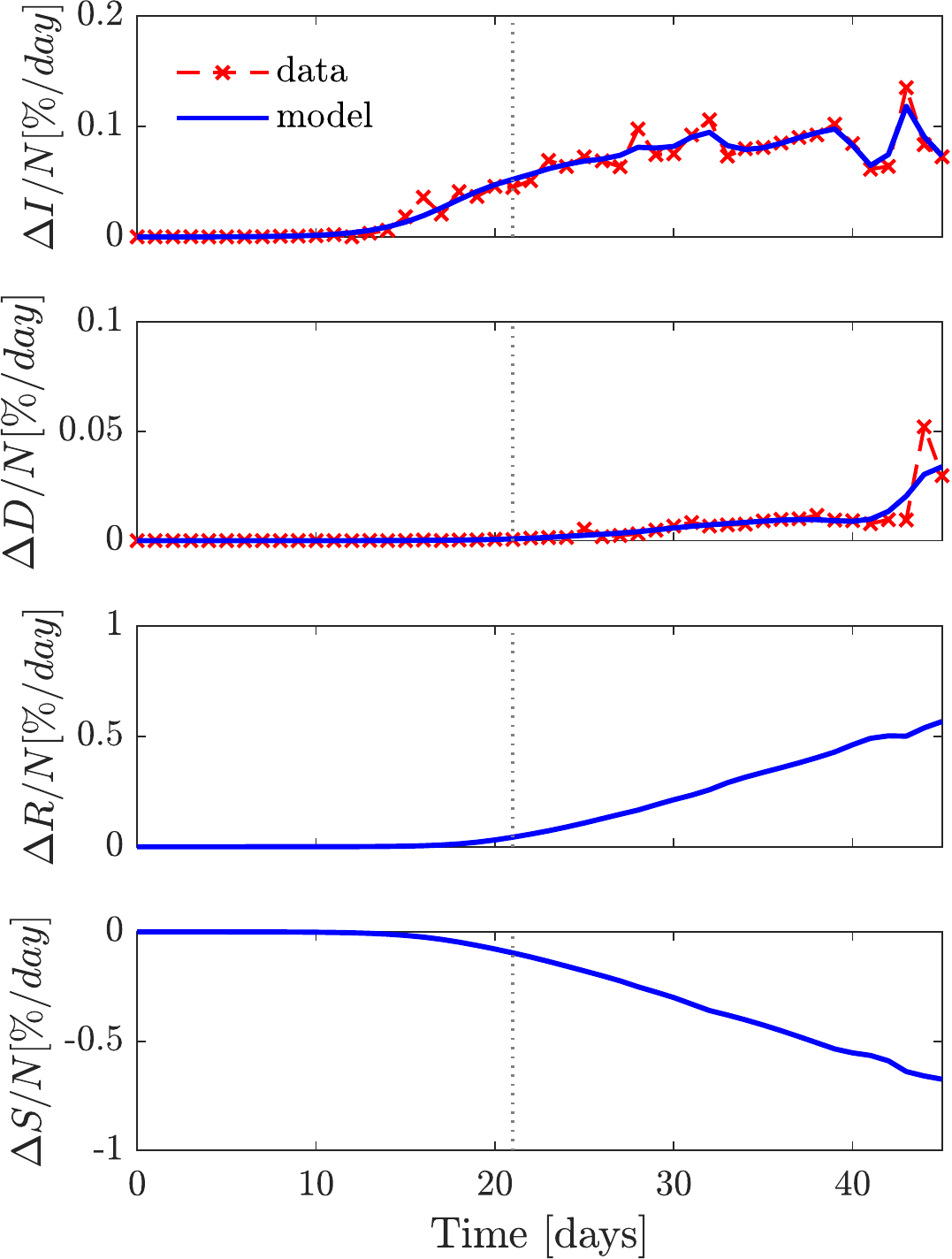}
         \caption{\subcaptionFIGbVAL}
         \label{fig:validation2_NY}
     \end{subfigure}
     \caption{\NYlabel \, \captionFIGVAL}
\end{figure}
%
%
\begin{figure}[ht]
     \centering
     \begin{subfigure}[t]{0.48\textwidth}
         \centering
        \includegraphics[width=\textwidth]{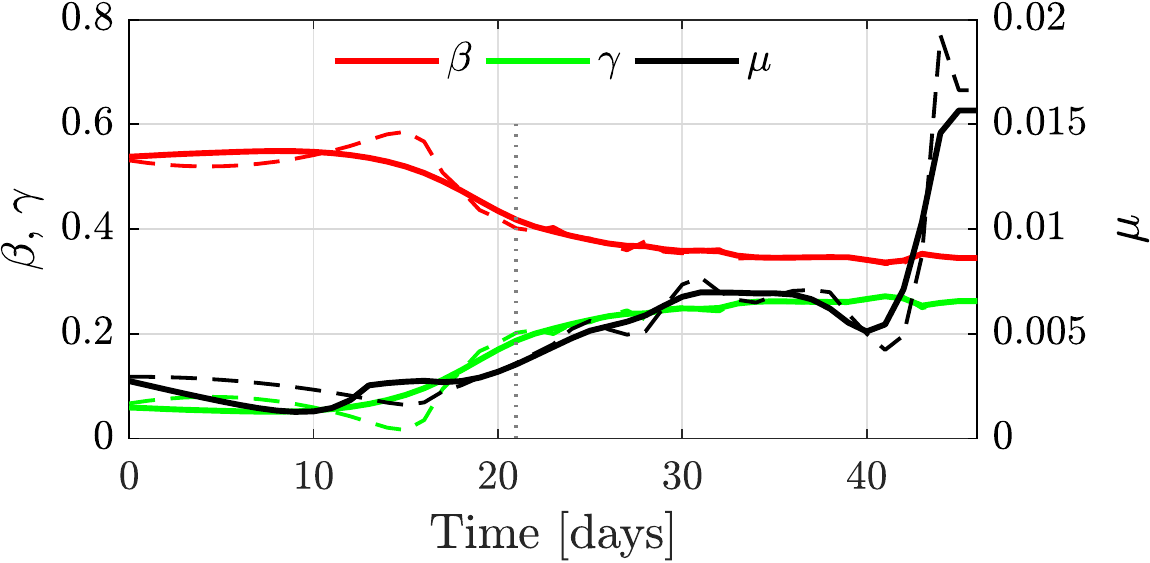}
         \caption{\subcaptionFIGaalpha}
         \label{fig:alpha_NY}
     \end{subfigure}
     \hfill
     \begin{subfigure}[t]{0.48\textwidth}
         \centering
        \includegraphics[width=\textwidth]{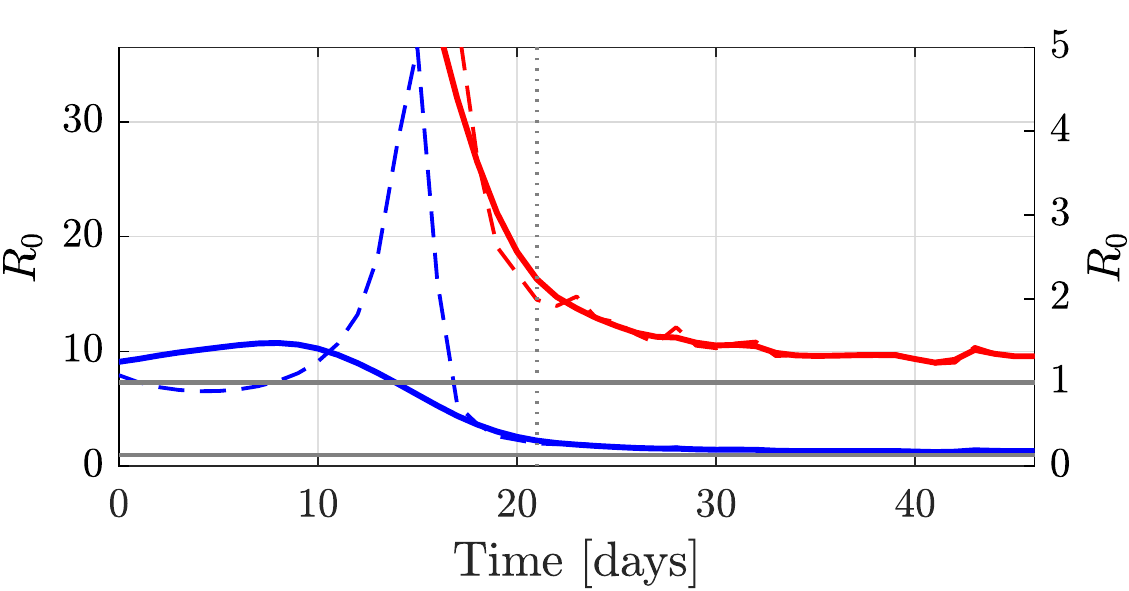}
         \caption{\subcaptionFIGbalpha}
         \label{fig:R0_NY}
     \end{subfigure}
     \caption{\NYlabel \,\captionFIGalpha}
\end{figure}
\begin{figure}[ht]
     \centering
     \begin{subfigure}[t]{0.48\textwidth}
         \centering
        \includegraphics[width=\textwidth]{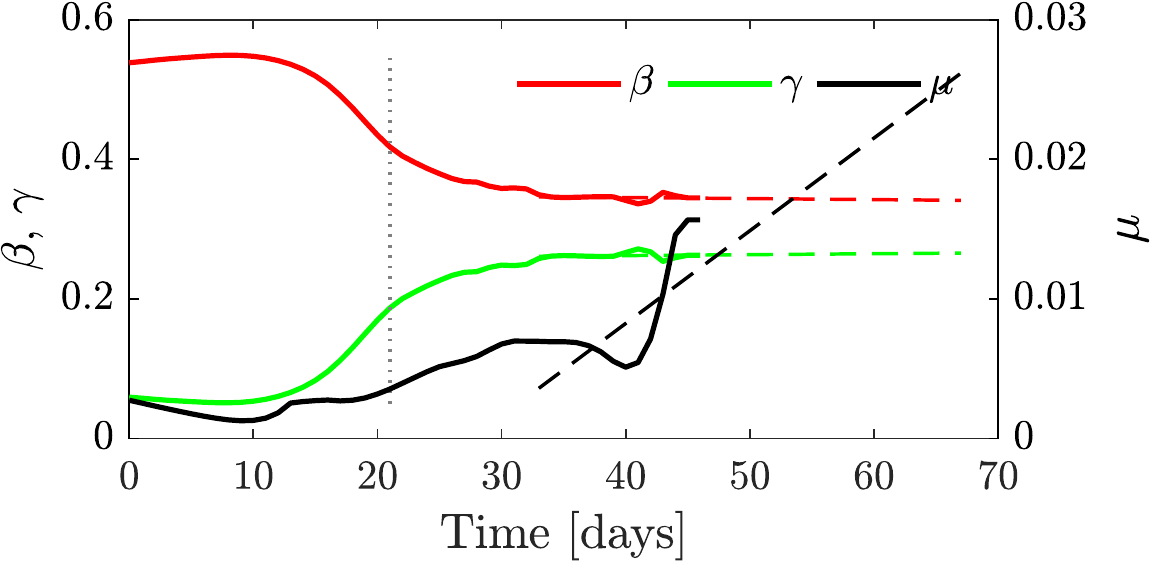}
         \caption{Extrapolated trends of the time-varying contact rate ($\beta$), recovery rate ($\gamma$), and death rate ($\mu$) with average slope over the last fourteen days (dashed lines).  The positive slope of the death rate is a consequence of an anomaly in the data on confirmed deaths. The cause of the anomaly is not known to the authors. }
         \label{fig:extrap_alpha_NY}
     \end{subfigure}
     \hfill
     \begin{subfigure}[t]{0.48\textwidth}
         \centering
        \includegraphics[width=\textwidth]{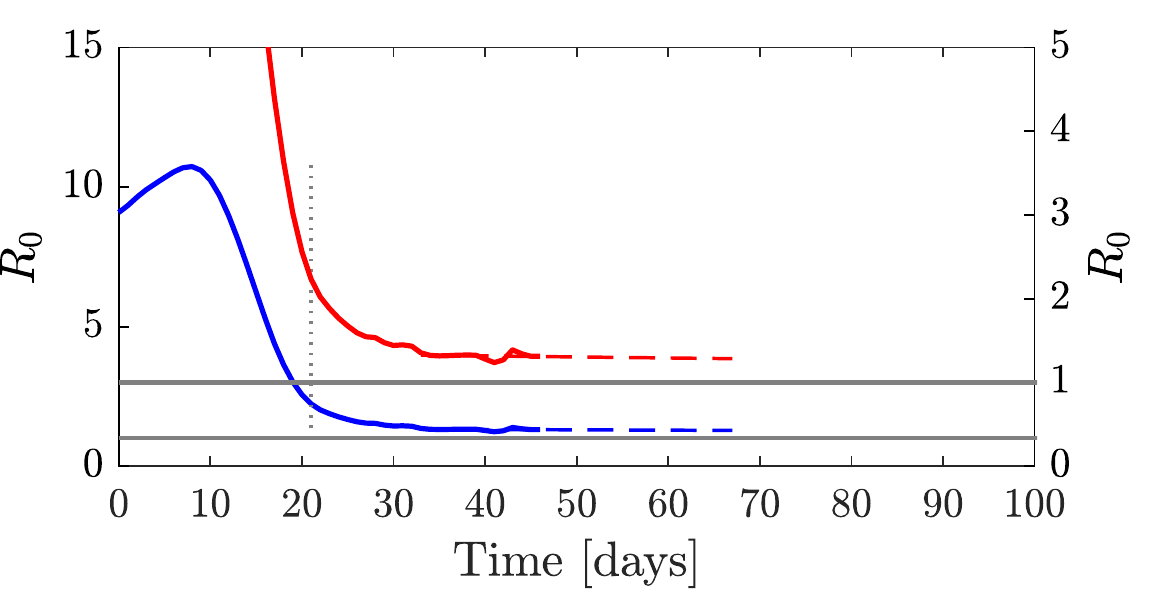}
         \caption{Extrapolated trend of the basic reproduction number with average slope over the last fourteen days (dashed lines).  }
         \label{fig:extrap_R0_NY}
     \end{subfigure}
     \caption{\NYlabel \,\captionFIGextrap}
\end{figure}

     \begin{figure}[h] 
         \centering
        \includegraphics[width=0.48\textwidth]{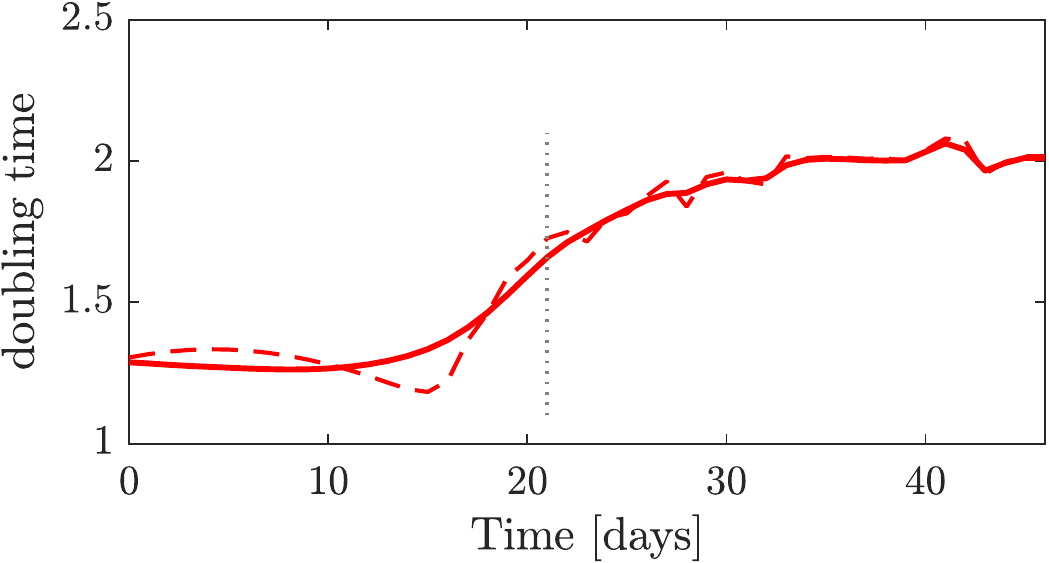}
         \caption{\NYlabel \, \captionFIGdoubling }
         \label{fig:doubling_NY}
     \end{figure}
     \begin{figure}[h] 
         \centering
        \includegraphics[width=0.48\textwidth]{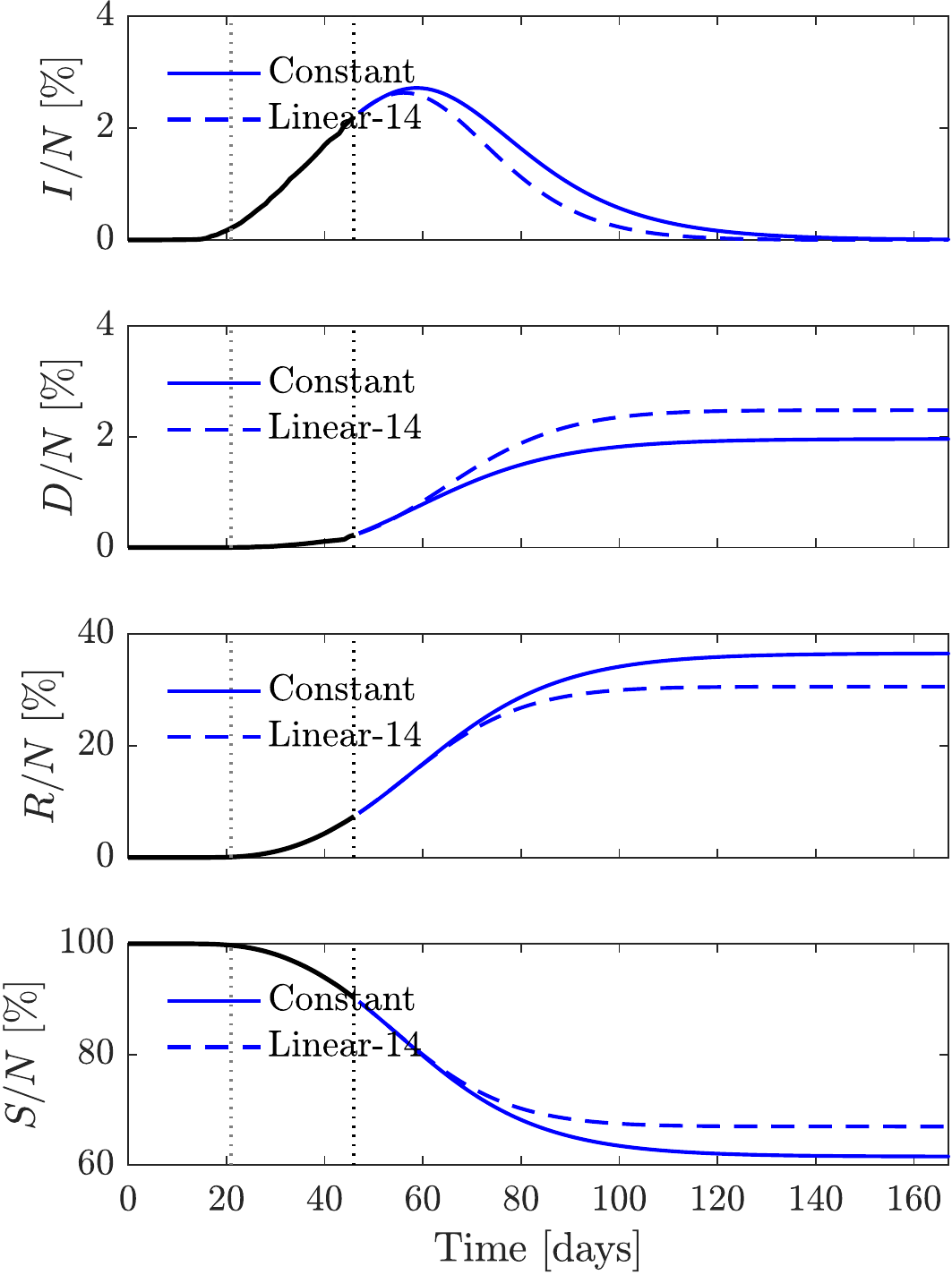}
         \caption{\NYlabel \, From the top: Blue lines  indicate the extrapolated trends of the percentage of infected, recovered, deaths, and susceptible. Estimates with average slope over the last  fourteen days (dashed lines), and with values of the parameters assumed to be constant and equal to the last day (solid lines). Black lines: The left vertical dotted line represents the day of  lockdown, the right vertical line is the last day of the training data set, hence, the starting day for extrapolation.  The black solid lines are taken from Fig.~\ref{fig:validation1_NY}. }
         \label{fig:extrap_state_NY}
     \end{figure}

     \clearpage
\subsection{China}

\begin{figure}[ht]
     \centering
     \begin{subfigure}[t]{0.48\textwidth}
         \centering
        \includegraphics[width=\textwidth]{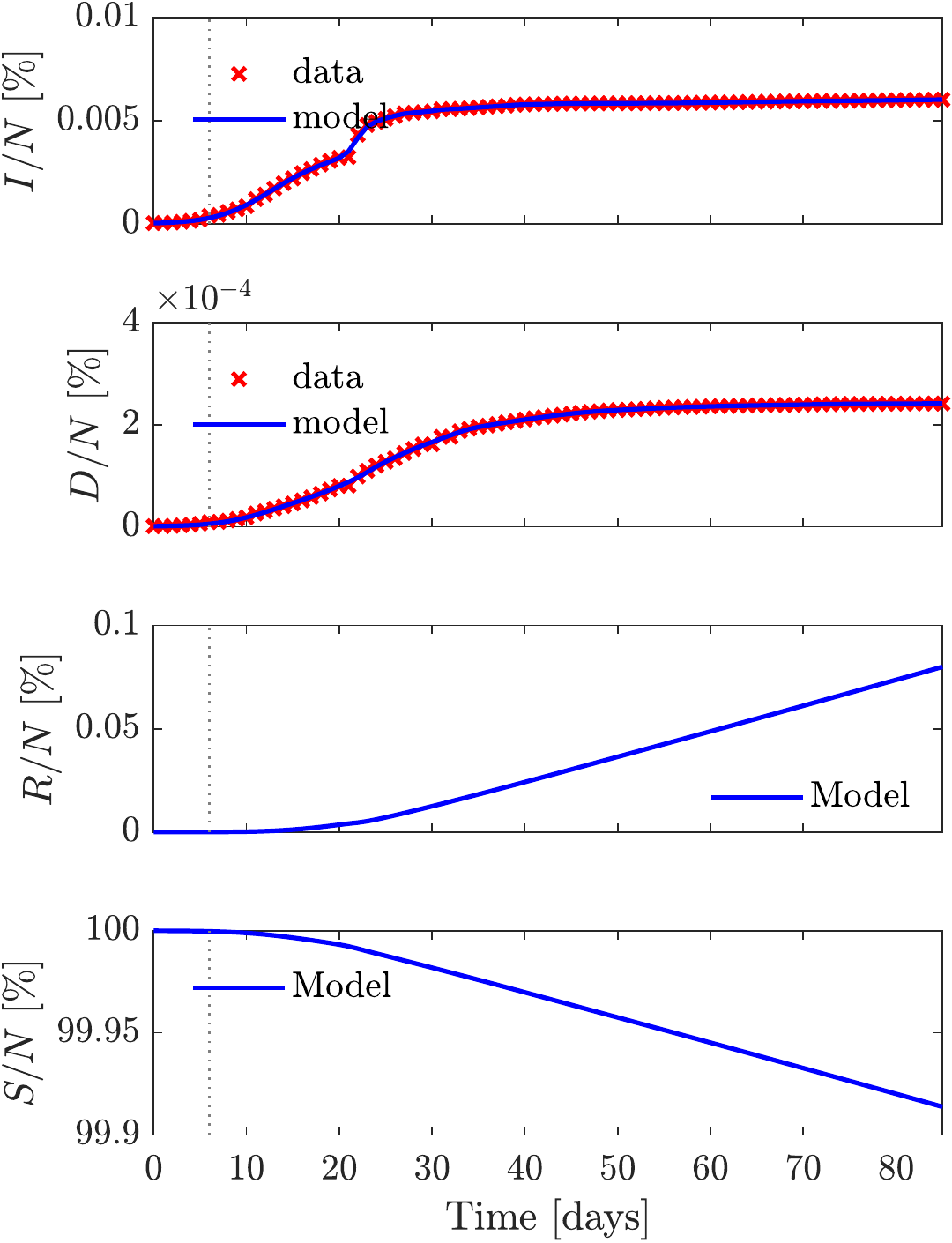}
         \caption{\subcaptionFIGaVAL}
         \label{fig:validation1_China}
     \end{subfigure}
     \hfill
     \begin{subfigure}[t]{0.48\textwidth}
         \centering
        \includegraphics[width=\textwidth]{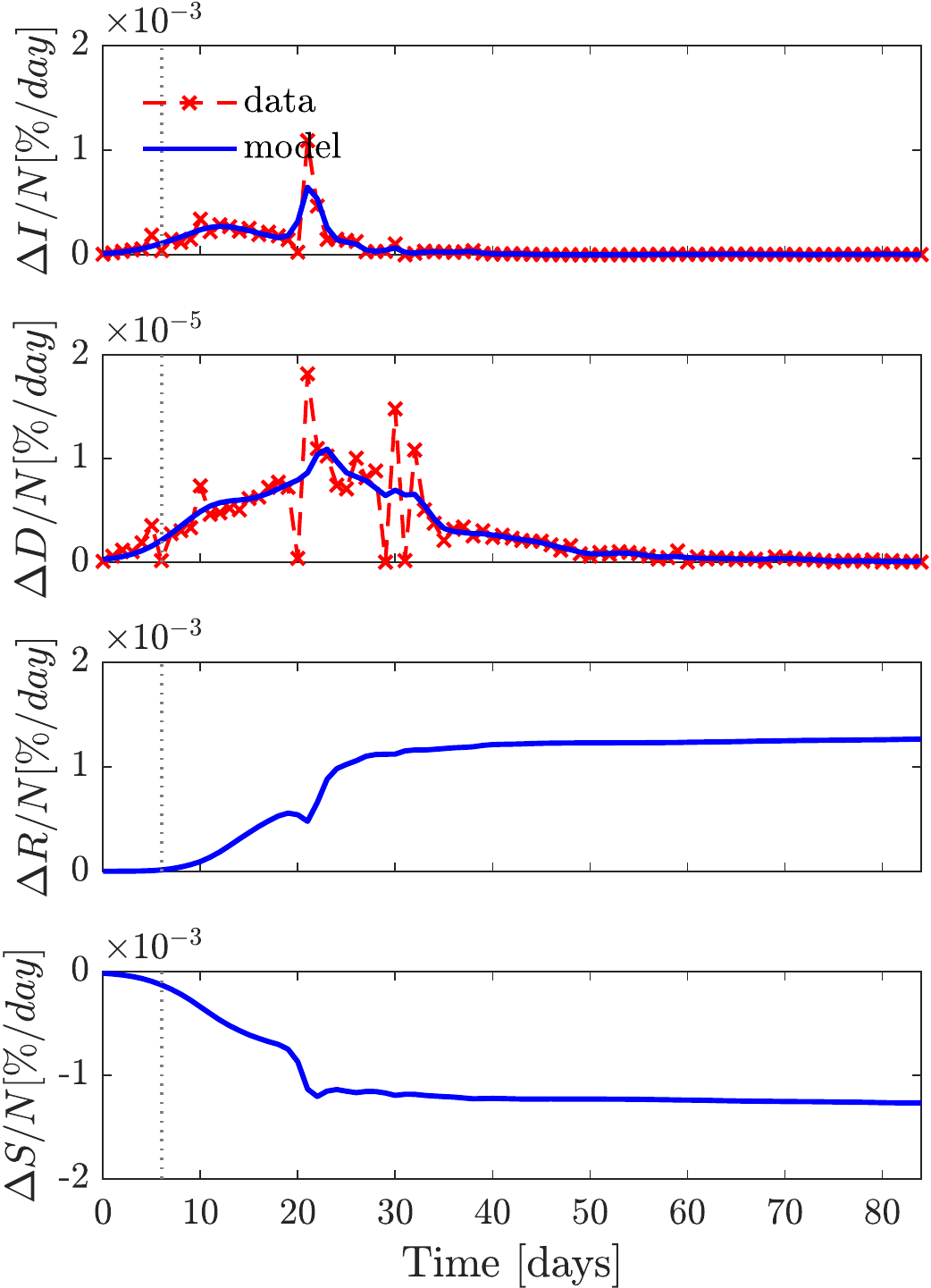}
         \caption{\subcaptionFIGbVAL}
         \label{fig:validation2_China}
     \end{subfigure}
     \caption{\Chinalabel \, \captionFIGVAL}
\end{figure}
%
%
\begin{figure}[ht]
     \centering
     \begin{subfigure}[t]{0.48\textwidth}
         \centering
        \includegraphics[width=\textwidth]{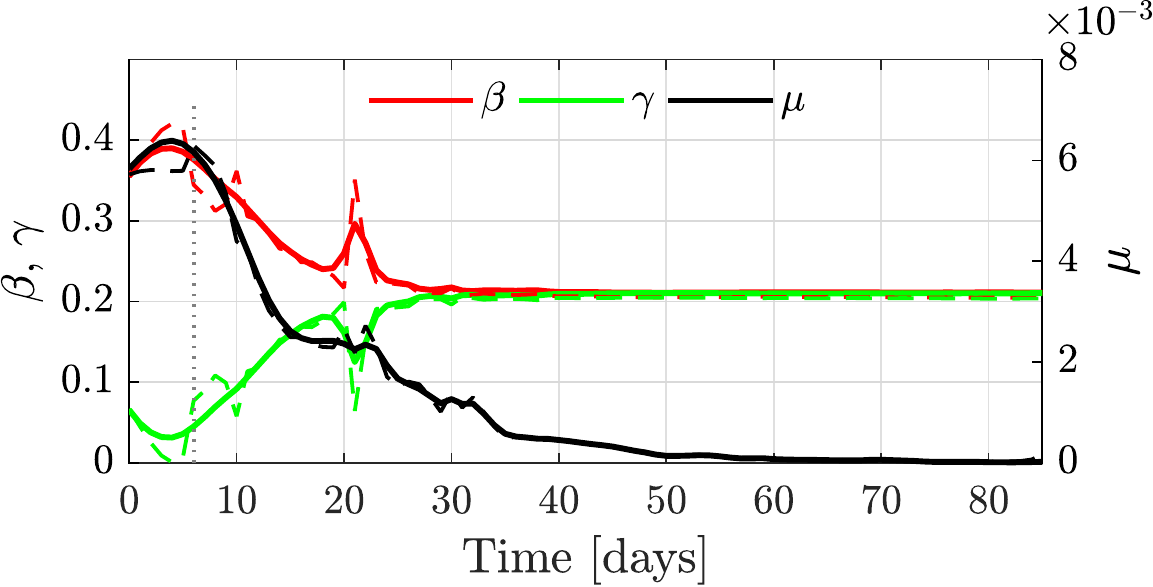}
         \caption{\subcaptionFIGaalpha}
         \label{fig:alpha_China}
     \end{subfigure}
     \hfill
     \begin{subfigure}[t]{0.48\textwidth}
         \centering
        \includegraphics[width=\textwidth]{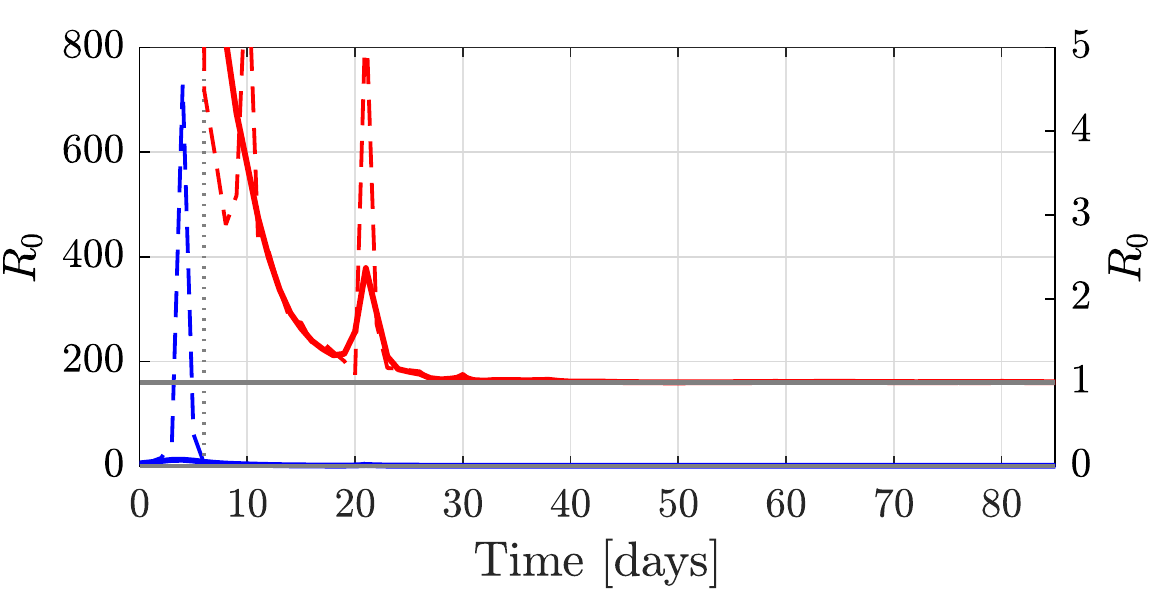}
         \caption{\subcaptionFIGbalpha}
         \label{fig:R0_China}
     \end{subfigure}
     \caption{\Chinalabel \,\captionFIGalpha}
\end{figure}
\begin{figure}[ht]
     \centering
     \begin{subfigure}[t]{0.48\textwidth}
         \centering
        \includegraphics[width=\textwidth]{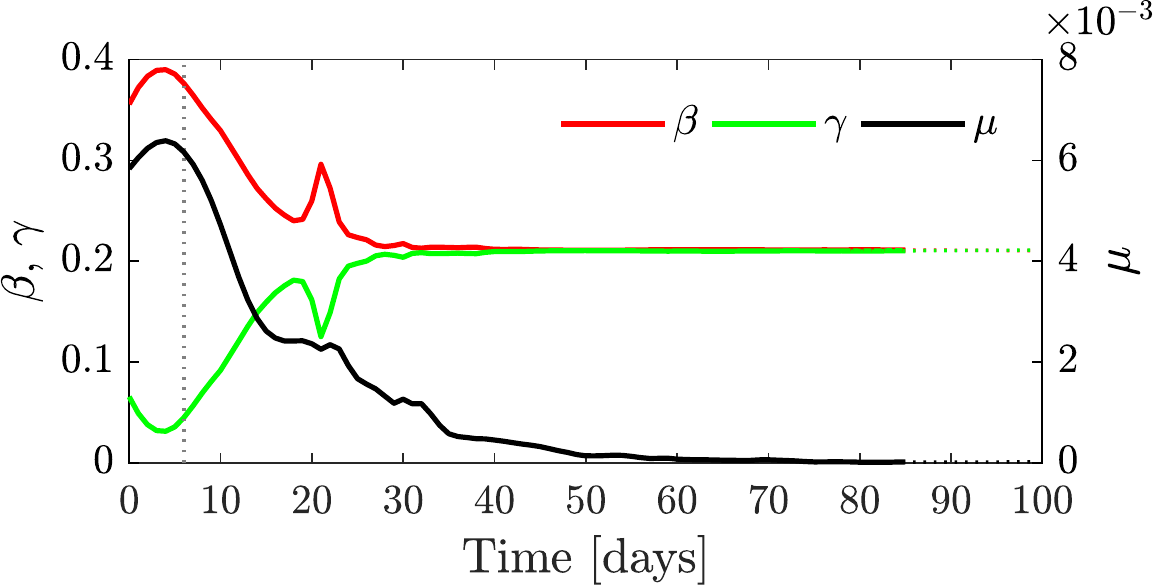}
         \caption{Extrapolated trends of the time-varying contact rate ($\beta$), recovery rate ($\gamma$), and death rate ($\mu$) with average slope over the last seven days (dotted lines). }
         \label{fig:extrap_alpha_China}
     \end{subfigure}
     \hfill
     \begin{subfigure}[t]{0.48\textwidth}
         \centering
        \includegraphics[width=\textwidth]{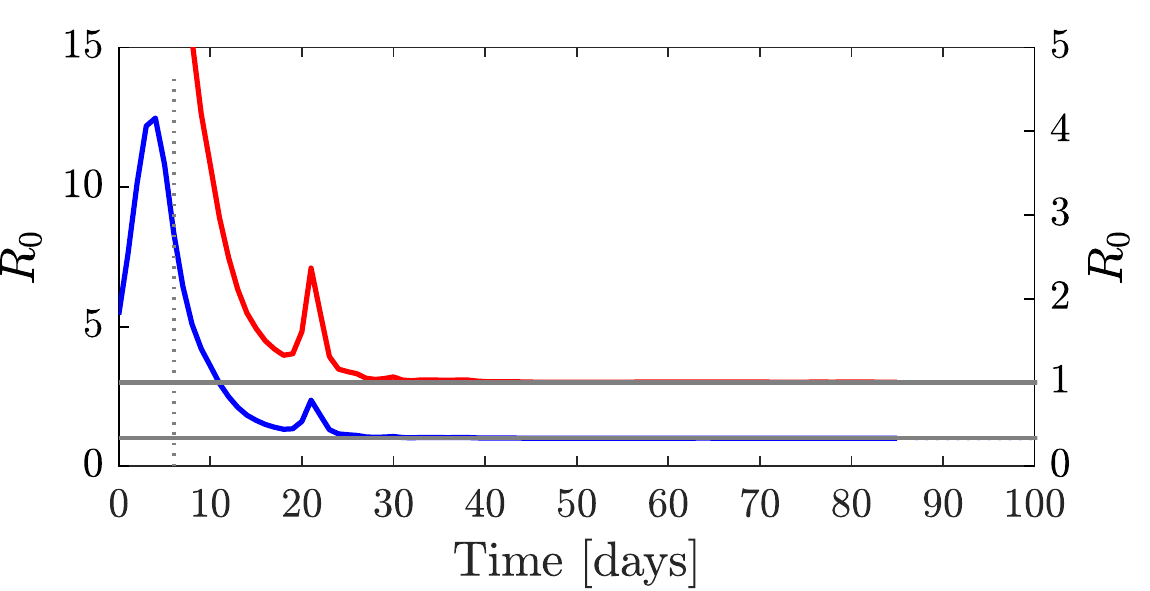}
         \caption{Extrapolated trend of the basic reproduction number with average slope over the last seven days (dotted lines). }
         \label{fig:extrap_R0_China}
     \end{subfigure}
     \caption{\Chinalabel \,\captionFIGextrap}
\end{figure}

     \begin{figure}[h] 
         \centering
        \includegraphics[width=0.48\textwidth]{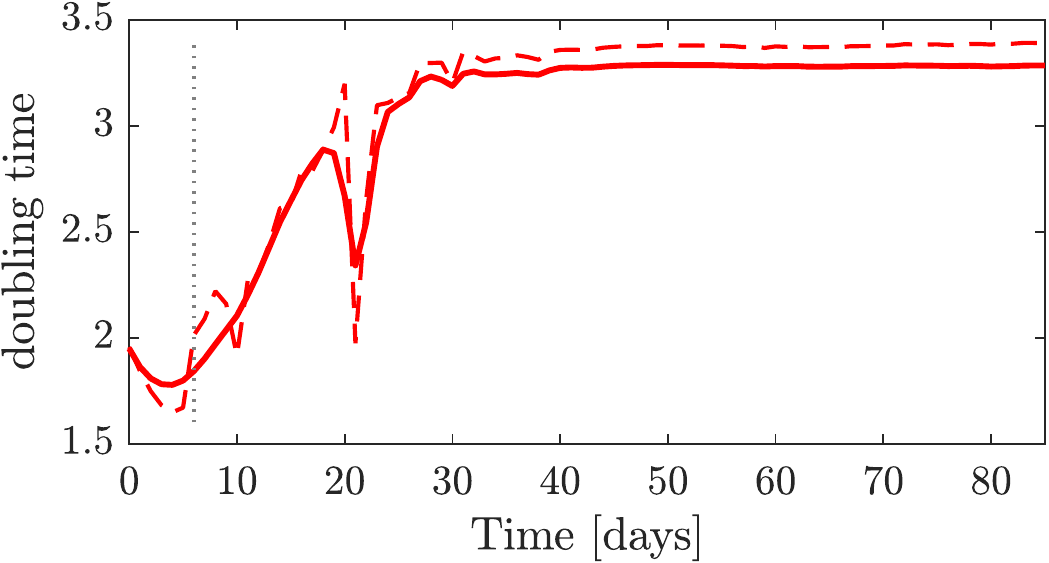}
         \caption{\Chinalabel \, \captionFIGdoubling }
         \label{fig:doubling_China}
     \end{figure}
     \begin{figure}[h] 
         \centering
        \includegraphics[width=0.48\textwidth]{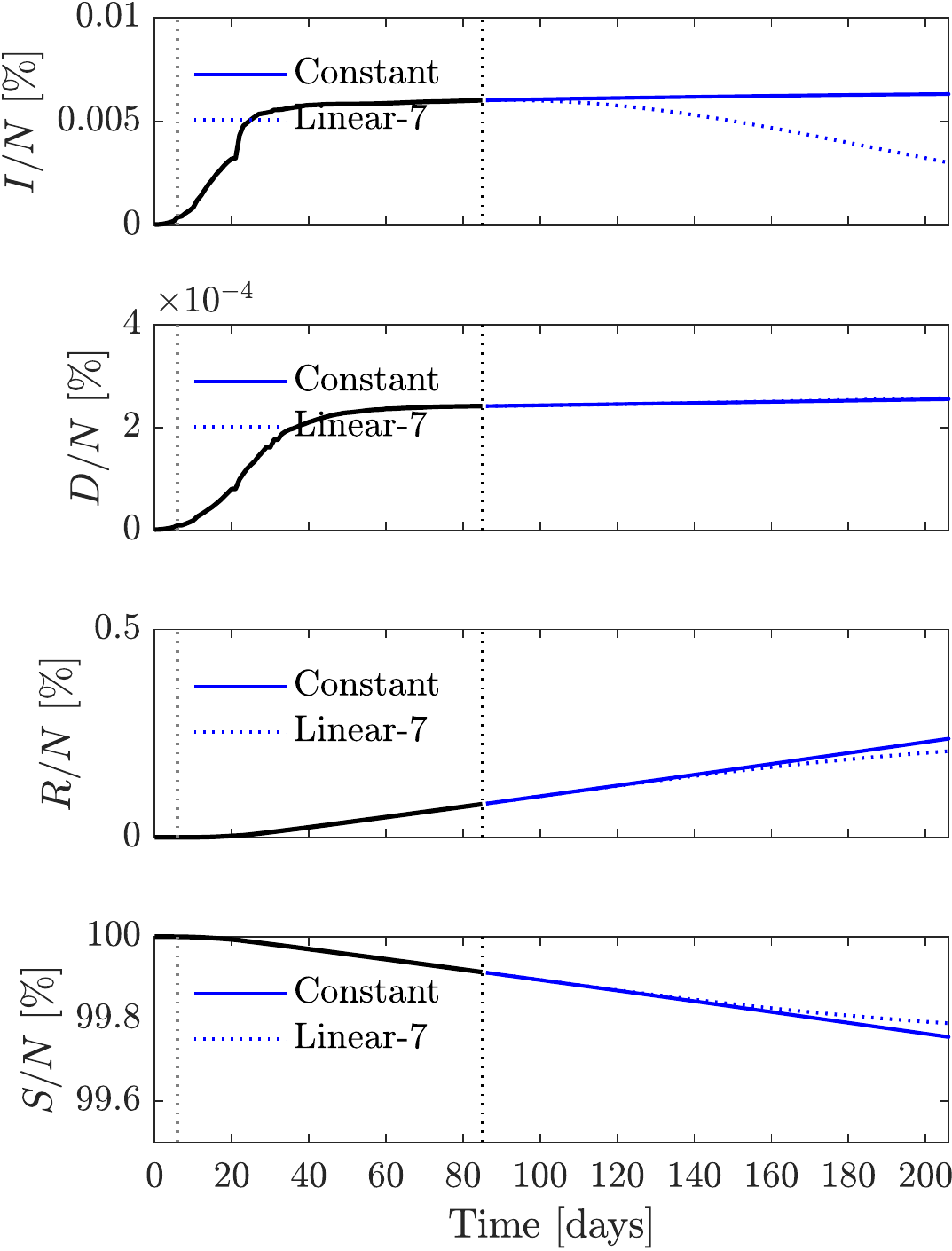}
         \caption{\Chinalabel \, From the top: Blue lines  indicate the extrapolated trends of the percentage of infected, recovered, deaths, and susceptible. Estimates with average slope over the last seven days (dotted lines)  and with values of the parameters assumed to be constant and equal to the last day (solid lines). Black lines: The left vertical dotted line represents the day of  lockdown, the right vertical line is the last day of the training data set, hence, the starting day for extrapolation. The black solid lines are taken from Fig.~\ref{fig:validation1_China}. }
         \label{fig:extrap_state_China}
     \end{figure}

     \clearpage

\subsection{World}

\begin{figure}[ht]
     \centering
     \begin{subfigure}[t]{0.48\textwidth}
         \centering
        \includegraphics[width=\textwidth]{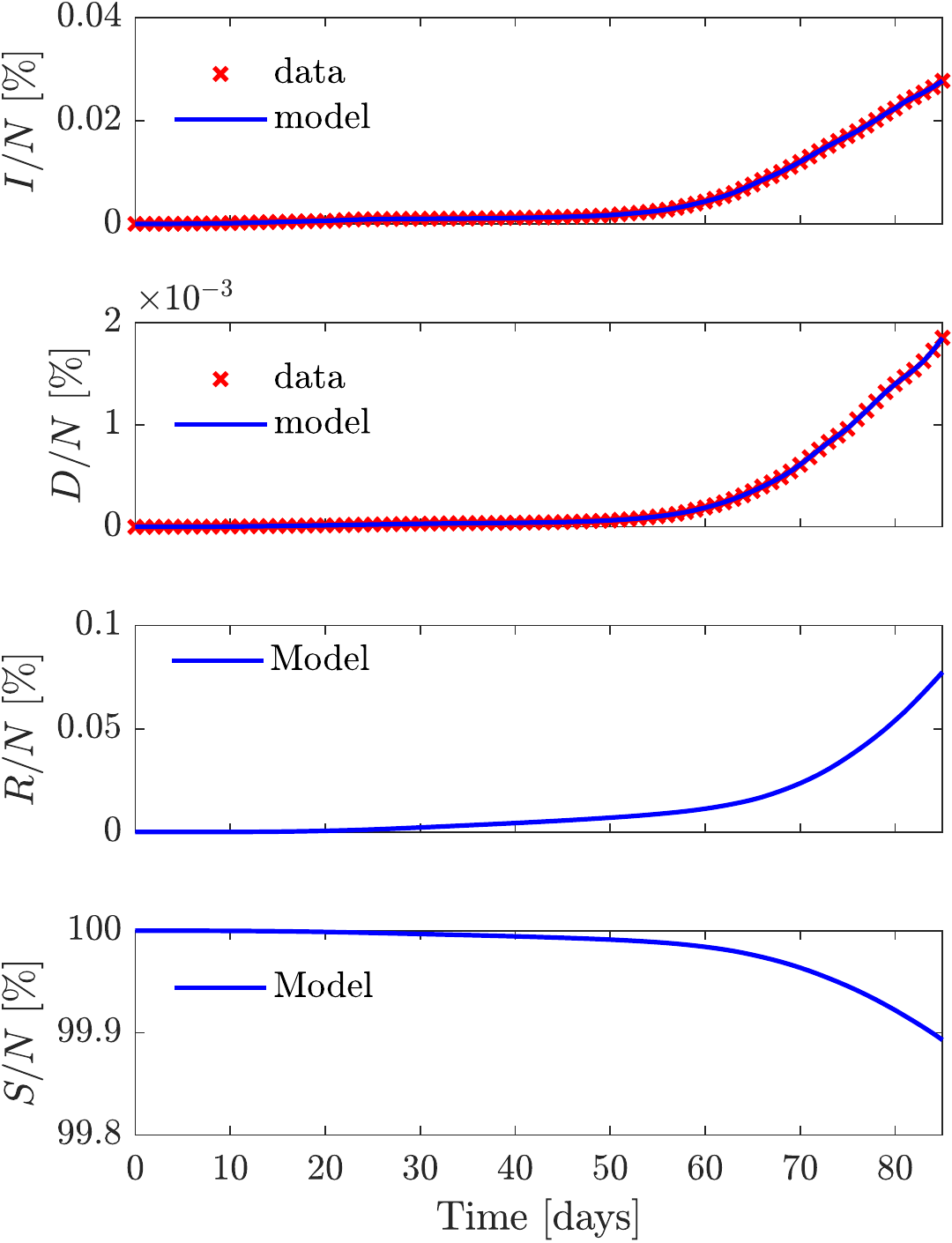}
         \caption{\subcaptionFIGaVAL}
         \label{fig:validation1_World}
     \end{subfigure}
     \hfill
     \begin{subfigure}[t]{0.48\textwidth}
         \centering
        \includegraphics[width=\textwidth]{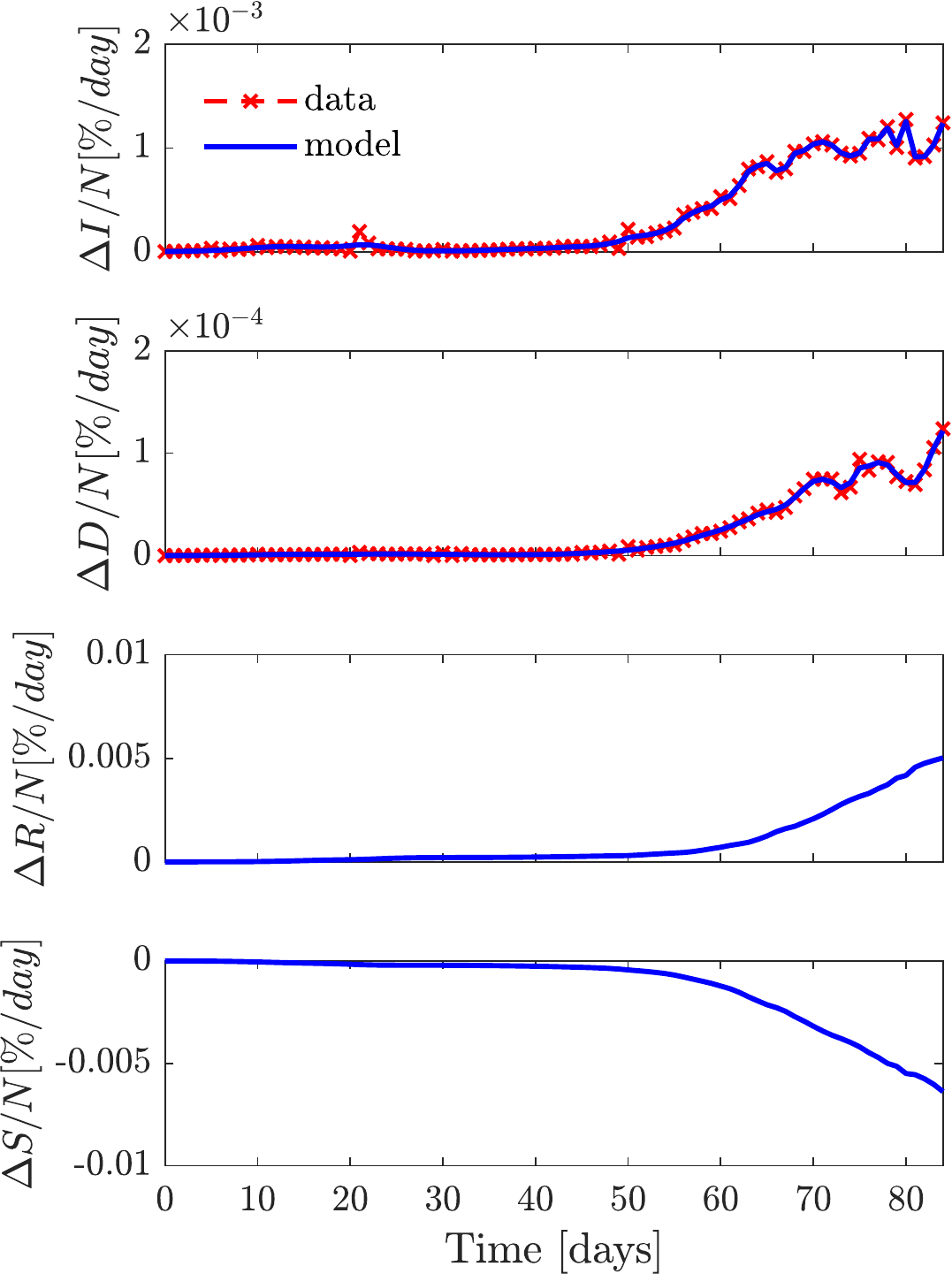}
         \caption{\subcaptionFIGbVAL}
         \label{fig:validation2_World}
     \end{subfigure}
     \caption{\Worldlabel \, First and second rows: Validation of first-principles machine learning epidemic modelling. Third and fourth rows: Inference of recovered and susceptible. }
\end{figure}
%
%
\begin{figure}[ht]
     \centering
     \begin{subfigure}[t]{0.48\textwidth}
         \centering
        \includegraphics[width=\textwidth]{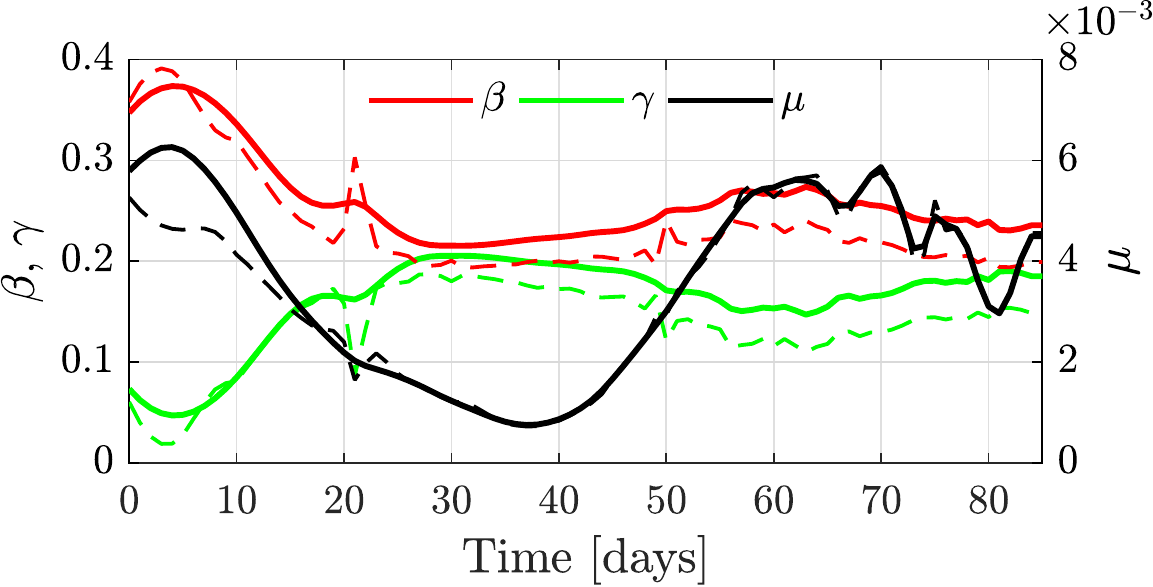}
         \caption{\subcaptionFIGaalpha}
         \label{fig:alpha_World}
     \end{subfigure}
     \hfill
     \begin{subfigure}[t]{0.48\textwidth}
         \centering
        \includegraphics[width=\textwidth]{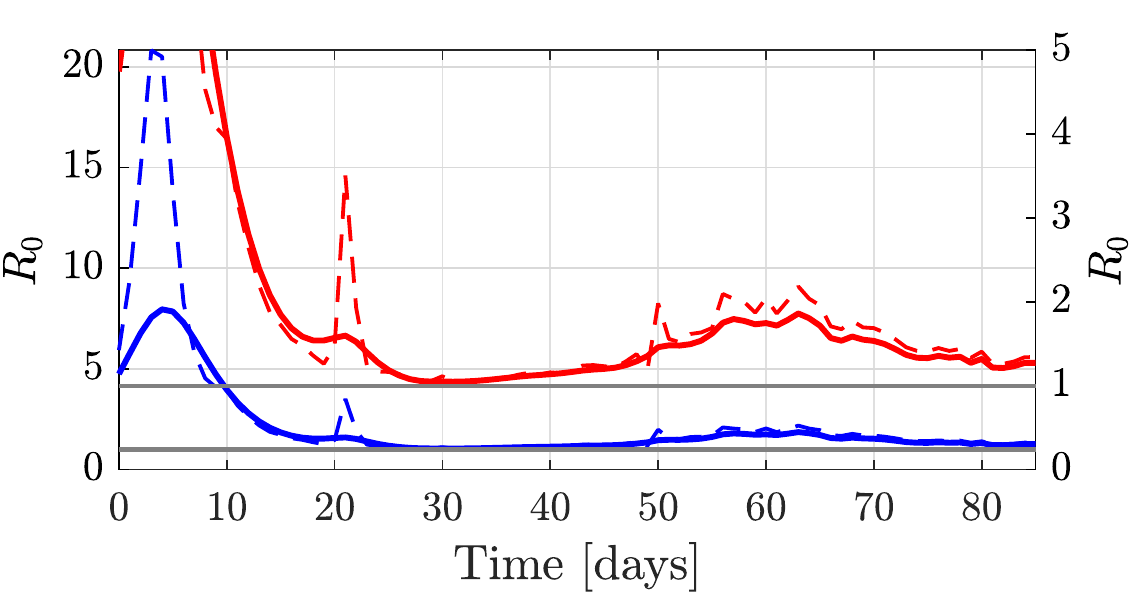}
         \caption{\subcaptionFIGbalpha}
         \label{fig:R0_World}
     \end{subfigure}
     \caption{\Worldlabel \,SIRD parameters. Neural network trained with the $\log$ (solid line) and without the $\log$ (dashed line) in the loss function~\eqref{eq:f30rjfl2}. }
\end{figure}
\begin{figure}[ht]
     \centering
     \begin{subfigure}[t]{0.48\textwidth}
         \centering
        \includegraphics[width=\textwidth]{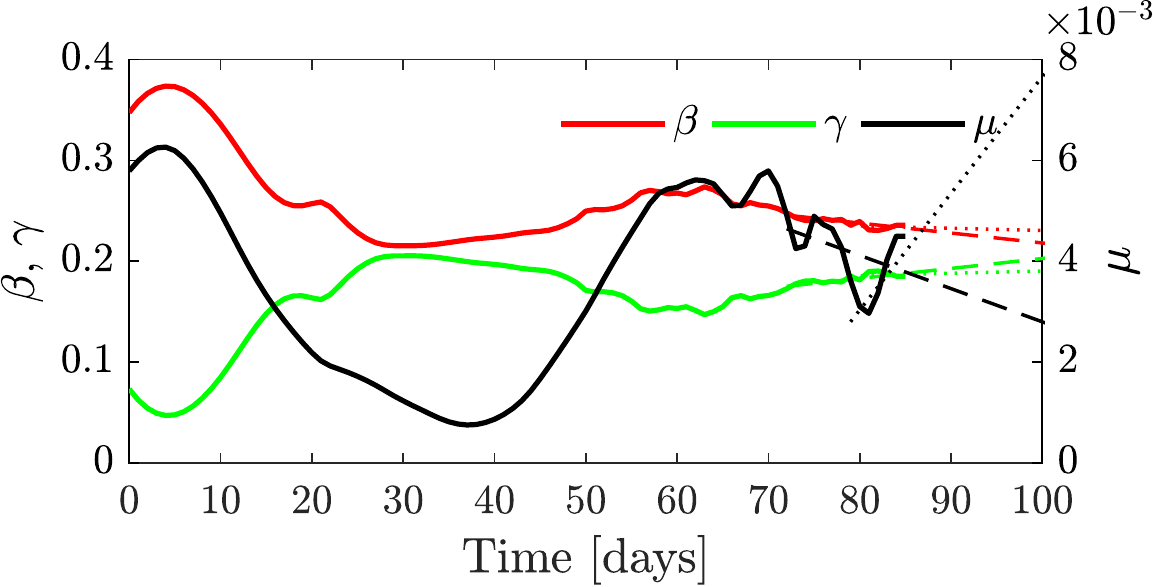}
         \caption{\subcaptionFIGaextrap   }
         \label{fig:extrap_alpha_World}
     \end{subfigure}
     \hfill
     \begin{subfigure}[t]{0.48\textwidth}
         \centering
       \includegraphics[width=\textwidth]{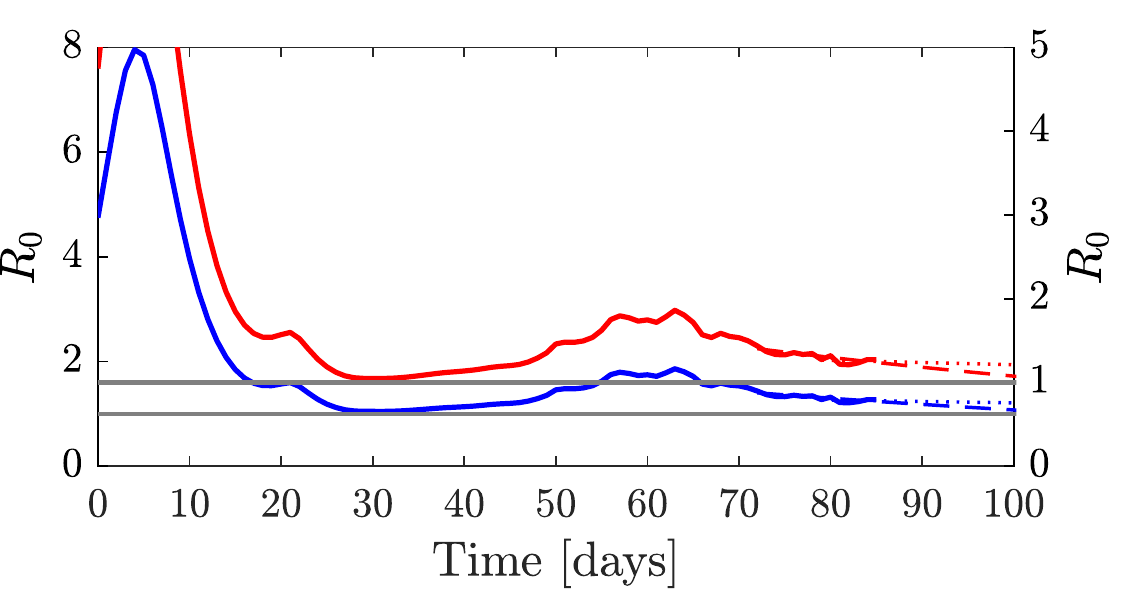}
         \caption{\subcaptionFIGbextrap }
         \label{fig:extrap_R0_World}
     \end{subfigure}
     \caption{\Worldlabel \,Extrapolated trends of the SIRD parameters. }
\end{figure}

     \begin{figure}[h] 
         \centering
        \includegraphics[width=0.48\textwidth]{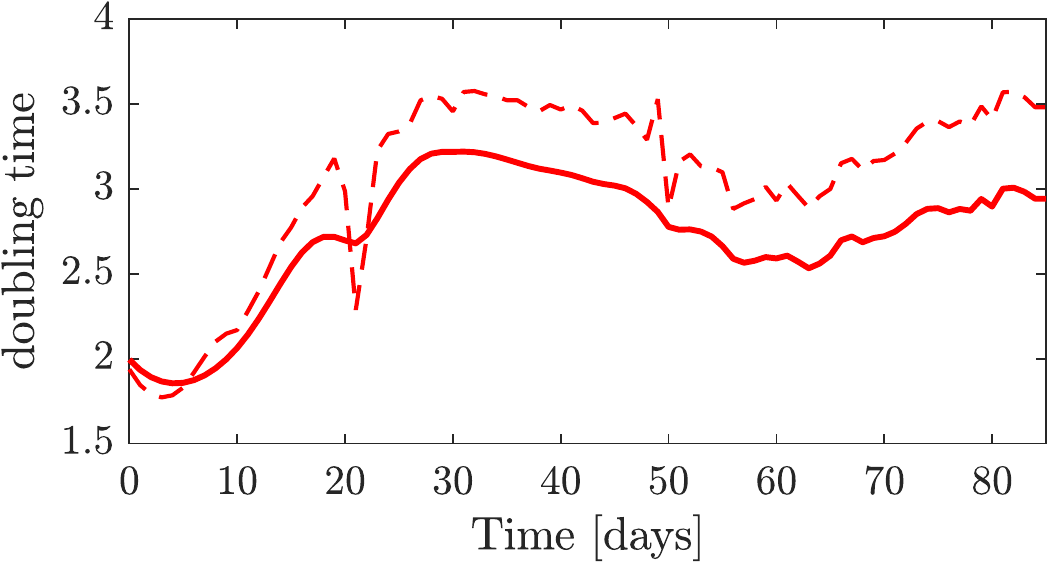}
         \caption{\Worldlabel \, \captionFIGdoubling }
         \label{fig:doubling_World}
     \end{figure}
     \begin{figure}[h] 
         \centering
        \includegraphics[width=0.48\textwidth]{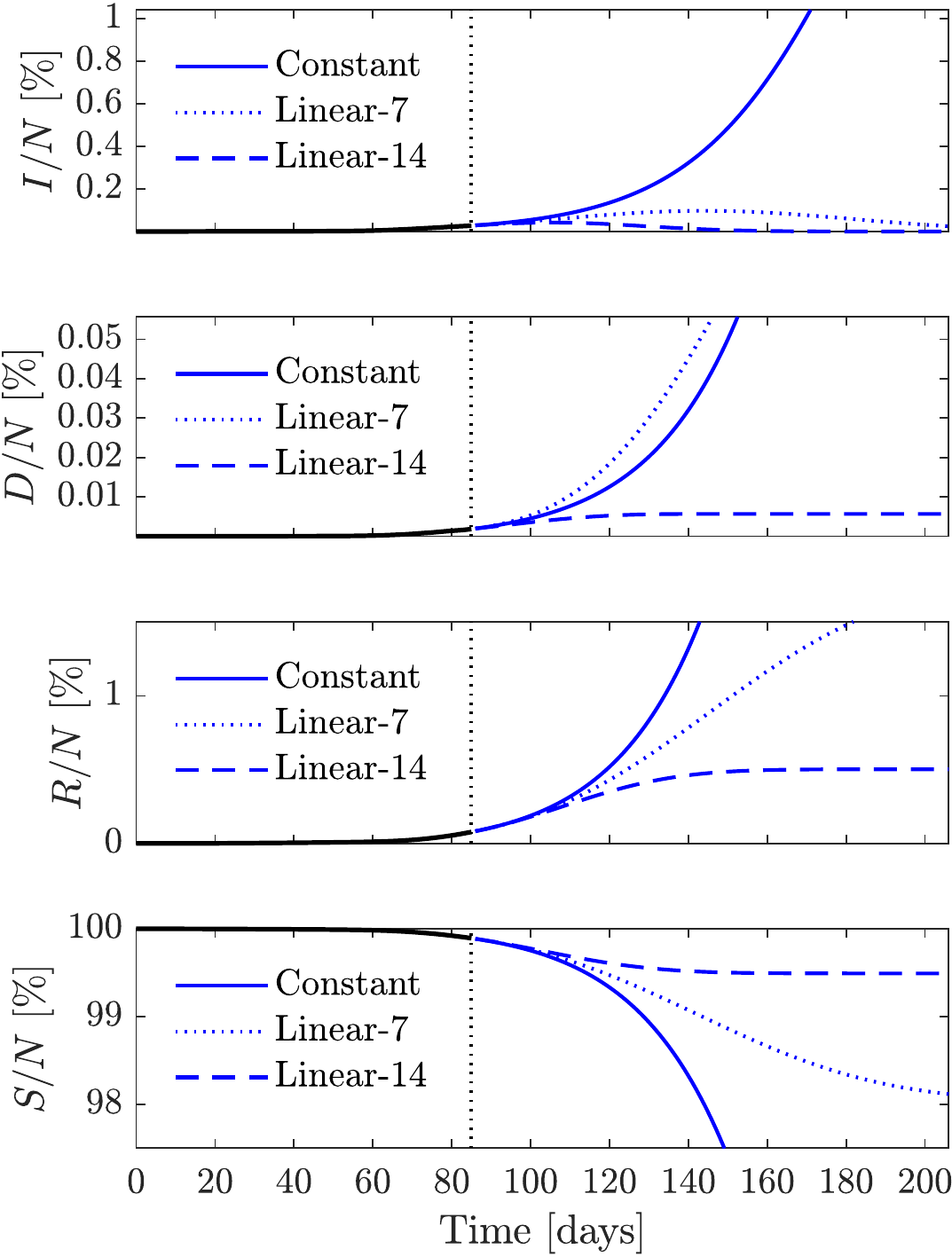}
         \caption{\Worldlabel \, From the top: Blue lines  indicate the extrapolated trends of the percentage of infected, recovered, deaths, and susceptible. Estimates with average slope over the last seven days (dotted lines),  fourteen days (dashed lines), and with values of the parameters assumed to be constant and equal to the last day (solid lines). Black lines: The vertical line is the last day of the training data set, hence, the starting day for extrapolation.  The black solid lines are taken from Fig.~\ref{fig:validation1_World}. }
         \label{fig:extrap_state_World}
     \end{figure}

     \clearpage


\setlength{\tabcolsep}{10pt}
\renewcommand{\arraystretch}{1.5}
\begin{table}[!ht]
    \centering
    \begin{tabular}{l || c  c | c  c | c }
Country & $I_{max}$ [\%]& & $D_{max}$ [\%]& &$R_0=1$ \\ &&&&&
  \\ \hline \hline 
United Kingdom&2.8415&18/07/2020&2.1331&10/02/2021&- \\
&0.18959&26/04/2020&0.031601&02/05/2020&25/04/2020\\
&0.19187&26/04/2020&0.028537&29/04/2020&25/04/2020\\ \hline
Italy&0.79865&11/07/2020&0.41861&10/02/2021&-\\
&0.31652&30/04/2020&0.068895&02/07/2020&04/05/2020\\
&0.32581&03/05/2020&0.046895&08/05/2020 &06/05/2020\\\hline
Germany&0.72874&04/08/2020&0.26889&10/02/2021&-\\
&0.17774&25/04/2020&0.037317&01/11/2020&26/04/2020\\
&0.17436&23/04/2020&0.016698&08/10/2020&23/04/2020\\\hline
France&6.9024&24/06/2020&2.9703&10/02/2021&-\\
&0.25882&23/04/2020&0.053117&12/06/2020&22/04/2020\\
&0.50498&15/05/2020&0.047987&07/05/2020&16/05/2020\\\hline
Spain&2.064&27/06/2020&0.64766&10/02/2021&-\\
&0.87499&06/06/2020&0.050432&02/05/2020&21/07/2020\\
&0.48603&03/05/2020&0.046109&26/04/2020&05/05/2020\\\hline
Belgium&2.9745&01/07/2020&2.6482&10/02/2021&-\\
&1.7878&21/06/2020&0.13973&24/05/2020&19/08/2020\\
&0.41385&02/05/2020&0.16903&17/07/2020&06/05/2020\\\hline
USA&2.8442&08/07/2020&1.5018&10/02/2021&-\\
&0.3925&16/05/2020&0.54629&22/12/2020&12/06/2020\\
&0.24041&25/04/2020&0.04707&27/08/2020&26/04/2020\\\hline
New York City&2.7161&30/04/2020&1.9625&10/02/2021&-\\
&-&-&-&-&-\\
&2.6298&27/04/2020&2.4807&16/12/2020&21/02/2021\\\hline
China&0.0063773&06/11/2020&0.00027578&10/02/2021&-\\
&0.0060284&23/04/2020&0.0002657&10/02/2021&23/04/2020\\
&0.0089155&10/02/2021&0.0002416&22/04/2020&-\\\hline
World&2.053&20/08/2020&0.86118&10/02/2021&-\\
&0.097413&15/06/2020&0.16569&10/02/2021&29/07/2020\\
&0.042022&08/05/2020&0.005671&13/06/2020&08/05/2020\\
    \end{tabular}
        \caption{First column: Countries analysed. Second and third columns: Estimated maximum percentages of infected ($I_{max}$) and deaths ($D_{max}$) with dates. Fourth column: Estimate date on which the basic reproduction number becomes unity. For each country, the first / second / third row reports the estimate based on the extrapolation with constant parameters / linear parameters with average slope over a short window / linear parameters with average slope over a long window. The three different extrapolations provide an estimate of the range where the actual value lies. The results are consistent with the first principles and working assumptions of the SIRD model and the data (Sec.~\ref{sec:ffrijmrfi4}).}
    \label{tab:results_peaks}    
\end{table}

\clearpage

\bibliographystyle{unsrt}
\bibliography{MyCollection}
\end{document}